\documentclass[12pt,a4paper,dvips]{article}
\usepackage{calc,xspace} \usepackage{cite,mcite}
\usepackage{rotating,graphicx} \usepackage{physics}
\usepackage{rotating}
\usepackage{a4p} \usepackage{physics}
\usepackage{thisl3_titlePH,ifthen,Lep}

\DeclareGraphicsExtensions{.eps,.ps,.eps.gz,.ps.gz}
\journalname{European Physical Journal C}
\date{October 3, 2005}
\preprint{2005-044}
\Lep{2}

\newlength{\capindent}
\newlength{\capwidth}
\newlength{\figwidth}
\newcommand{\icaption}[2][!*!,!]{\hspace*{\capindent}%
  \begin{minipage}{\capwidth}
    \ifthenelse{\equal{#1}{!*!,!}}%
      {\caption{#2}}%
      {\caption[#1]{#2}}
  \end{minipage}}
\newcommand{\ff}{\ensuremath{f\kern 0.15em\overline{\kern -0.25em f}}}

\newcommand{\bitem}{\begin{itemize}}
\newcommand{\eitem}{\end{itemize}}

\newcommand{\SM}{Standard Model}

\newcommand{\Nsel}{\ensuremath{N_\mathrm{sel}}}% number of selected events
\newcommand{\Nf}{\ensuremath{N_\mathrm{f}}}% number of forward  events
\newcommand{\Nb}{\ensuremath{N_\mathrm{b}}}% number of backward events

\newcommand{\AfbSM}{\ensuremath{A_\mathrm{fb}^\mathrm{SM}}}

\newcommand{\sigsm}{\ensuremath{\sigma_{\mathrm{SM}}}}

\newcommand{\CoM}{centre-of-mass}
\newcommand{\sqrtsp}{\ensuremath{\sqrt{s^{\prime}}}}
\newcommand{\mumu}{\ensuremath{\mu^{+}\mu^{-}}}
\newcommand{\cost}{\ensuremath{\cos\theta}}
\newcommand{\minv}{\ensuremath{m_{\ffbar}}}
\newcommand{\epsdif}{\ensuremath{\varepsilon(\cost,\minv)}}

\begin{document}
       
%%%%%%%%%%%%%%%%%%%%%%%%%%%%%%%%%%%%%%%%%%%%%%%%%%%%%%%%%%%%%%%%%%%%%%%%%%%%%%%%
\begin{titlepage}
%%%%%%%%%%%%%%%%%%%%%%%%%%%%%%%%%%%%%%%%%%%%%%%%%%%%%%%%%%%%%%%%%%%%%%%%%%%%%%%%
%
  \title{\boldmath Measurement of Hadron and Lepton-Pair Production \\
    in $\epem$ Collisions at $\sqrt{s}=192-208\GeV$ at LEP}
\author{{\Large L3 Collaboration}}
\begin{abstract}
Hadron production and lepton-pair production in $\epem$ collisions are
studied with data collected with the L3 detector at LEP at
centre-of-mass energies $\sqrt{s}=192-208\GeV$.  Using a total
integrated luminosity of 453~\pb, 36057 hadronic events and 12863
lepton-pair events are selected. The cross sections for hadron
production and lepton-pair production are measured for the full sample
and for events where no high-energy initial-state-radiation photon is
emitted prior to the collisions. Lepton-pair events are further
investigated and forward-backward asymmetries are measured. Finally,
the differential cross sections for electron-positron pair-production
is determined as a function of the scattering angle. An overall good
agreement is found with Standard Model predictions.
\vspace{1cm}

\centerline{\it{Dedicated to the memory of Dr.\,Stephan Wynhoff}}

\end{abstract}
\submitted
\end{titlepage}

%%%%%%%%%%%%%%%%%%%%%%%%%%%%%%%%%%%%%%%%%%%%%%%%%%%%%%%%%%%%%%%%%%%%%%%%%%%%%%%%
\section{Introduction}
%%%%%%%%%%%%%%%%%%%%%%%%%%%%%%%%%%%%%%%%%%%%%%%%%%%%%%%%%%%%%%%%%%%%%%%%%%%%%%%%

The study of fermion-pair production in $\epem$ collisions constitutes
an important part of the LEP scientific program. It allows a test of the
Standard Model of electroweak interactions~\cite{SM} at energies never
achieved before. At the same time, the large rates of these processes
and the simplicity of the final states provide a useful resource to control
detector performance and calibration. In addition, fermion pairs
constitute an irreducible background for many measurements and for the
search for new physics beyond the Standard Model. Therefore, its production
mechanism must be studied and controlled. Finally, LEP
explores a new energy range above the Z resonance and possible
deviations of fermion-pair production measurements from their precise
theoretical expectations could give access to effects of new physics
beyond the Standard Model at a scale too large to be directly observed.

This paper describes the study of fermion-pair production through
the processes:
\begin{displaymath}
 \epem\rightarrow\mbox{hadrons}\,(\gamma)\, , ~~~~
 \epem\rightarrow\mumu(\gamma)\, , ~~~~
 \epem\rightarrow\tautau(\gamma)\,\,\, {\rm and}~~
 \epem\rightarrow\epem(\gamma)\, ,
\end{displaymath}
where the symbol $(\gamma)$ indicates the possible presence of
additional photons. These reactions proceed through $s$-channel
$\epem$ annihilation mediated by a photon or a Z boson. The
$\epem\ra\epem(\gamma)$ process receives additional contributions from
$t$-channel exchange amplitudes, which increase for decreasing
scattering angles, $\theta$. The scattering angle is defined as the
angle between the directions of the incoming electron and the outgoing
fermion.

The L3 collaboration studied fermion-pair production at the Z
resonance~\cite{l3-202} and for centre-of-mass energies $\sqrt{s} =
130 \GeV - 189 \GeV$~\cite{l3-90+l3-117,l3-196}. This paper extends
these studies to the high-energy and high-luminosity data sample
collected at LEP at $\sqrt{s} = 191.6 \GeV - 209.2 \GeV$,
corresponding to an integrated luminosity of 453~\pb.  The study of
the $\epem\rightarrow\nu\bar{\nu}(\gamma)$ process is discussed in
Reference~\citen{l3nunu}.  Measurements of hadron and lepton-pair
production above the Z resonance were also performed by the other LEP
collaborations~\cite{others}.

For a substantial fraction of the events, initial-state-radiation
(ISR) photons lower the initial \CoM\ energy to an \emph{effective}
\CoM\ energy, \sqrtsp.  The case $\sqrtsp\approx m_{\rm Z}$, where
$m_{\rm Z}$ is the mass of the Z boson, is called \emph{radiative
return to the Z}.  The value of $s^{\prime}$ can be computed from the
sum of the energies of all ISR photons, $E_{\gamma}$, and of their
momentum vectors, $\vec{P_\gamma}$, as:
\begin{equation} \label{eq:spri}
  s^{\prime} \, = \, s - 2 E_\gamma\sqrt{s} + E_{\gamma}^2 - \vec{
  P_\gamma}^2 \, .
\end{equation}

Events from fermion-pair production are divided into two categories:
\emph{inclusive} events and \emph{high-energy} events.  The former
include radiative return to
the Z.  The latter comprises events with small ISR effects, where
$\sqrt{s^\prime}\approx\sqrt{s}$. The quantity $\sqrt{s^{\prime}}$ is
a natural choice to assign events to these two categories for
$s$-channel processes. In the presence of $t$-channel contributions in
the $\epem\ra\epem(\gamma)$ process, the acollinearity angle, $\zeta$, is a
more appropriate choice. It is calculated as the complement to
$180^{\circ}$ of the angle between the directions of the final-state
electrons. In the following, the criteria listed in Table~\ref{tab:0}
are used to assign events to the two classes.
\begin{table}[h]
  \begin{center}
    \begin{tabular}{|l|c|c|c|c|}
      \cline{2-5} \multicolumn{1}{c|}{} &
      \multicolumn{4}{c|}{$\epem\ra$} \\ \cline{2-5}
      \multicolumn{1}{c|}{} & hadrons\,($\gamma$) & $\mumu(\gamma)$ &
      $\tautau(\gamma)$ & $\epem(\gamma)$ \\ \hline Inclusive events &
      $\sqrt{s^{\prime}}> 60\GeV$ &
      \multicolumn{2}{c|}{$\sqrt{s^{\prime}}> 75\GeV$} & $\zeta
      <120^{\circ}$ \\ \hline High-energy events &
      \multicolumn{3}{c|}{$\sqrt{s^{\prime}}> 0.85\sqrt{s}$} & $\zeta
      <25^{\circ}$ \\ \hline
      ISR/FSR interference &
      \multicolumn{3}{c|}{Excluded} & Included\\ \hline 
      Low-mass fermion pairs&
      \multicolumn{3}{c|}{Excluded} & Included\\ \hline 
    \end{tabular}\\ 
    \caption{The signal definition: criteria used to classify  events into the
    inclusive and high-energy samples and the
    channel-by-channel treatment of the interference of initial- and
    final-state radiation photons and additional low-mass fermion pairs.
    \label{tab:0}}
  \end{center}
\end{table}
Measurements for the $s$-channel processes are performed in a
limited finducial volume and then extrapolated to the full
angular region. Measurements for the $\epem\ra\epem(\gamma)$ process
are instead given in a limited angular region, with no extrapolation.
Events with low values of $\sqrtsp$ in the $s$-channel processes
and large values of $\zeta$ in the $\epem\ra\epem(\gamma)$ process are
excluded in order to obtain a high experimental signal-to-background ratio and
reduce uncertainties on radiative corrections. Experimental
uncertainties on the determination of $\sqrtsp$ and $\zeta$ introduce
an additional background, due to event migration, denoted as ISR
contamination.

The effective centre-of-mass energy is not well defined in presence of
interference between initial- and final-state photon radiation. This
effect is excluded from the signal definition for the $
\epem\rightarrow\mbox{hadrons}\,(\gamma)$, $
\epem\rightarrow\mumu(\gamma)$ and $ \epem\rightarrow\tautau(\gamma)$
processes, as discussed in Reference~\citen{l3-196}. The signal
definition for the $\epem\ra\epem(\gamma)$ process includes effects
of this interference.

Besides the emission of ISR photons, the production of initial-state
fermion pairs could also lower the value of $\sqrt{s^{\prime}}$. This
effect was previously investigated and found to have a negligible
impact on the selection efficiencies~\cite{l3-196}. In the following,
it is excluded from the signal definition of the $
\epem\rightarrow\mbox{hadrons}\,(\gamma)$, $
\epem\rightarrow\mumu(\gamma)$ and $ \epem\rightarrow\tautau(\gamma)$
processes and included in the signal definition of the
$\epem\ra\epem(\gamma)$ process.

Table~\ref{tab:0} summarises the treatment of the interference between
initial- and final-state photon radiation and the production of
initial-state fermion pairs for the different channels.

This paper presents results of the measurements of cross sections for
hadron and lepton-pair production for both inclusive and high-energy
events.  The forward-backward asymmetries of lepton-pair production,
\Afb, are studied for both the inclusive and the high-energy samples.
Finally, electron-positron pair-production is further investigated and
its differential cross sections as a function of the scattering angle,
$\rm d\sigma/d\cos\theta$, are measured for high-energy events.

For the high-energy sample, \Afb\ is defined through the
parametrisation of the differential cross section:
\begin{equation}
  \label{eq:cost-afbfit}
  \frac{1}{\sigma}\frac{\mathrm{d}\sigma}{\mathrm{d}\!\cost} =
  \frac{3}{8}(1+\cos^{2}\theta) + \Afb \cost +
  \frac{1}{\varepsilon(\cost)\sigma}
  \int_{\sqrtsp}^{\sqrt{s}}\mathrm{d}\minv\,\epsdif\,\frac{\partial^2
  \sigma_{\rm intf}}{ \partial \!\cos \theta \, \partial m_{\rm f
  \overline{f}}} \: ,
\end{equation}
where $\sigma_{\rm intf}$ is the contribution to the cross section
from the interference between initial- and final-state photon radiation and
$\varepsilon$ is the efficiency as a one- or two-dimensional function
of $\cos\theta$ and of the fermion-pair mass, $m_{\rm f
\overline{f}}$. It is computed from Monte Carlo simulations.  For the
inclusive sample, ISR distorts the angular distribution such that the
Born approximation of Equation~(\ref{eq:cost-afbfit}) is not
appropriate. Instead, for the $\epem\ra\mu^+\mu^-(\gamma)$ and the
$\epem\ra\tau^+\tau^-(\gamma)$ processes, the forward-backward
asymmetry is obtained from the differential cross section and
extrapolated to the full solid angle using the ZFITTER
program~\cite{ZFITTER}.  These corrections are about
$2\%$.  The forward-backward asymmetry
of the $\epem\ra\epem(\gamma)$ process is obtained by counting
forward- and backward-scattered events in a given fiducial volume.

Section~\ref{sec:data} describes the data sample and the measurement
of the integrated luminosity. Section~\ref{sec:mc} describes the Monte
Carlo simulation of signal and background processes as well as
the theoretical predictions for fermion-pair
production.  The analysis methods and the event selections for the
individual channels closely follow those used at lower centre-of-mass
energies~\cite{l3-196}. They are summarised, together with the results for
cross sections and asymmetries and a discussion of the systematic 
uncertainties, in Section~\ref{sec:hadron} for
the $\epem\rightarrow\mbox{hadrons}\,(\gamma)$ process, in
Section~\ref{sec:muon} for the $ \epem\rightarrow\mumu(\gamma)$
process, in Section~\ref{sec:tau} for the $
\epem\rightarrow\tautau(\gamma)$ process and, finally, in
Section~\ref{sec:electron} for the $ \epem\rightarrow\epem(\gamma)$
process.  Section~\ref{sec:last} contains the summary and the conclusions.

%%%%%%%%%%%%%%%%%%%%%%%%%%%%%%%%%%%%%%%%%%%%%%%%%%%%%%%%%%%%%%%%%%%%%%%%%%%%%%%%
\section{Data sample}
%%%%%%%%%%%%%%%%%%%%%%%%%%%%%%%%%%%%%%%%%%%%%%%%%%%%%%%%%%%%%%%%%%%%%%%%%%%%%%%%
\label{sec:data}

Data collected at LEP using the L3 detector~\cite{l3,L3-EGAP,L3-SLUM}
in the years 1999 and 2000 are investigated. In the year 1999, LEP was
operated at four centre-of-mass energies between $191.6\GeV$ and
$201.9\GeV$ which are treated separately in the following. In the year
2000, in order to enhance the discovery potential for the Standard
Model Higgs boson, the LEP centre-of-mass energy was varied between
$202.5\GeV$ and $209.2\GeV$. These data are divided into three energy
ranges. The seven average centre-of-mass energies considered in this
analysis are listed in Table~\ref{tab:roots}. The precise
determination of the LEP centre-of-mass energy is discussed in
Reference~\citen{lep-energy}.

A total integrated luminosity of 453~\pb\ is considered. Its breakdown
for the different values of $\sqrt{s}$ is given in
Table~\ref{tab:roots} for the four final states under
investigation. Differences between the channels are due to different
data-quality requirements.

The integrated luminosity is measured using small-angle Bhabha
scattering events recorded by two BGO calorimeters located close to
the beam line on opposite sides of the interaction region, and collected by a
dedicated trigger~\cite{L3-SLUM}. Events with two back-to-back energy
clusters are selected and a tight fiducial volume cut, $34 \mbox{
mrad} < \theta < 54 \mbox{ mrad}$, $|90^{\circ}-\phi| > 11.25^{\circ}$
and $|270^{\circ}-\phi| > 11.25^{\circ}$~\footnote{The azimuthal
angle, $\phi$, is measured from an axis pointing toward the centre of
the LEP ring.}, is imposed on the coordinates of the highest-energy
cluster.  The highest-energy cluster on the opposite side
should be contained in a larger fiducial volume, $32 \mbox{ mrad} <
\theta < 65 \mbox{ mrad}$, $|90^{\circ}-\phi| > 3.75^{\circ}$ and
$|270^{\circ}-\phi| > 3.75^{\circ}$.

Systematic uncertainties on the measurement of the integrated
luminosity originate from the event selection criteria, the precise
knowledge of the detector geometry and position, and the limited Monte
Carlo statistics used to determine the selection efficiency. For 20\%
of the data collected in 2000, some trigger instabilities required the
use of additional information from the cross section of hadron
production in photon-photon collisions, resulting in a further
systematic uncertainty.  The total experimental systematic
uncertainties for the years 1999 and 2000 are 0.14\% and 0.18\%,
respectively.  An additional theoretical uncertainty of 0.12\% affects
the determination of the integrated luminosity. These uncertainties
are negligible with respect to the statistical and systematic
uncertainties of the measurements described below.

%%%%%%%%%%%%%%%%%%%%%%%%%%%%%%%%%%%%%%%%%%%%%%%%%%%%%%%%%%%%%%%%%%%%%%%%%%%%%%%%
\section{Monte Carlo samples and theoretical predictions}
%%%%%%%%%%%%%%%%%%%%%%%%%%%%%%%%%%%%%%%%%%%%%%%%%%%%%%%%%%%%%%%%%%%%%%%%%%%%%%%%
\label{sec:mc}

The efficiencies and background levels of each selection, as well as
some systematic uncertainties, are
determined by means of Monte Carlo simulations. The following event
generators are used: BHAGENE~\cite{BHAGENE} and BHWIDE~\cite{BHWIDE}
for the $\epem\ra\epem(\gamma)$ process; BHLUMI~\cite{BHLUMI} for
Bhabha scattering in the fiducial volume used in the determination of
the integrated luminosity; TEEGG~\cite{TEEGG} for the
$\epem\rightarrow\epem\gamma$ process where one fermion is close to
the beam line and the photon in the detector; KK2f~\cite{kk2f} for the
$\epem\rightarrow\mbox{hadrons}\,(\gamma)$,
$\epem\rightarrow\mumu(\gamma)$ and $\epem\rightarrow\tautau(\gamma)$
processes; PYTHIA~\cite{PYTHIA} for the $\epem\ra\mathrm{ZZ}$ and
$\epem\ra\mathrm{Z}\epem$ processes; KORALW~\cite{KORALW} for W-boson
pair production, $\epem\rightarrow\mathrm{W^{+}W^{-}}$;
EXCALIBUR~\cite{EXCALIBUR} for the four-fermion processes
$\epem\rightarrow\mathrm{q\bar{q}^{\prime}}\e\nu$ and
$\epem\rightarrow\epem\epem$; GGG~\cite{GGG} for the
$\epem\rightarrow\gamma\gamma(\gamma)$ process; PHOJET~\cite{PHOJET}
and DIAG36~\cite{DIAG} for hadron and lepton production in
photon-photon collisions, $\epem\ra\epem\mbox{hadrons}$ and
$\epem\ra\epem\ell^+\ell^-$, respectively.

The hadronisation process is described with the PYTHIA Monte Carlo,
whose parameters are tuned with data collected with the L3 detector at
the Z resonance~\cite{l3-290-qcd-report}. The HERWIG~\cite{HERWIG} and
ARIADNE~\cite{ARIADNE} Monte Carlo programs, also tuned on the same
data~\cite{l3-290-qcd-report}, are used for systematic studies in the
$\epem\ra\mbox{hadrons}$ channel.

Monte Carlo events are generated for each centre-of-mass energy.  The L3 detector
response is simulated using the GEANT~\cite{geant} program which takes
into account the effects of energy loss, multiple scattering and
showering in the detector. The GHEISHA~\cite{gheisha} package is used for the
simulation of hadronic interactions. Time-dependent detector
efficiencies, as monitored during the data-taking period, are included
in the simulations. This ``real-detector'' simulation assures the
control of the selection efficiencies. However, time-dependent
second-order effect might escape the monitoring procedure and introduce a
difference between data and the Monte Carlo description of some
selection variables. The selection cuts described in the following are
chosen so as to minimise these small discrepancies.
The region of maximal discrimination between the signals and the
backgrounds is scanned in a window of width several times the resolution of
the selection variables. A value of the cut is retained 
for which the data and Monte Carlo differences are minimal. The
observed differences are then 
retained as an estimation of systematic uncertainties on the detector
modelling. These are discussed in detail in the following sections.

The measurements are compared to the predictions of the \SM\ as
calculated using the ZFITTER~\cite{ZFITTER} program for the $
\epem\rightarrow\mbox{hadrons}\,(\gamma)$, $
\epem\rightarrow\mumu(\gamma)$ and $ \epem\rightarrow\tautau(\gamma)$
processes and the TOPAZ0~\cite{TOPAZ0} program for the
$\epem\ra\epem(\gamma)$ process. The following input parameters are
used~\cite{PDG}: $m_{\rm Z}=91.1876 \pm 0.0021$~{\GeV}, $m_{\rm
t}=174.3 \pm 5.1$~{\GeV} for the top-quark mass,
$\alpha_{s}(m_Z^2)=0.1187 \pm 0.0020$ for the strong coupling and
$\Delta\alpha_{\mathrm{had}}^{(5)}=0.02763 \pm 0.00036$ for the
hadronic contribution to the running of the electromagnetic
coupling. The Higgs-boson mass is chosen as $m_{\rm H} =
150^{+110}_{-\phantom{0}36}$~{\GeV}, compatible with the lower and
higher mass limits of $114.4$~{\GeV}~\cite{mhlow} and
$285$~{\GeV}~\cite{mhhigh}, respectively.  The theoretical
uncertainties on the \SM\ predictions are estimated to be below 1\%
except for the predictions for large-angle Bhabha scattering where
they reach 1.5\%~\cite{BHABHA-THEORY}. The values of $s^\prime$ used
in the calculations account for the energies of ISR photons through
Equation~(\ref{eq:spri}), where $E_\gamma$ and $\vec{P_\gamma}$ also
include the four-momenta of low-mass fermion pairs.  The stability of
the predictions with respect to the Standard Model input parameters is
checked by changing these within their uncertainties. The variations
of the predictions are below 0.1\%, resulting in a negligible
additional systematic uncertainty.

%%%%%%%%%%%%%%%%%%%%%%%%%%%%%%%%%%%%%%%%%%%%%%%%%%%%%%%%%%%%%%%%%%%%%%%%%%%%%%%%
\section{The {\boldmath $\epem\rightarrow\mbox{hadrons}\,(\gamma)$} process}
%%%%%%%%%%%%%%%%%%%%%%%%%%%%%%%%%%%%%%%%%%%%%%%%%%%%%%%%%%%%%%%%%%%%%%%%%%%%%%%%
\label{sec:hadron}

%%%%%%%%%%%%%%%%%%%%%%%%%%%%%%%%%%%%%%%%%%%%%%%%%%%%%%%%%%%%%%%%%%%%%%%%%%%%%%%%
\subsection{Event selection}
%%%%%%%%%%%%%%%%%%%%%%%%%%%%%%%%%%%%%%%%%%%%%%%%%%%%%%%%%%%%%%%%%%%%%%%%%%%%%%%%

High-multiplicity events from the
$\epem\rightarrow\mbox{hadrons}\,(\gamma)$ process are selected in
the fiducial volume of the L3 calorimeters,
$|\cos\theta|<0.995$~\cite{l3-196,jessica}. These events are collected
by redundant triggers based on the energy deposition in the
calorimeters, the presence of pairs of back-to-back charged tracks in the
tracker and the multiplicity of hits in the scintillator
time-of-flight system. The overall trigger efficiency is measured
from data to be close to 100\%, with a negligible
uncertainty. 

Background from lepton-pair production is rejected by
requiring the event to have at least 12 calorimetric
clusters. Background from hadron production in photon-photon
collisions is reduced by two criteria: the hadronic energy calculated
by excluding isolated clusters, $E_{\mathrm{had}}$, must be greater
than $0.4\sqrt{s} $ and the longitudinal energy-imbalance must be less
than $0.8E_{\mathrm{tot}}$, where $E_{\mathrm{tot}}$ is the total
energy reconstructed in the detector. Events from W-boson pair
production with semi-leptonic decay are removed by requiring the
transverse energy-imbalance to be less than $0.3 E_{\mathrm{tot}}$.
Hadronic decays of W bosons are reduced by applying the JADE
algorithm~\cite{jade} with a resolution parameter
$y_{\mathrm{cut}}=0.01$ and removing events with at least four jets,
each with an energy greater than 15 \GeV.

Two methods are used to derive the four-momentum of ISR photons and
calculate {\sqrtsp} through Equation~(\ref{eq:spri}). The first method
uses a kinematic fit assuming the emission of either zero, one, or two
photons along the beam line. The hypothesis which best fits the data
is retained and the photon four-momenta are derived from the fit.  In
the second method, each event is clustered into two jets using the
JADE algorithm.  A single photon is assumed to be emitted along the
beam line and its energy is estimated from the reconstructed polar
angles of the jets, $\theta_{1}$ and $\theta_{2}$, as:
\begin{equation} \label{eq:egam}
E_{\gamma} ~ = ~ \sqrt{s}\frac{|\sin(\theta_{1}+\theta_{2})|}
           {\sin\theta_{1}+\sin\theta_{2}+|\sin(\theta_{1}+\theta_{2})|}
           \,.
\end{equation}
In about 15\% of selected events, an isolated high-energy cluster is
detected in the electromagnetic calorimeter.  It is assumed to be an
ISR photon, and its energy and momentum are added to those determined
by either method before applying Equation~(\ref{eq:spri}).

The first method is used to derive the following results, while the
second is used as a cross check and to assess the systematic
uncertainty on the $\sqrtsp$ determination.

\begin{figure}[p]
  \begin{center}
    \includegraphics[width=0.45\textwidth]{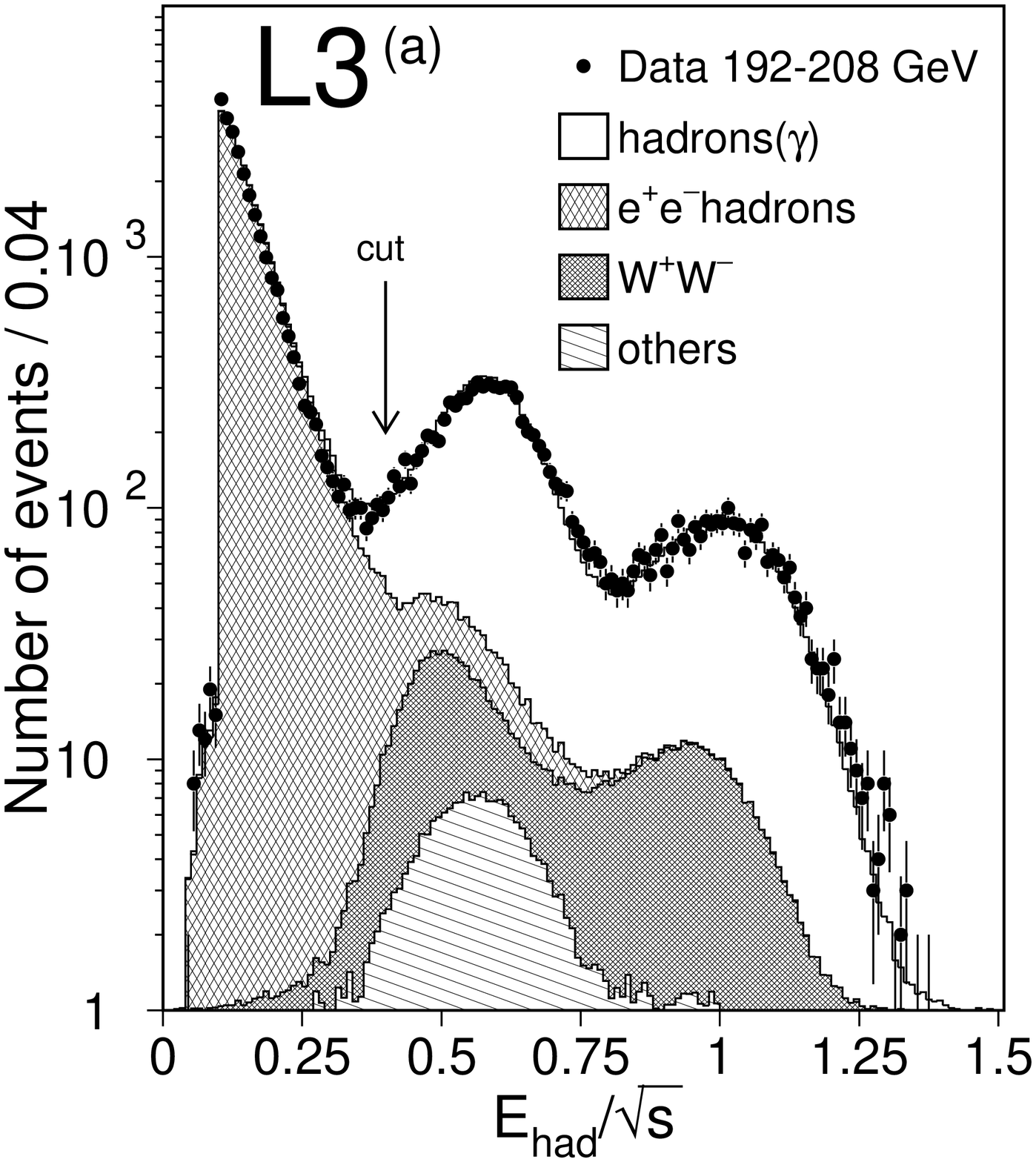}
    \includegraphics[width=0.45\textwidth]{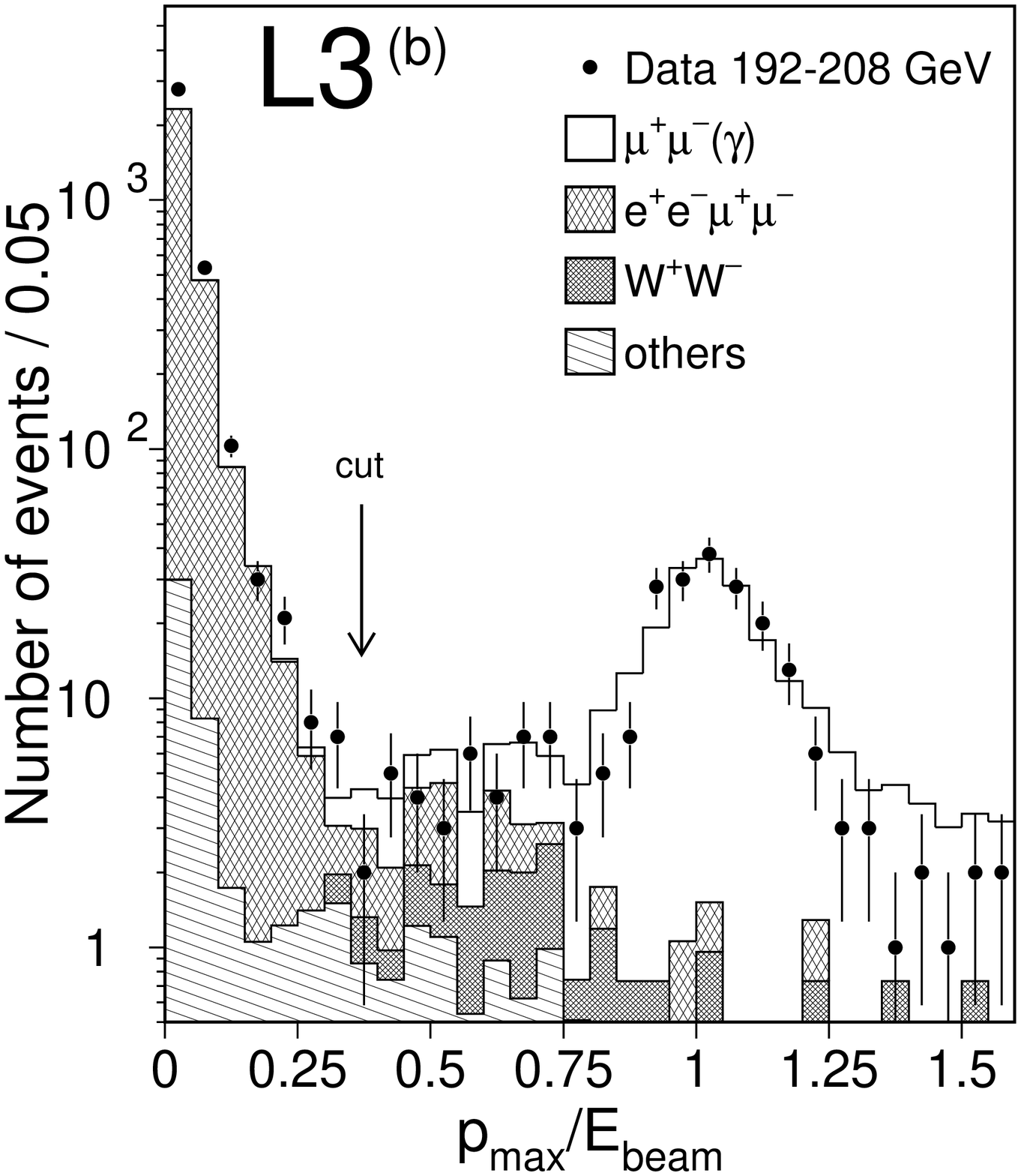}
    \includegraphics[width=0.45\textwidth]{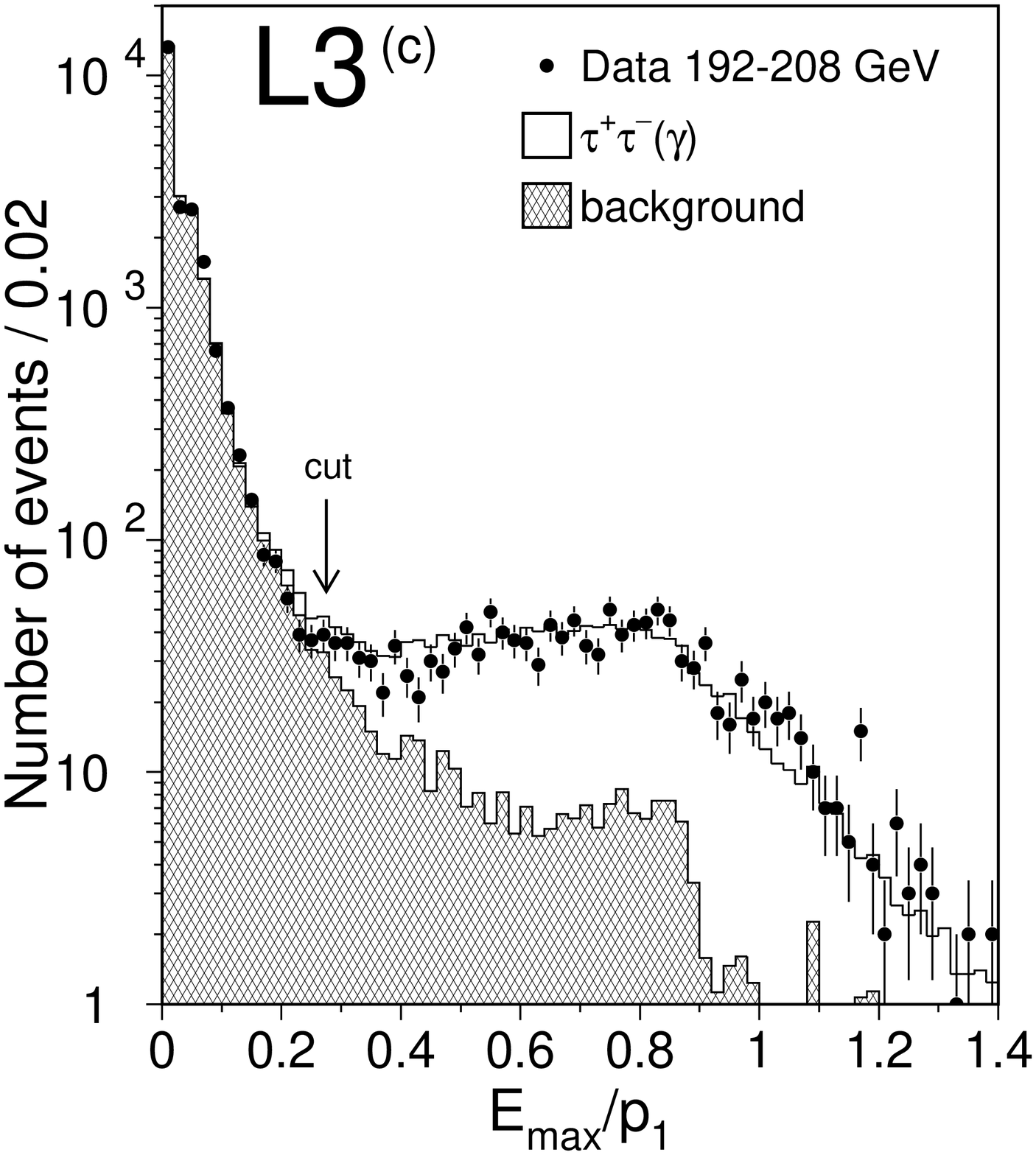}
    \includegraphics[width=0.45\textwidth]{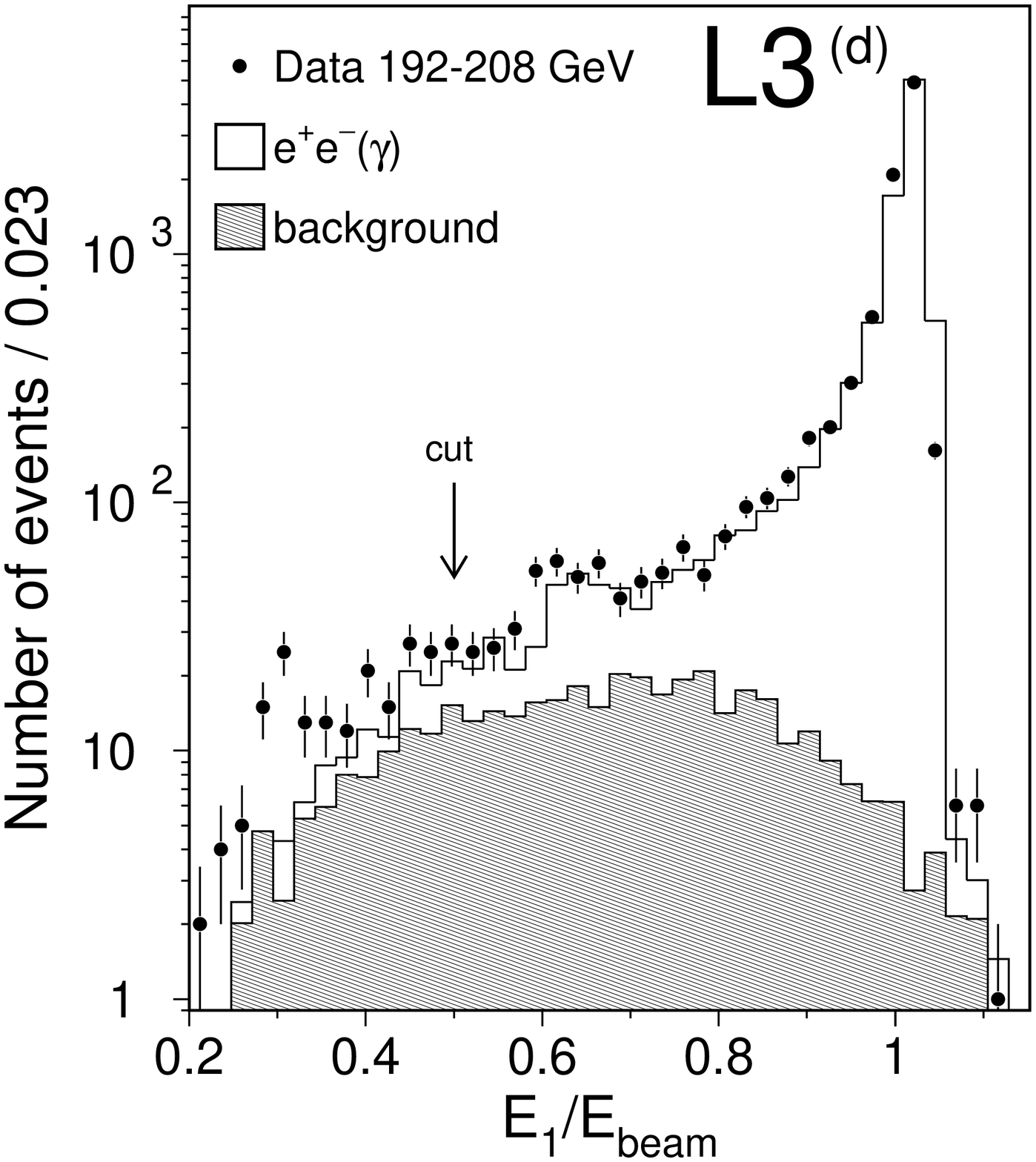}
      \caption[]{Distributions for data and Monte Carlo of a) the
        hadronic energy normalised to the {\CoM} energy for the
        {$\epem\rightarrow\mbox{hadrons}\,(\gamma)$} analysis, b) the
        highest muon momentum normalised to the beam energy for the
        $\epem\rightarrow\mumu(\gamma)$ analysis, c) the highest
        tau-jet energy normalised to the corresponding estimated
        momentum for the $\epem\rightarrow\tautau(\gamma)$ analysis and
        d) the highest electron energy normalised to the beam energy
        for the $\epem\rightarrow\epem(\gamma)$ analysis. The arrows
        indicate the positions of the selection cuts. All other cuts
        are applied.
      \label{fig:sele}}
  \end{center}
\end{figure}

\begin{figure}[p]
  \begin{center}
    \includegraphics[width=0.48\textwidth]{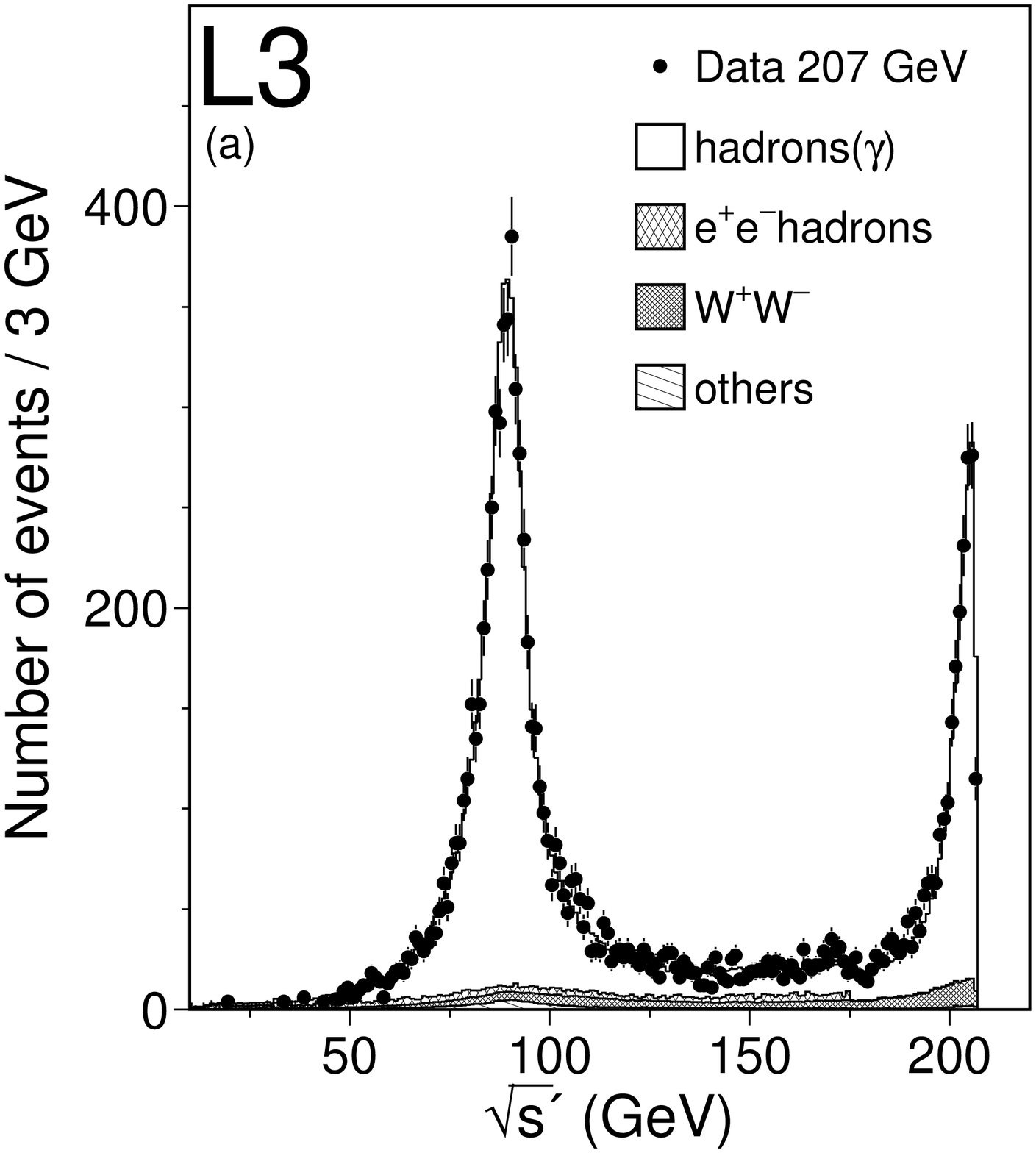}
    \includegraphics[width=0.48\textwidth]{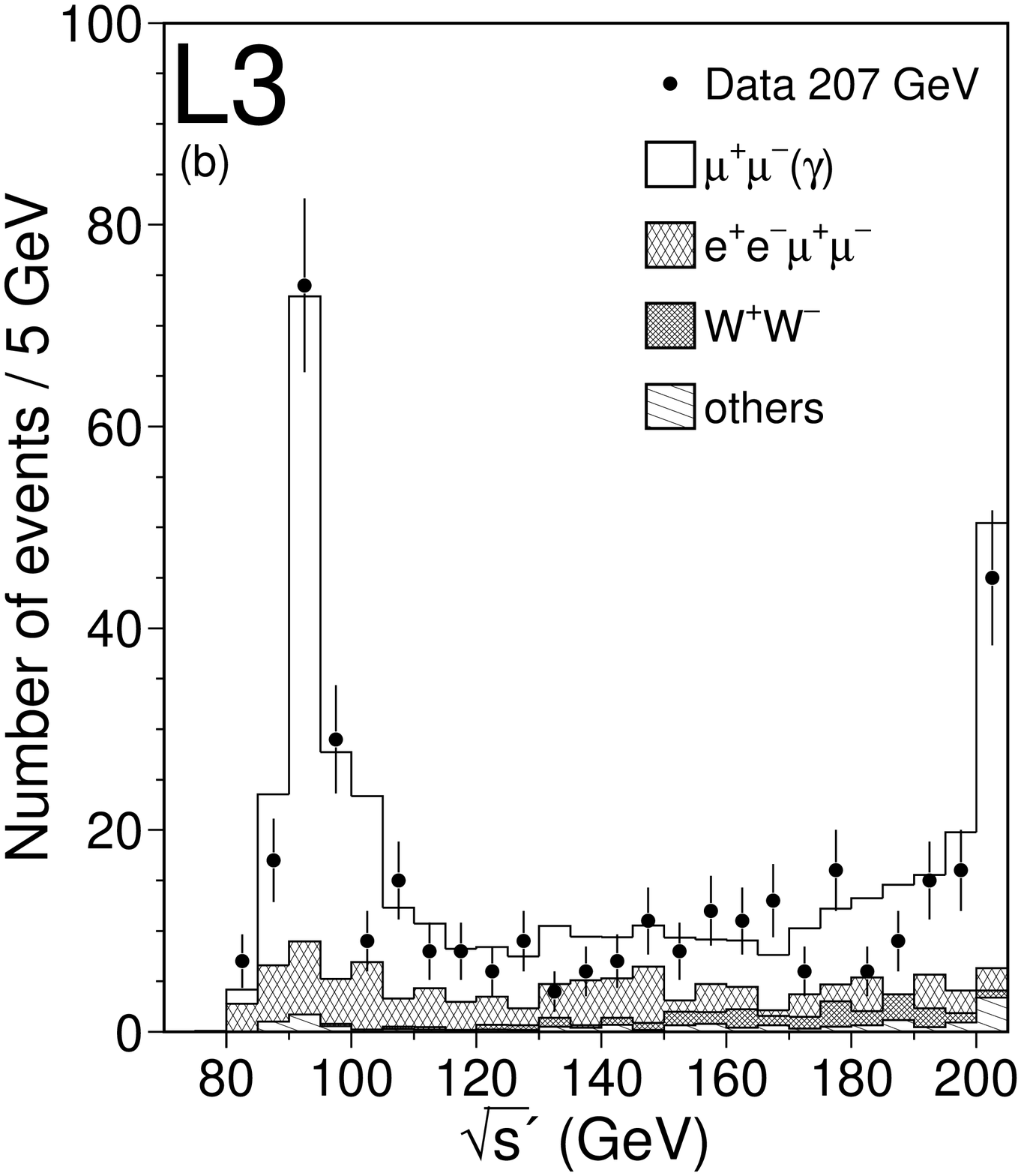}
    \includegraphics[width=0.48\textwidth]{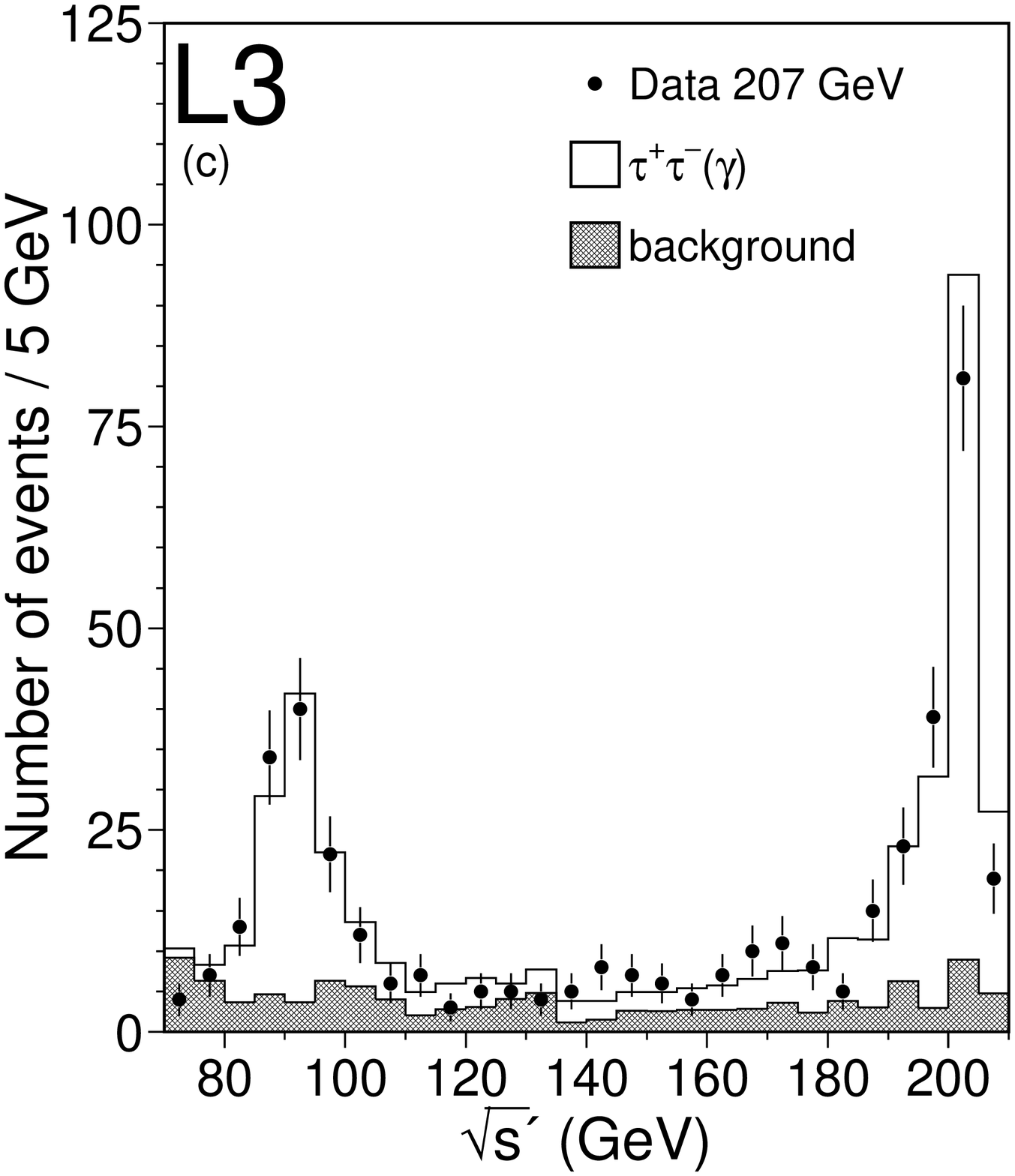}
    \includegraphics[width=0.48\textwidth]{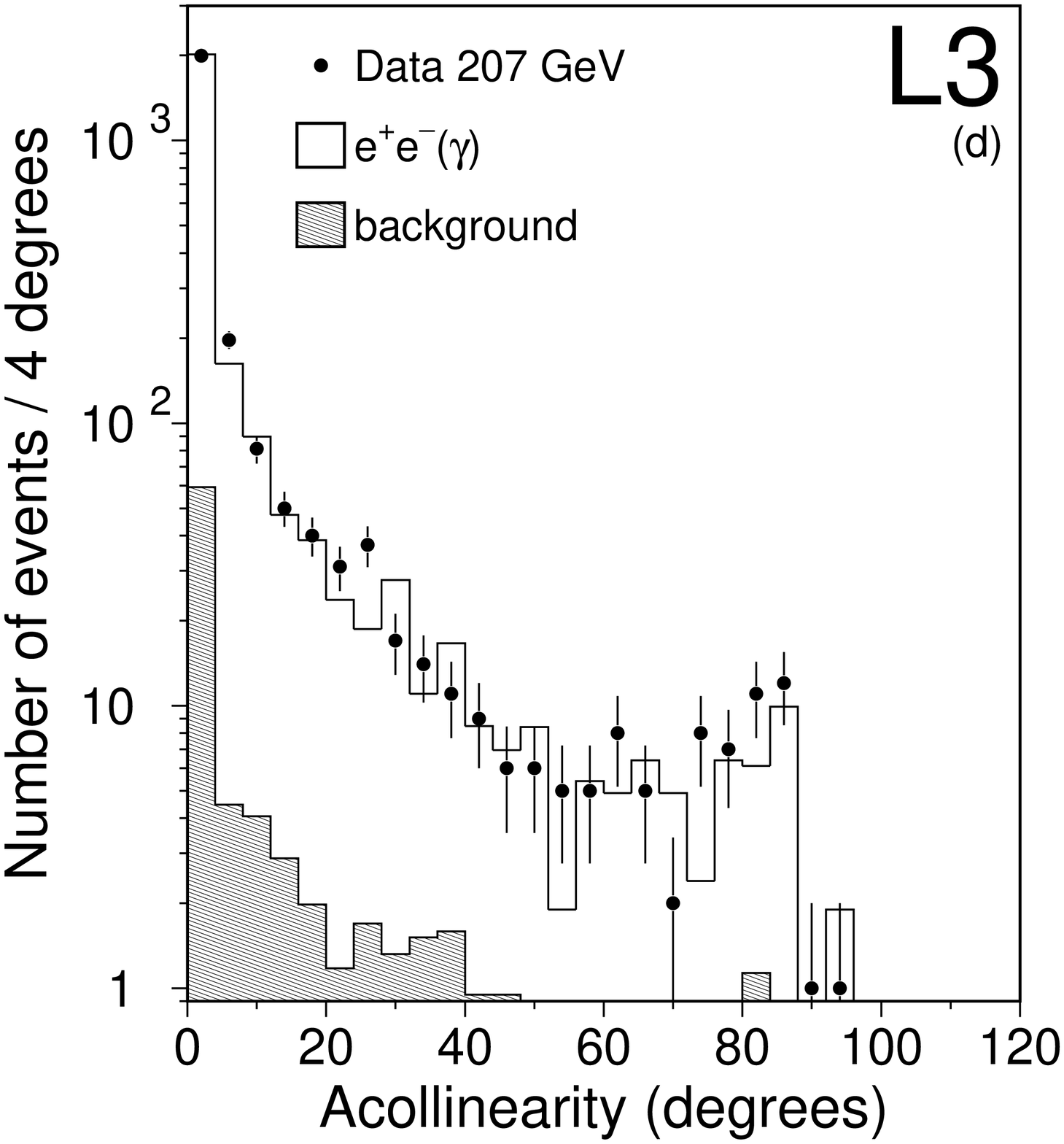}
      \caption[]{Distributions for data and Monte Carlo at
        $\sqrt{s}=207\GeV$ of the reconstructed effective {\CoM}
        energy, {\sqrtsp}, for the a)
        $\epem\rightarrow\mbox{hadrons}\,(\gamma)$, b)
        $\epem\rightarrow\mumu(\gamma)$ and c)
        $\epem\rightarrow\tautau(\gamma)$ channels and d) of the
        reconstructed acollinearity angle, $\zeta$, for the
        {$\epem\rightarrow\epem(\gamma)$} channel.}
      \label{fig:spri}
  \end{center}
\end{figure}

Figure~\ref{fig:sele}a shows the distributions for data and Monte
Carlo of $E_{\rm had}/\sqrt{s}$ for the full data sample. The three
peaks correspond, from left to right, to hadron production in
photon-photon collisions, to the radiative return to the Z and to
high-energy events.  Figure~\ref{fig:spri}a shows the data and Monte
Carlo distributions of the values of \sqrtsp\ reconstructed with the
second method for $\sqrt{s}=207\GeV$. The two peaks correspond to the
radiative return to the Z and to high-energy events.

Selection efficiencies and background contributions for the different
values of $\sqrt{s}$ are listed in Tables~\ref{tab:seleinc}
and~\ref{tab:sele} for the inclusive and high-energy samples,
respectively. The largest residual backgrounds are from W-boson pair
production, hadron production in photon-photon collisions and, for the
high-energy sample, ISR contamination. Other minor sources of
background are tau-pair production and four-fermion events from
Z-boson pair production and the $\epem\rightarrow
\mathrm{Z}\epem$ process.

%%%%%%%%%%%%%%%%%%%%%%%%%%%%%%%%%%%%%%%%%%%%%%%%%%%%%%%%%%%%%%%%%%%%%%%%%%%%%%%%
\subsection{Results}
%%%%%%%%%%%%%%%%%%%%%%%%%%%%%%%%%%%%%%%%%%%%%%%%%%%%%%%%%%%%%%%%%%%%%%%%%%%%%%%%

The numbers of observed events and the measurements of the cross
sections of the $\epem\rightarrow\mbox{hadrons}\,(\gamma)$ process for
the inclusive and high-energy samples are presented in
Table~\ref{tab:xsecincl}, together with their statistical and
systematic uncertainties. The corresponding Standard Model predictions
are also given. Figure~\ref{fig:ha_xsec} compares the cross section
measurements to the \SM\ predictions. Good agreement is observed. 

A $\chi^2$ test of the compatibility of data and Standard Model
prediction yields values of $\rm\chi^2/d.o.f.$ of 1.4 and 0.7 for the
inclusive and high-energy samples, respectively. These and all
following calculations of $\rm\chi^2/d.o.f.$  include only statistical uncertainties.

%%%%%%%%%%%%%%%%%%%%%%%%%%%%%%%%%%%%%%%%%%%%%%%%%%%%%%%%%%%%%%%%%%%%%%%%%%%%%%%%
\subsection{Systematic uncertainties}
%%%%%%%%%%%%%%%%%%%%%%%%%%%%%%%%%%%%%%%%%%%%%%%%%%%%%%%%%%%%%%%%%%%%%%%%%%%%%%%%

The statistical uncertainty on the cross sections of the
$\epem\rightarrow\mbox{hadrons}\,(\gamma)$ process varies between
1.1\% and 2.1\% for the inclusive sample and 2.5\% and 4.6\% for the
high-energy sample, depending on $\sqrt{s}$, with the exclusion of the
low-luminosity highest-energy point.  The overall systematic
uncertainties for the inclusive sample are comparable to the
statistical uncertainties, at about 1.1\%, while for the high-energy
sample, with a value of about 0.8\%, they are less than a third of the
statistical uncertainties~\cite{l3-196,jessica}.

The systematic uncertainty on the hadronisation process, which amounts
to 0.47\% for the inclusive sample and 0.63\% for the high-energy
sample, is derived by using the HERWIG and ARIADNE Monte Carlo
programs instead of the default PYTHIA Monte Carlo. The analysis is
repeated by using these alternative Monte Carlo simulations.  Their
average is calculated and half of its difference with respect to the
original measurement is assigned as systematic uncertainty. Limited
signal and background Monte Carlo statistics imply systematic
uncertainties of $0.07-0.17$\% and $0.14-0.50$\% for the inclusive and
high-energy samples, respectively, depending on the centre-of-mass
energy.  The systematic uncertainty from calorimeter calibration,
which amounts to 0.48\% for the inclusive sample and 0.26\% for the
high-energy sample, is assessed by repeating the analysis changing the
calorimeter calibration constants within the uncertainties of their
determination from Z-peak data. The Monte Carlo treatment of the
interference between initial- and final-state radiation contributes
systematic uncertainties of 0.10\% and 0.20\% for the inclusive and
high-energy samples, respectively.  The \sqrtsp\ reconstruction
uncertainty is estimated from the differences of the cross sections
obtained with each of the two methods as 0.36\% for the inclusive
sample and 0.15\% for the high-energy sample. The impact of the
event-selection procedure is studied by varying the selection
criteria, in order to assess the effects of possible discrepancies
between data and the Monte Carlo simulation, and by using a different
strategy to remove events from W-boson pair production. Uncertainties
of 0.22\% and 0.07\% are obtained for the inclusive and high-energy
samples, respectively.  Uncertainties in the modelling of hadron
production in photon-photon interactions propagate to a systematic
uncertainty of 0.05\% for both the inclusive and high-energy samples.
For the cross section measurement of the high-energy sample, three quarters
of the systematic uncertainty is correlated between energy
points.

%%%%%%%%%%%%%%%%%%%%%%%%%%%%%%%%%%%%%%%%%%%%%%%%%%%%%%%%%%%%%%%%%%%%%%%%%%%%%%%%
\section{The {\boldmath $\epem\rightarrow\mumu(\gamma)$} process}
%%%%%%%%%%%%%%%%%%%%%%%%%%%%%%%%%%%%%%%%%%%%%%%%%%%%%%%%%%%%%%%%%%%%%%%%%%%%%%%%
\label{sec:muon}

%%%%%%%%%%%%%%%%%%%%%%%%%%%%%%%%%%%%%%%%%%%%%%%%%%%%%%%%%%%%%%%%%%%%%%%%%%%%%%%%
\subsection{Event selection}
%%%%%%%%%%%%%%%%%%%%%%%%%%%%%%%%%%%%%%%%%%%%%%%%%%%%%%%%%%%%%%%%%%%%%%%%%%%%%%%%

Muon-pair candidates are selected from low-multiplicity events with
two identified muons~\cite{l3-196,alessandro}. These events are mainly
collected by a trigger based on several possible combinations of
tracks in different regions of the muon spectrometer. The trigger
efficiency is enhanced by including events with back-to-back tracks
in the central tracker and events with isolated photons in the
calorimeters, susceptible to originate from ISR. The combined trigger
efficiency is determined from data and varies between 97.8\% and
99.9\%, according to the data-taking conditions, with statistical
uncertainties between 0.1\% and 0.8\%.

The muon candidates are required to have at least two track-segments
reconstructed in the fiducial volume, $|\cos\theta|<0.9$, of the muon
spectrometer. In addition, for 15\% of the events, only one muon is
identified in the muon spectrometer while the other is reconstructed
from the signature of a minimum-ionising particle in the calorimeters
matched to a track in the central tracker and, possibly, a single
track-segment in the muon spectrometer.

Background from lepton-pair production in photon-photon collisions and
from tau-pair production is suppressed by requiring the momentum of
the most energetic muon, $p_{\rm max}$, to satisfy $p_{\rm
max}>0.4E_{\rm beam}$, where $E_{\rm beam}$ is the beam energy. These
backgrounds are further removed by requiring the acollinearity angle to
be less than $90^\circ$. Background from cosmic rays is rejected by
three criteria: at least one of the muons must originate from the
interaction point; at least one of the muons must have a signal in the
scintillator time-of-flight system in time with the beam crossing;
finally, if both muons have such a scintillator hit, these must be
simultaneous. The residual background from cosmic rays is estimated
from complementary subsamples of data.

The value of \sqrtsp\ is derived from Equation~(\ref{eq:spri}).  If
one or more isolated high-energy photons are detected in the event,
their energies are directly used. If no such photons are detected, the
hypothesis that a single ISR photon is emitted along the beam line is
made, and Equation~(\ref{eq:egam}) is used to derive its energy from
the muon polar angles.

Figure~\ref{fig:sele}b shows the distribution for data and Monte Carlo
of $p_{\rm max}/E_{\rm beam}$ for the full data sample, while
Figure~\ref{fig:spri}b shows the distributions of \sqrtsp\
reconstructed at $\sqrt{s}=207\GeV$.

Selection efficiencies and background contributions are listed in
Tables~\ref{tab:seleinc} and~\ref{tab:sele} for the inclusive and
high-energy samples, respectively. The largest residual backgrounds
are from lepton-pair production in photon-photon collisions, W-boson
pair production and, for the high-energy sample, ISR
contamination. Other minor sources of background are tau-pair
production, Z-boson pair production and cosmic rays.

%%%%%%%%%%%%%%%%%%%%%%%%%%%%%%%%%%%%%%%%%%%%%%%%%%%%%%%%%%%%%%%%%%%%%%%%%%%%%%%%
\subsection{Results}
%%%%%%%%%%%%%%%%%%%%%%%%%%%%%%%%%%%%%%%%%%%%%%%%%%%%%%%%%%%%%%%%%%%%%%%%%%%%%%%%

The numbers of observed events and the measurements of the cross
sections of the $\epem\rightarrow\mumu(\gamma)$ process for the
inclusive and high-energy samples are presented in
Table~\ref{tab:xsecincl}, together with their statistical and
systematic uncertainties. The corresponding Standard Model predictions
are also given.  Figure~\ref{fig:le_xsec_afb}a compares the measured
cross section for the inclusive and high-energy samples with the \SM\
predictions as a function of $\sqrt {s}$. Good agreement is observed,
with values of $\rm\chi^2/d.o.f.$ of 1.9 and 1.4 for the inclusive and
high-energy samples, respectively.

The forward-backward asymmetry is determined for the inclusive and
high-energy samples with the results presented in Table~\ref{tab:afb},
together with the numbers of events selected in the forward and
backward hemispheres. The determination of \Afb\ takes into account
both the charge confusion per event, measured in data to be between
0.2\% and 0.5\%, and the asymmetries induced by the accepted
background. Figure~\ref{fig:le_xsec_afb}b presents the values of \Afb\
measured as a function of $\sqrt {s}$. They are in good agreement with
the Standard Model predictions, also shown, with values of
$\rm\chi^2/d.o.f.$ of 0.2 and 0.8 for the inclusive and high-energy
samples, respectively.

%%%%%%%%%%%%%%%%%%%%%%%%%%%%%%%%%%%%%%%%%%%%%%%%%%%%%%%%%%%%%%%%%%%%%%%%%%%%%%%%
\subsection{Systematic uncertainties}
%%%%%%%%%%%%%%%%%%%%%%%%%%%%%%%%%%%%%%%%%%%%%%%%%%%%%%%%%%%%%%%%%%%%%%%%%%%%%%%%

For the high-energy sample, the systematic uncertainties on the
measurement of the $\epem\rightarrow\mumu(\gamma)$ cross sections and
forward-backward asymmetries are in the ranges $2.7-4.0\%$ and $3.4-10\%$,
respectively~\cite{l3-196,alessandro}, depending on $\sqrt{s}$. These
uncertainties are at least three times smaller than the corresponding
statistical uncertainties.

The limited signal and background Monte Carlo statistics imply
systematic uncertainties of $1.8-3.0$\% and $2.4-5.5$\%, depending on
$\sqrt{s}$, for the cross section and \Afb\ measurements,
respectively. The uncertainty in detector modelling, assessed by varying
the selection criteria, is dominated by the simulation of $p_{\rm
max}$ and the control of the fiducial volume. Depending on the running
conditions and the detector ageing, this uncertainty varies between
1.7\% and 2.4\% for the cross sections and 1.0\% and 7.0\% for \Afb.
The uncertainty on the trigger efficiency has a small impact on the
cross section with a systematic uncertainty between 0.2\% and 0.8\%,
depending on the year of data taking. The charge confusion per event
has a relative uncertainty of about 20\%, which results in a small
additional uncertainty on \Afb, of about 0.2\%.

For the high-energy measurement of the cross sections, between one
third and half of the systematic uncertainty is correlated between
the energy points. For the asymmetries, these figures increase to one half
and two thirds.

%%%%%%%%%%%%%%%%%%%%%%%%%%%%%%%%%%%%%%%%%%%%%%%%%%%%%%%%%%%%%%%%%%%%%%%%%%%%%%%%
\section{The {\boldmath $\epem\rightarrow\tautau(\gamma)$} process}
%%%%%%%%%%%%%%%%%%%%%%%%%%%%%%%%%%%%%%%%%%%%%%%%%%%%%%%%%%%%%%%%%%%%%%%%%%%%%%%%
\label{sec:tau}

%%%%%%%%%%%%%%%%%%%%%%%%%%%%%%%%%%%%%%%%%%%%%%%%%%%%%%%%%%%%%%%%%%%%%%%%%%%%%%%%
\subsection{Event selection}
%%%%%%%%%%%%%%%%%%%%%%%%%%%%%%%%%%%%%%%%%%%%%%%%%%%%%%%%%%%%%%%%%%%%%%%%%%%%%%%%

Tau candidates are identified in the fiducial volume
$|\cos\theta|<0.94$ as narrow, low multiplicity, jets containing at
least one charged particle~\cite{l3-196,sandra}. Several classes of
triggers collect these events with an efficiency, measured from data,
close to 100\% with a negligible uncertainty: low- and large-angle
charged-track triggers, the muon triggers, a scintillator
time-of-flight multiplicity trigger and calorimeter-based energy
triggers. 

Events with two tau candidates are selected. If both jets
contain electrons\footnote{Here and in the following, the term
``electron'' denotes both electrons and positrons.} or muons the
events are rejected.  The momentum of the most energetic tau jet,
$p_1$, is estimated from its polar angle, $\theta_1$, and the polar
angle of the other tau jet, $\theta_2$, imposing energy and momentum
conservation as:
\begin{equation}
p_1 = \sqrt{s} \frac{ \sin\theta_{2} } {
\sin\theta_{1}+\sin\theta_{2}+\sin(\theta_{1} + \theta_{2}) }\,.
\end{equation}
The momentum of the other tau jet is estimated analogously.
Background from hadronic events is removed by requiring at most 16
calorimetric clusters and at most 9 tracks in the central
tracker. Residual background from the $\epem\ra\epem(\gamma)$ process
is suppressed by requiring that the energies of the most energetic and
the second most energetic electromagnetic cluster in the event are
less than 85\% and 50\% of the estimated momenta of the corresponding
tau jets. Similarly, background from muon-pair production is further
reduced by requiring the momentum of each muon in the event to be less
than 85\% of the estimated momentum of the corresponding tau
jet. Background from photon-photon collisions is reduced by requiring
the most energetic jet to have an energy $E_{\rm max}$ such that
$E_{\rm max}>0.275 p_{1}$. Leptonic final states from W-boson pair
production are rejected by requiring the acollinearity angle to be
less than $15^\circ$.  Background
from cosmic rays is suppressed using information from the
time-of-flight system and by requiring any muons in the event to
originate from the interaction point.

The value of \sqrtsp\ is derived from Equation~(\ref{eq:spri}) by
using the four-momenta of all detected isolated high-energy
photons. If the event contains no such photons, the energy of a single
ISR photon directed along the beam line is calculated from the tau-jet
polar angles with Equation~(\ref{eq:egam}).

Figure~\ref{fig:sele}c shows the data and Monte Carlo distributions of
$E_{\rm max}/p_{1}$ for the full data sample, while
Figure~\ref{fig:spri}c shows the distributions of \sqrtsp\
reconstructed at $\sqrt{s}=207\GeV$.

The charge of the tau candidates is determined from the sum of the
charges of the tracks constituting the jets or of the identified
electrons or muons. Only event with an unambiguous charge assignment
are retained for the study of \Afb. These comprise 72\% of the
inclusive sample and 75\% of the high-energy sample.

Tables~\ref{tab:seleinc} and~\ref{tab:sele} list the selection
efficiencies and background contributions for the inclusive and
high-energy samples, respectively.  The largest residual backgrounds
are from tau production in photon-photon collisions and other sources
such as the $\epem\ra\epem(\gamma)$ process, muon-pair production and
W-boson pair production. ISR contamination contributes to the
background to the high-energy sample.

%%%%%%%%%%%%%%%%%%%%%%%%%%%%%%%%%%%%%%%%%%%%%%%%%%%%%%%%%%%%%%%%%%%%%%%%%%%%%%%%
\subsection{Results}
%%%%%%%%%%%%%%%%%%%%%%%%%%%%%%%%%%%%%%%%%%%%%%%%%%%%%%%%%%%%%%%%%%%%%%%%%%%%%%%%

The numbers of observed events and the measurements of the cross
sections of the $\epem\rightarrow\tautau(\gamma)$ process for the
inclusive and high-energy samples are presented in
Table~\ref{tab:xsecincl}, together with their statistical and
systematic uncertainties. The measurements are compared in
Figure~\ref{fig:le_xsec_afb}a with the \SM\ predictions as a function
of $\sqrt {s}$.  All measurements are in good agreement with the
Standard Model predictions: $\rm\chi^2/d.o.f.$ of 0.3 and 0.5 are
observed for the inclusive and high-energy cross sections,
respectively.

The forward-backward asymmetry is determined for the inclusive and
high-energy samples with the results presented in
Figure~\ref{fig:le_xsec_afb}b and Table~\ref{tab:afb}, which also
lists the numbers of events selected in the forward and backward
hemispheres. The determination of \Afb\ takes into account the charge
confusion per event, measured in data, which is of the order of 2\%.
Good agreement with the Standard Model predictions is found, with
values of $\rm\chi^2/d.o.f.$ of 0.8  and 0.9 for the inclusive and
high-energy sample, respectively.

%%%%%%%%%%%%%%%%%%%%%%%%%%%%%%%%%%%%%%%%%%%%%%%%%%%%%%%%%%%%%%%%%%%%%%%%%%%%%%%%
\subsection{Systematic uncertainties}
%%%%%%%%%%%%%%%%%%%%%%%%%%%%%%%%%%%%%%%%%%%%%%%%%%%%%%%%%%%%%%%%%%%%%%%%%%%%%%%%

For the high-energy sample, the systematic uncertainties on the
measurement of the $\epem\rightarrow\tautau(\gamma)$ cross sections
and forward-backward asymmetries are 
$2.0-3.4$\% and $9.6-17\%$, depending on the centre-of-mass energy and
excluding the highest-energy point, respectively~\cite{l3-196,sandra}.
These uncertainties are considerably lower than the corresponding
statistical uncertainties.

The main systematic uncertainties on the cross section determination
are the detector modelling and the limited signal and background Monte
Carlo statistics. The former, of 1.4\%, receives equal contributions
from the simulation of the variables used for background rejection
on the basis of calorimetric information and from the control of the fiducial
volume. The limited Monte Carlo statistics implies systematic
uncertainties of $1.4-3.1$\%, depending on $\sqrt{s}$.

The systematic uncertainties on \Afb\ are dominated by the limited
Monte Carlo statistics, with an additional contribution of about 3.0\%
from the detector modelling. The charge confusion per event is
determined with a relative uncertainty up to 50\%, which has a
negligible contribution to the systematic uncertainties.

The systematic uncertainties on the high-energy measurements of the
cross sections and asymmetries are mostly uncorrelated between
the energy points.

%%%%%%%%%%%%%%%%%%%%%%%%%%%%%%%%%%%%%%%%%%%%%%%%%%%%%%%%%%%%%%%%%%%%%%%%%%%%%%%%
\section{The {\boldmath $\epem\rightarrow\epem(\gamma)$ } process}
%%%%%%%%%%%%%%%%%%%%%%%%%%%%%%%%%%%%%%%%%%%%%%%%%%%%%%%%%%%%%%%%%%%%%%%%%%%%%%%%
\label{sec:electron}

%%%%%%%%%%%%%%%%%%%%%%%%%%%%%%%%%%%%%%%%%%%%%%%%%%%%%%%%%%%%%%%%%%%%%%%%%%%%%%%%
\subsection{Event selection}
%%%%%%%%%%%%%%%%%%%%%%%%%%%%%%%%%%%%%%%%%%%%%%%%%%%%%%%%%%%%%%%%%%%%%%%%%%%%%%%%

Electron candidates are identified as clusters in the BGO
electromagnetic calorimeter in the range
$|\cos\theta|<0.98$~\cite{l3-196,patrick}. Two triggers collect these
events: a charged-track trigger which requires two back-to-back tracks
and a calorimeter-based energy trigger. The combined efficiency of the
two triggers, measured from the data, is close to 100\%, with a
negligible uncertainty, for $|\cos\theta|<0.72$ and is $99.0\pm0.1\%$ for
$|\cos\theta|<0.98$. 

The electron-candidate clusters must be
associated with tracks which contain at least 20\% of the expected
number of hits in a three-degree azimuthal wedge around the electron
direction. Backgrounds from tau-pair production, lepton production in
photon-photon collisions and fully-leptonic decays of W-boson pairs
are removed by selection criteria on the energy of the clusters.  For
the barrel region, $|\cos\theta|<0.72$, the energy of the most energetic cluster must
satisfy $E_1> 0.5 E_{\rm beam}$, while the energy of the other cluster
must satisfy $E_2>20\GeV$. For the endcap regions, $0.81<|\cos\theta|<0.98$, these
criteria are relaxed to $E_1>0.4 E_{\rm beam}$ and $E_2>10\GeV$.
Events with clusters in the region between the BGO barrel and either one
of the BGO endcaps, $0.72<|\cos\theta|<0.81$, instrumented with a lead and
scintillating-fibre calorimeter~\cite{L3-EGAP}, are rejected.  

The absolute value of the cosine of the centre-of-mass scattering
angle, $|\cos\theta^\star|$, is determined from the polar angles of
the electron candidates as:   
\begin{equation}
  \label{eq:cos}
  |\cos\theta^\star|\equiv
  \frac{\left|\sin\theta_1-\theta_2\right|}{\sin\theta_1+\sin\theta_2}.
\end{equation}
Only events in the fiducial volume $|\cos\theta|<0.72$ are used to
measure the cross section and \Afb, while the measurement of the
differential cross section covers the fiducial volume
$|\cos\theta^\star|<0.9$.

Figure~\ref{fig:sele}d shows the data and Monte Carlo distributions of
$E_1 /E_{\rm beam }$ for the full data sample, while
Figure~\ref{fig:spri}d presents the $\zeta$ distribution for
$\sqrt{s}=207\GeV$.

Selection efficiencies and background contributions are listed in
Tables~\ref{tab:seleinc} and~\ref{tab:sele}  for the inclusive and
high-energy samples, respectively.  The largest residual backgrounds
are from tau-pair production and W-boson pair production. Minor
sources of background are electron production in photon-photon
collisions, and the $\epem\rightarrow\epem\gamma$,
$\epem\ra\mathrm{Z}\epem$ and $\epem\rightarrow\gamma\gamma(\gamma)$
processes. ISR contamination is negligible.

%%%%%%%%%%%%%%%%%%%%%%%%%%%%%%%%%%%%%%%%%%%%%%%%%%%%%%%%%%%%%%%%%%%%%%%%%%%%%%%%
\subsection{Results}
%%%%%%%%%%%%%%%%%%%%%%%%%%%%%%%%%%%%%%%%%%%%%%%%%%%%%%%%%%%%%%%%%%%%%%%%%%%%%%%%

Table~\ref{tab:xsecincl} presents the numbers of observed events and
the measurements of the cross sections of the
$\epem\rightarrow\epem(\gamma)$ process for the inclusive and
high-energy samples for $|\cos\theta|<0.72$, together with their
statistical and systematic uncertainties.
Figure~\ref{fig:ee_xsec_afb}a compares these cross sections with
Standard Model predictions. Good agreement is observed with
$\rm\chi^2/d.o.f.$ of 1.4 and 1.3 for the inclusive and high-energy
samples, respectively.

Table~\ref{tab:bhabha1}, continued in Table~\ref{tab:bhabha2},
presents the differential cross section as a function of
$|\cos\theta^\star|$, together with the numbers of observed events and
the background fractions, along with the selection efficiencies for
the high-energy sample. Only
events with $|\cos\theta^\star|<0.9$ are considered. The differential
cross section is compared in Figure~\ref{fig:ee_dsig} to the Standard
Model predictions of the BHWIDE Monte Carlo, also given in
Tables~\ref{tab:bhabha1} and~\ref{tab:bhabha2}. Good agreement is
found with $\rm\chi^2/d.o.f.=1.0$.

The measured values of the forward-backward asymmetry are listed in
Table~\ref{tab:afb}, together with the numbers of events selected in
the forward and backward hemispheres. This measurements include a
correction for the charge confusion per event, estimated from data,
which varies between 4.5\% and 8.9\% according to the polar angle. A
comparison with Standard Model predictions, also listed in
Table~\ref{tab:afb} and presented in Figure~\ref{fig:ee_xsec_afb}b,
shows good agreement, with
$\rm\chi^2/d.o.f.$ of 1.4 and 1.1 for the inclusive and high-energy
samples, respectively.

%%%%%%%%%%%%%%%%%%%%%%%%%%%%%%%%%%%%%%%%%%%%%%%%%%%%%%%%%%%%%%%%%%%%%%%%%%%%%%%%
\subsection{Systematic uncertainties}
%%%%%%%%%%%%%%%%%%%%%%%%%%%%%%%%%%%%%%%%%%%%%%%%%%%%%%%%%%%%%%%%%%%%%%%%%%%%%%%%

The systematic uncertainties on the $\epem\rightarrow\epem(\gamma)$
cross sections and \Afb\ are between a factor two and
five smaller than the corresponding statistical
uncertainties~\cite{l3-196,patrick}. Excluding the luminosity-limited 
highest
centre-of-mass energy, statistical uncertainties on the cross sections
of $2.0-3.8$\% and $2.0-4.0$\% are observed for the inclusive and
high-energy samples, respectively, while systematic uncertainties are
about 1.3\% and 0.5\%, respectively. The high-energy asymmetry is
measured with a statistical precision of $2.0-4.0$\%, depending on the
centre-of-mass energy, while its systematic uncertainty is about 1\%.

The systematic uncertainties on the high-energy cross sections and
\Afb\ are dominated by the modelling of the tracker response and of edge
effects in the control of the fiducial volume, 0.43\%. Another
important contribution arises from the
limited signal, $0.19-0.31$\%, and background, 0.30\%, Monte Carlo
statistics. The simulation of the calorimeter response for the
most-energetic electron contributes 0.13\% to the total systematic
uncertainty while the simulation of the least-energetic electron
contributes 0.15\%. The systematic uncertainty on \Afb\ contains an
additional contribution of about 0.1\% arising from the relative
uncertainty on the charge-confusion, determined in data as $17-24$\%.

The determination of the differential cross section is also limited by
the statistical uncertainties. It has the same sources of systematic
uncertainty discussed above. At $\sqrt{s}=207\GeV$, the detector
modelling and control of the fiducial volume contributes $0.2- 1.5\%$
and the limited background Monte Carlo statistics contributes
$0.2-3.3\%$, depending on the polar angle. The effect of the limited
signal Monte Carlo statistics raises to $0.5-8.1$\%, depending on the
polar angle. While the overall increase of this source is due to the
increased number of bins in which the Monte Carlo is divided, the
largest amount corresponds to the region
$0.72<|\cos\theta^\star|<0.81$, where some extrapolation factors
account for the transition region between the barrel and endcap BGO
calorimeters which is not used to identify electrons.

The systematic uncertainties on the high-energy measurements of the
cross sections and asymmetries are mostly correlated between the energy
points. For the measurements of the differential cross sections,
systematic uncertainties are mostly uncorrelated  between energy points
and between different angular ranges.

%%%%%%%%%%%%%%%%%%%%%%%%%%%%%%%%%%%%%%%%%%%%%%%%%%%%%%%%%%%%%%%%%%%%%%%%%%%%%%%%
\section{Summary and conclusions}
%%%%%%%%%%%%%%%%%%%%%%%%%%%%%%%%%%%%%%%%%%%%%%%%%%%%%%%%%%%%%%%%%%%%%%%%%%%%%%%%
\label{sec:last}

A detailed study of the properties of fermion-pair production in
$\epem$ collisions at LEP has been performed.  The cross sections for
hadron and lepton-pair production, as well as the forward-backward
asymmetries for lepton-pair production, are measured both for the
inclusive and high-energy samples. In addition, the high-energy samples of
electron-positron pair-production are used to measure the differential cross
sections as a function of the scattering angle.  

\begin{figure}[p]
  \begin{center}
    \includegraphics[width=0.85\textwidth]{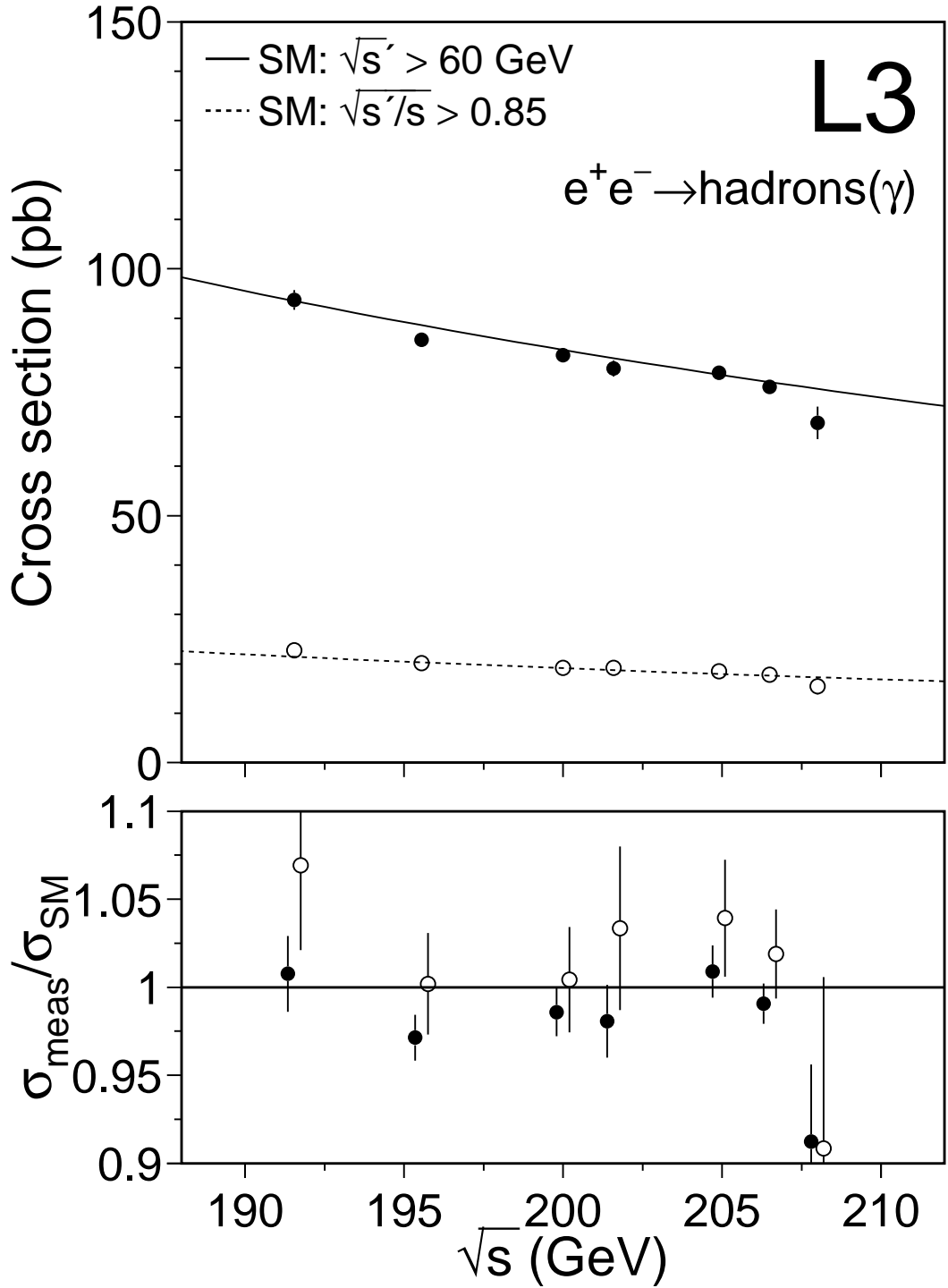}
    \parbox{\capwidth}{
      \caption[]{ Cross sections of the process
        {$\epem\rightarrow\mbox{hadrons}\,(\gamma)$} for the inclusive
        sample, solid symbols, and the high-energy sample, open
        symbols.  The {\SM} predictions are shown as a solid line for
        the inclusive sample and as a dashed line for the high-energy
        sample. The lower plot shows the ratio of measured and
        predicted cross sections; for clarity, 
        symbols denoting the two final states are slightly
        shifted. The bars correspond to the sum 
        in quadrature of statistical and systematic uncertainties.}
      \label{fig:ha_xsec}}
  \end{center}
\end{figure}

\begin{figure}[p]
  \begin{center}
    \includegraphics[height=0.4\textheight]{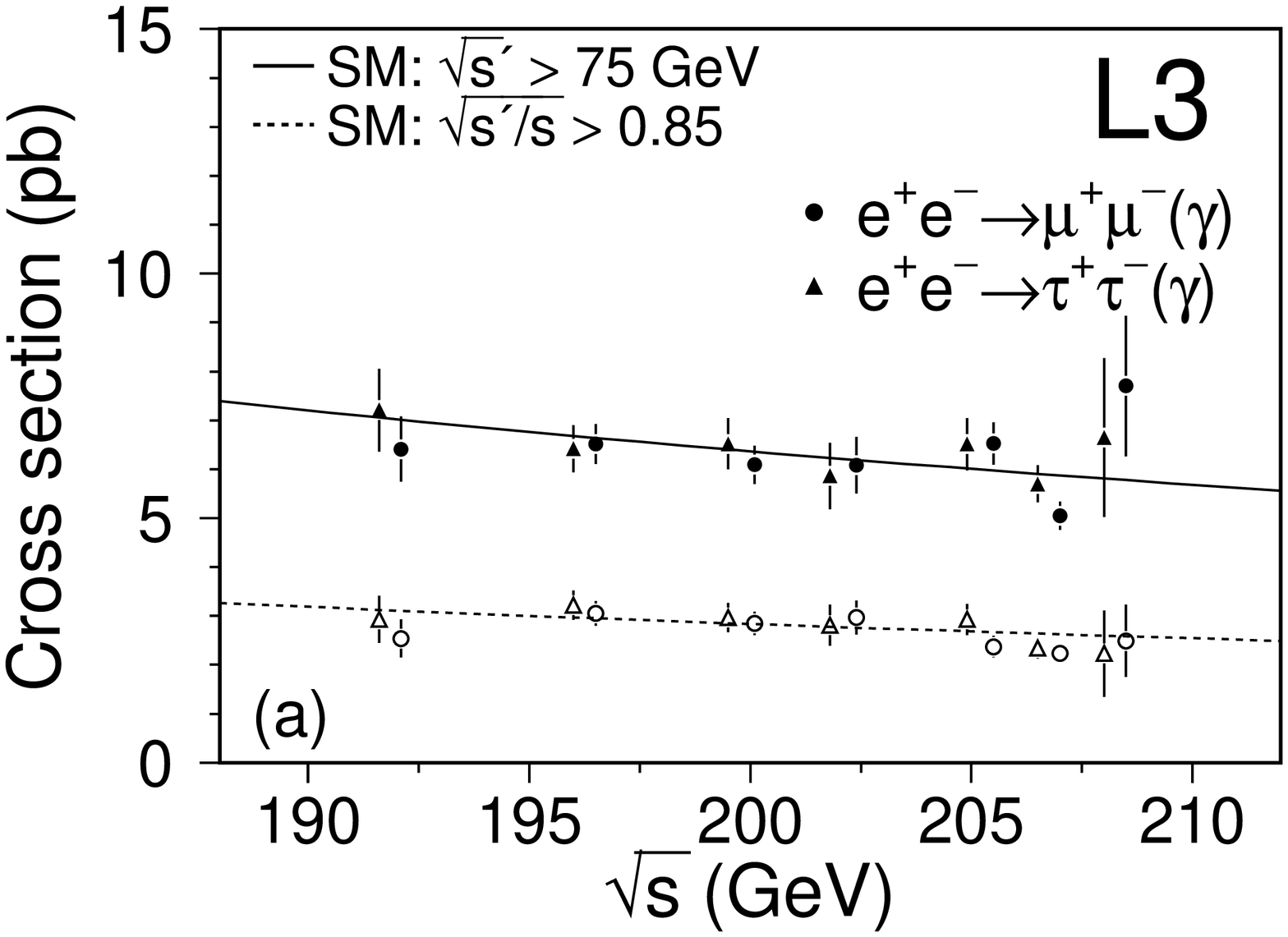}
    \includegraphics[height=0.4\textheight]{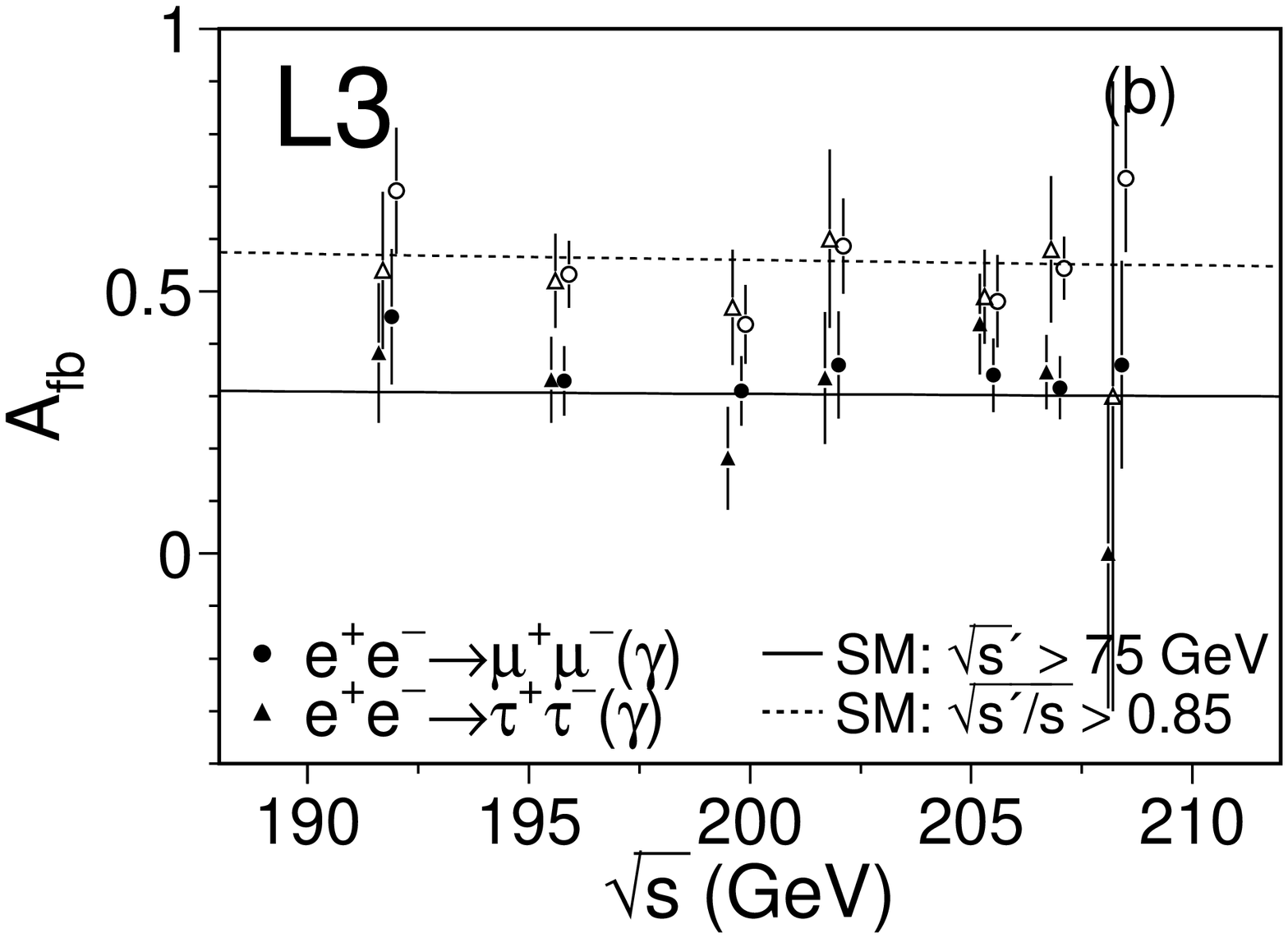}
    \parbox{\capwidth}{
      \caption[]{ a) Cross sections and b) forward-backward
        asymmetries, \Afb, of the {$\epem\rightarrow\mumu(\gamma)$}
        and {$\epem\rightarrow\tautau(\gamma)$} processes for the
        inclusive sample, solid symbols, and the high-energy sample,
        open symbols. The {\SM} predictions are shown as solid lines
        for the inclusive sample and as dashed lines for the
        high-energy sample. For clarity, the solid and open
        symbols are slightly shifted. The bars correspond to the sum
        in quadrature of statistical and systematic uncertainties.}
      \label{fig:le_xsec_afb}}
  \end{center}
\end{figure}

\begin{figure}[p]
  \begin{center}
    \includegraphics[height=0.4\textheight]{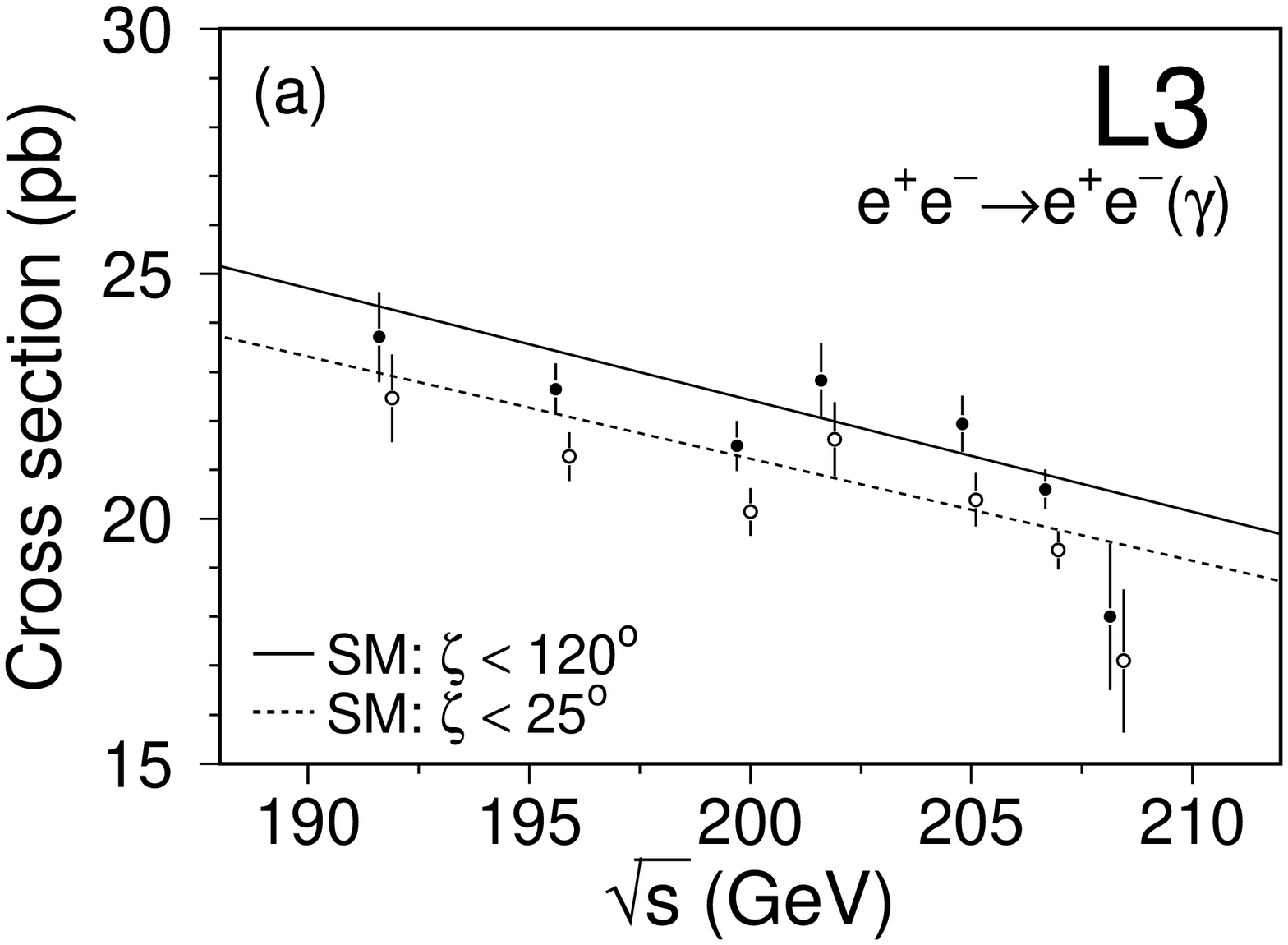}
    \includegraphics[height=0.4\textheight]{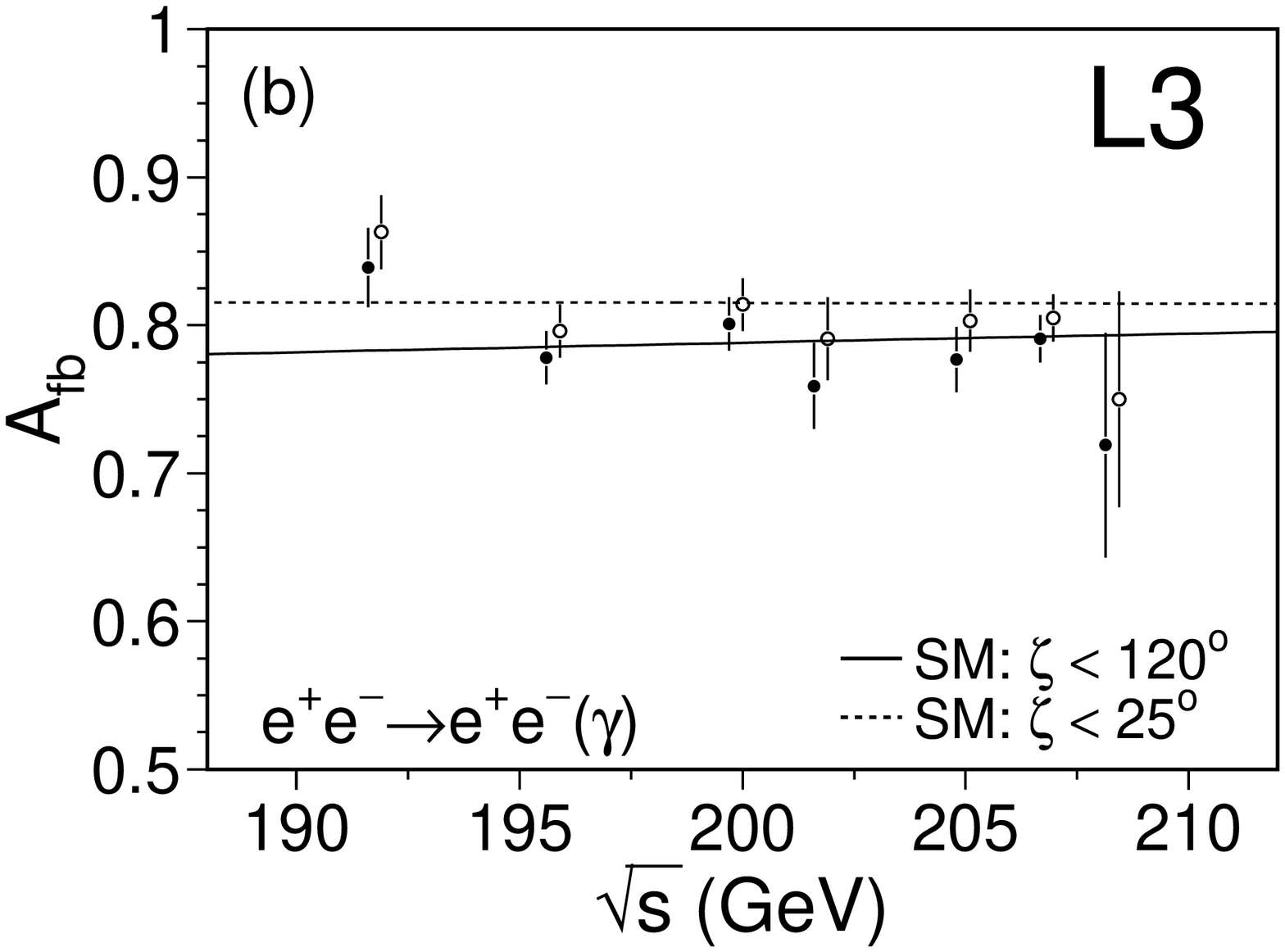}
      \caption[]{ a) Cross sections and b) forward-backward
        asymmetries, \Afb, of the {$\epem\rightarrow\epem(\gamma)$}
        process for $|\cos\theta|<0.72$ for the
        inclusive sample, solid symbols, and the high-energy sample,
        open symbols. The {\SM} predictions are shown as solid lines
        for the inclusive sample and as dashed lines for the
        high-energy sample. For clarity, the solid and open
        symbols are slightly shifted. The bars correspond to the sum
        in quadrature of statistical and systematic uncertainties.
      \label{fig:ee_xsec_afb}}
  \end{center}
\end{figure}

\begin{figure}[p]
  \begin{center}
    \includegraphics[width=1.\textwidth]{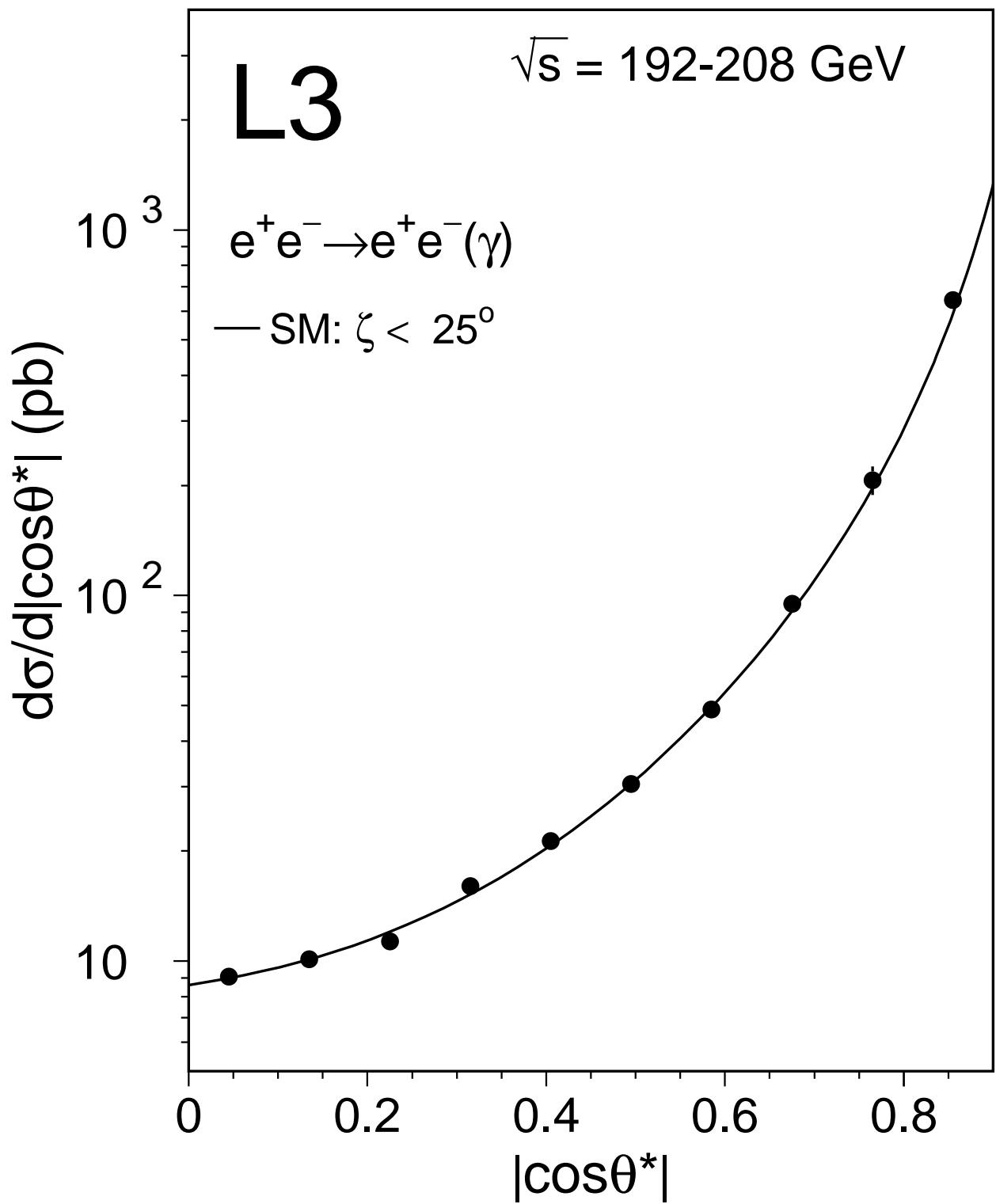}
      \caption[]{Differential cross section of the
        $\epem\rightarrow\epem(\gamma)$ process for the high-energy
        sample at $\sqrt{s} = 192-208 \GeV$, corresponding to an
        average centre-of-mass energy
        $\langle\sqrt{s}\rangle=201.4\GeV$. The line indicates the
        \SM\ prediction.
      \label{fig:ee_dsig}}
  \end{center}
\end{figure}

These results are summarised in
Figures~\ref{fig:ha_xsec}$-$\ref{fig:ee_dsig} and
Tables~\ref{tab:xsecincl}$-$\ref{tab:bhabha2} together with their
statistical and systematic uncertainties. To a good approximation,
systematic uncertainties are not correlated between the different
final states, both for the cross section measurements and for the
measurements of the forward-backward asymmetries. 

\begin{figure}[p]
  \begin{center}
    \includegraphics[width=0.9\textwidth]{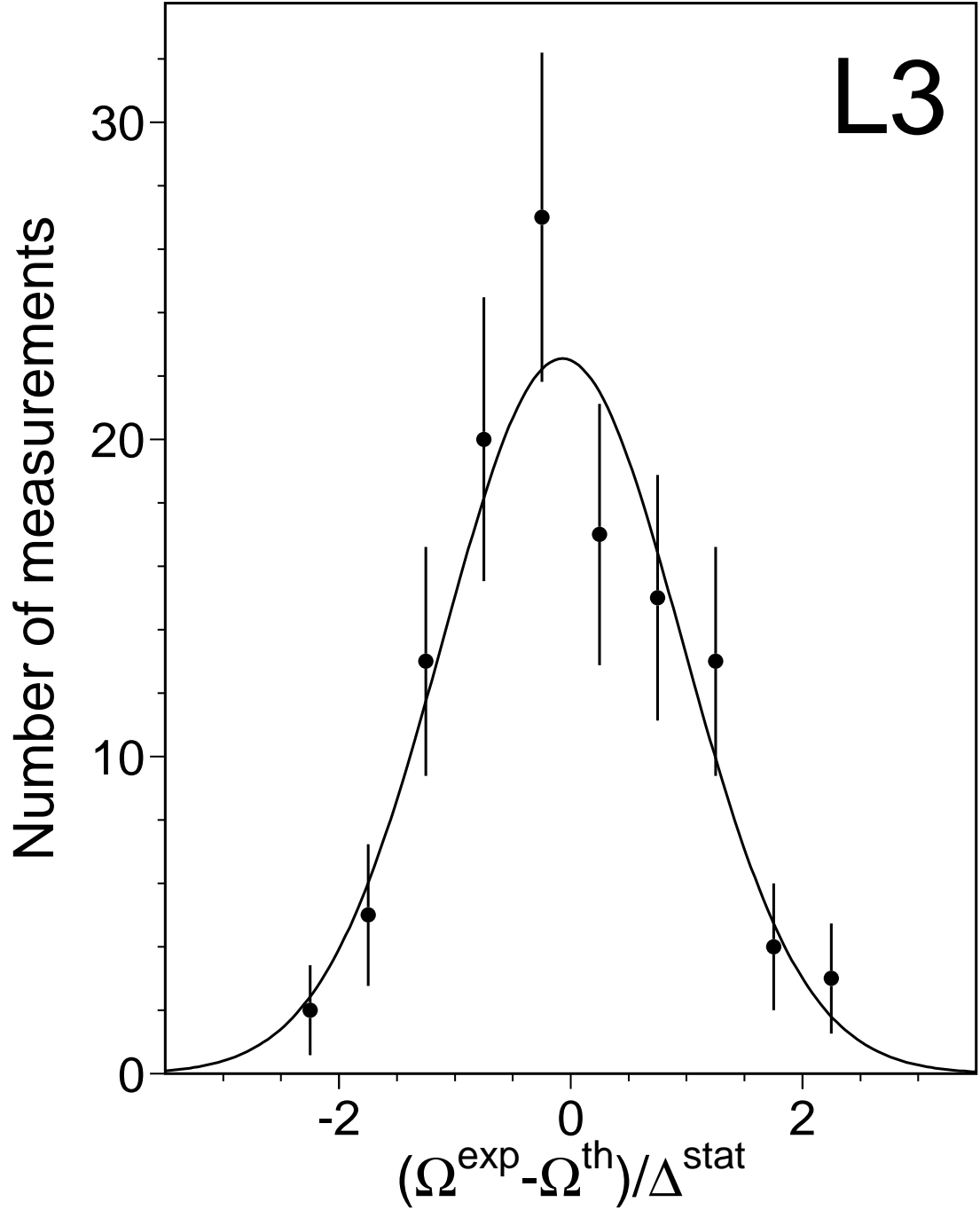}
      \caption[]{Distribution of the difference of the measured total
      and differential cross sections and forward-backward asymmetries
      and the corresponding Standard Model predictions divided by the
      statistical uncertainty of the measurements. Only the
      high-energy samples are considered. The line represents the
      results of a Gaussian fit to this distribution, which finds a
      mean of $-0.07\pm0.10$ and a width of $1.03\pm0.09$, in
      excellent agreement with the expected spread of the
      measurements.
      \label{fig:prob}}
  \end{center}
\end{figure}

The global agreement of the results presented in this paper with the
Standard Model expectations is presented in Figure~\ref{fig:prob}.
The 119 measurements of total and differential cross sections and of
forward-backward asymmetries for the high-energy samples are
considered. For each measurement, its difference with respect to the
Standard Model expectation is plotted, divided by the statistical
uncertainty. An excellent agreement with the expected Gaussian
statistical spread of the measurements is observed.

The results on the $\epem\ra\epem(\gamma)$ process give access to the
evolution of the electromagnetic coupling with the momentum transfer,
whose measurement is discussed in a companion letter~\cite{alphaQed}.
These measurements allow a search for manifestations of new physics at
a scale which would not be directly detected at
LEP~\cite{manifestations}.

\begin{figure}[p]
  \begin{center}
    \includegraphics[width=0.9\textwidth]{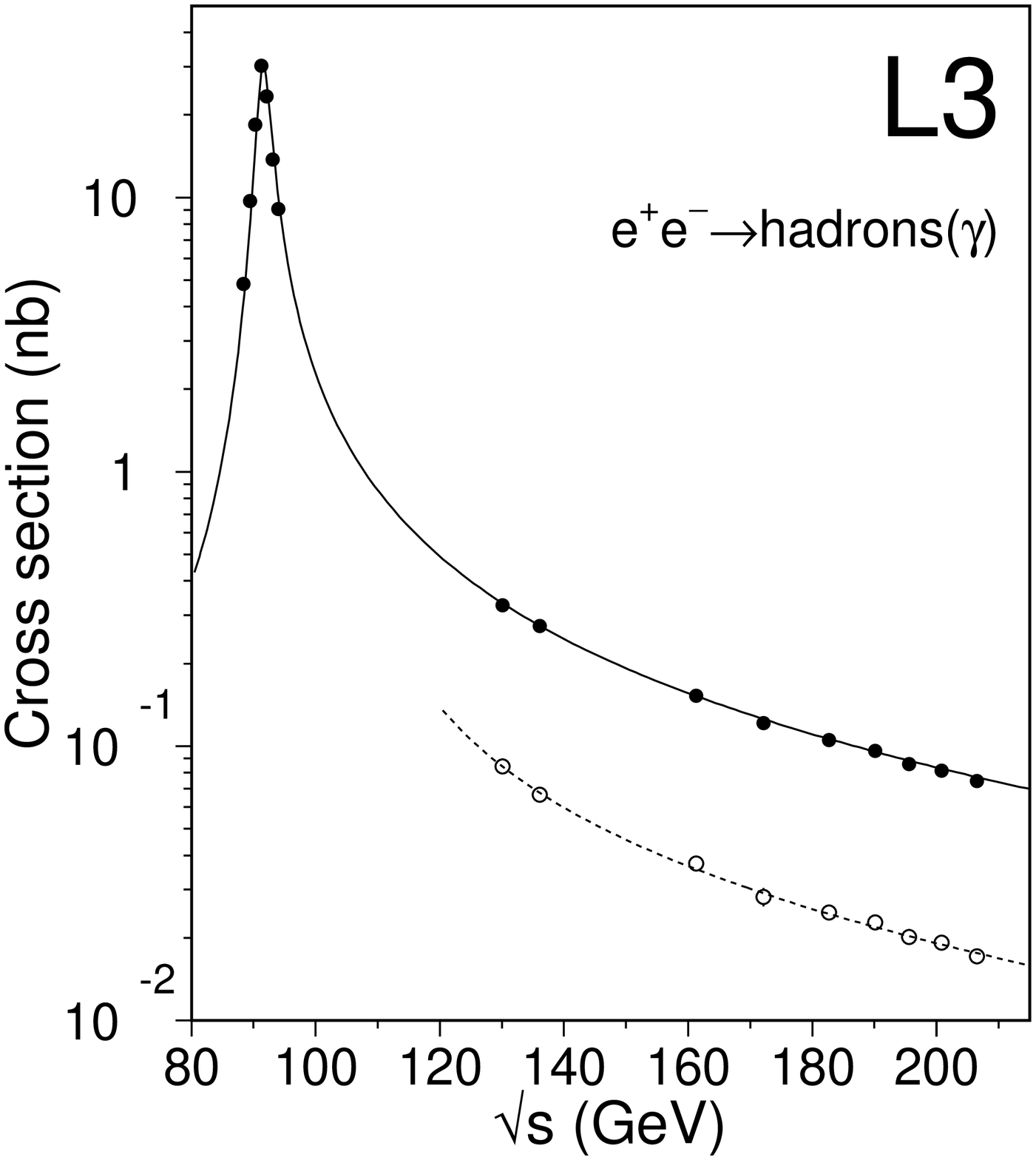}
      \caption[]{ Cross sections of the process
        {$\epem\rightarrow\mbox{hadrons}\,(\gamma)$}, for the inclusive
        sample, solid symbols, and the high-energy sample, open
        symbols.  The {\SM} predictions are shown as a solid line for
        the inclusive sample and as a dashed line for the high-energy
        sample.  The entire LEP data-sample is shown. The bars correspond to the sum
        in quadrature of statistical and systematic uncertainties.
      \label{fig:qqm}}
  \end{center}
\end{figure}

\begin{figure}[p]
  \begin{center}
    \includegraphics[width=0.9\textwidth]{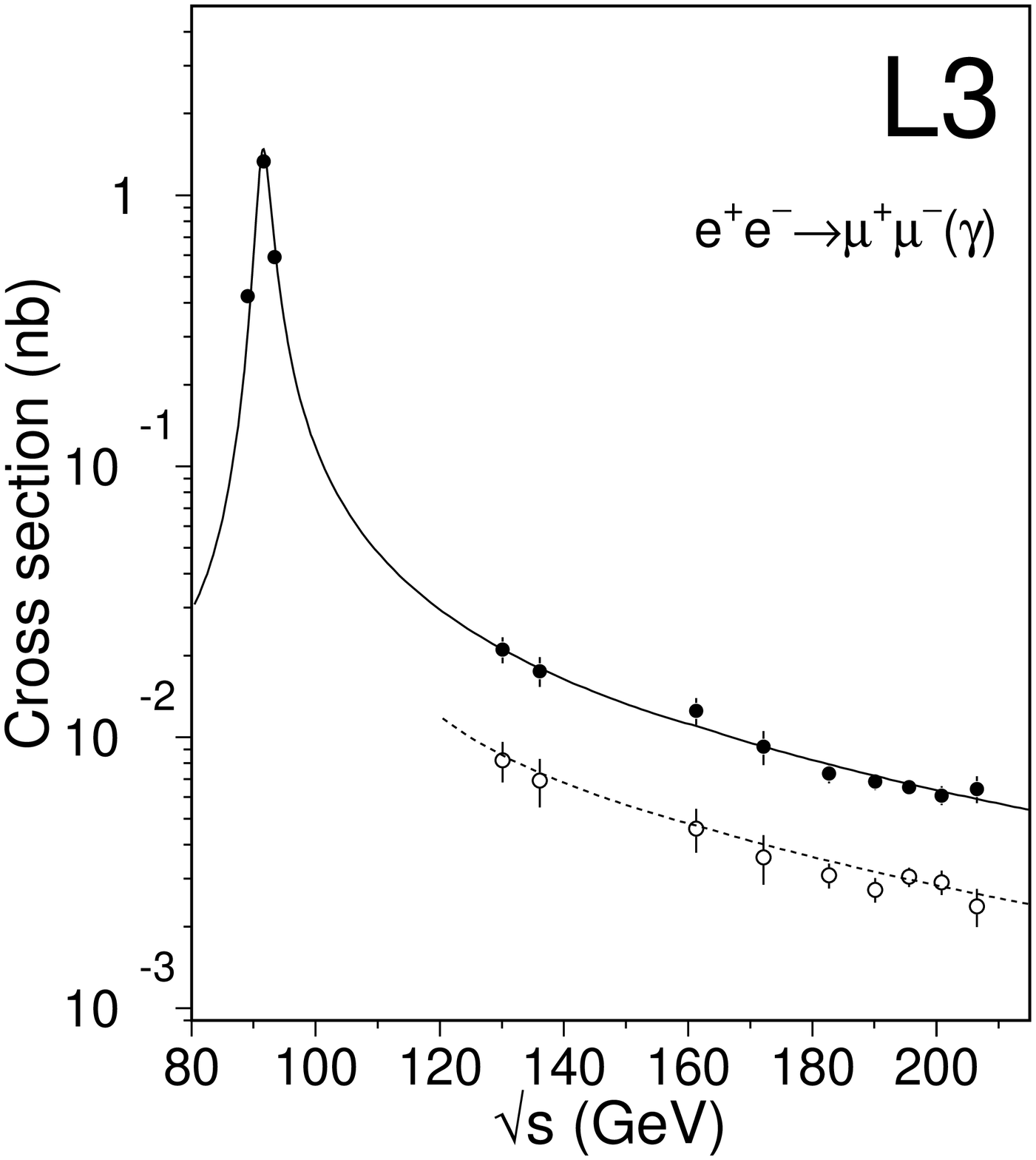}
      \caption[]{ Cross sections  of the
         {$\epem\rightarrow\mumu(\gamma)$} process for the inclusive
         sample, solid symbols, and the high-energy sample, open
         symbols. The {\SM} predictions are shown as solid lines for
         the inclusive sample and as dashed lines for the high-energy
         sample.  The entire LEP data-sample is shown. The bars correspond to the sum
        in quadrature of statistical and systematic uncertainties.
      \label{fig:mmm}}
  \end{center}
\end{figure}

\begin{figure}[p]
  \begin{center}
    \includegraphics[width=0.9\textwidth]{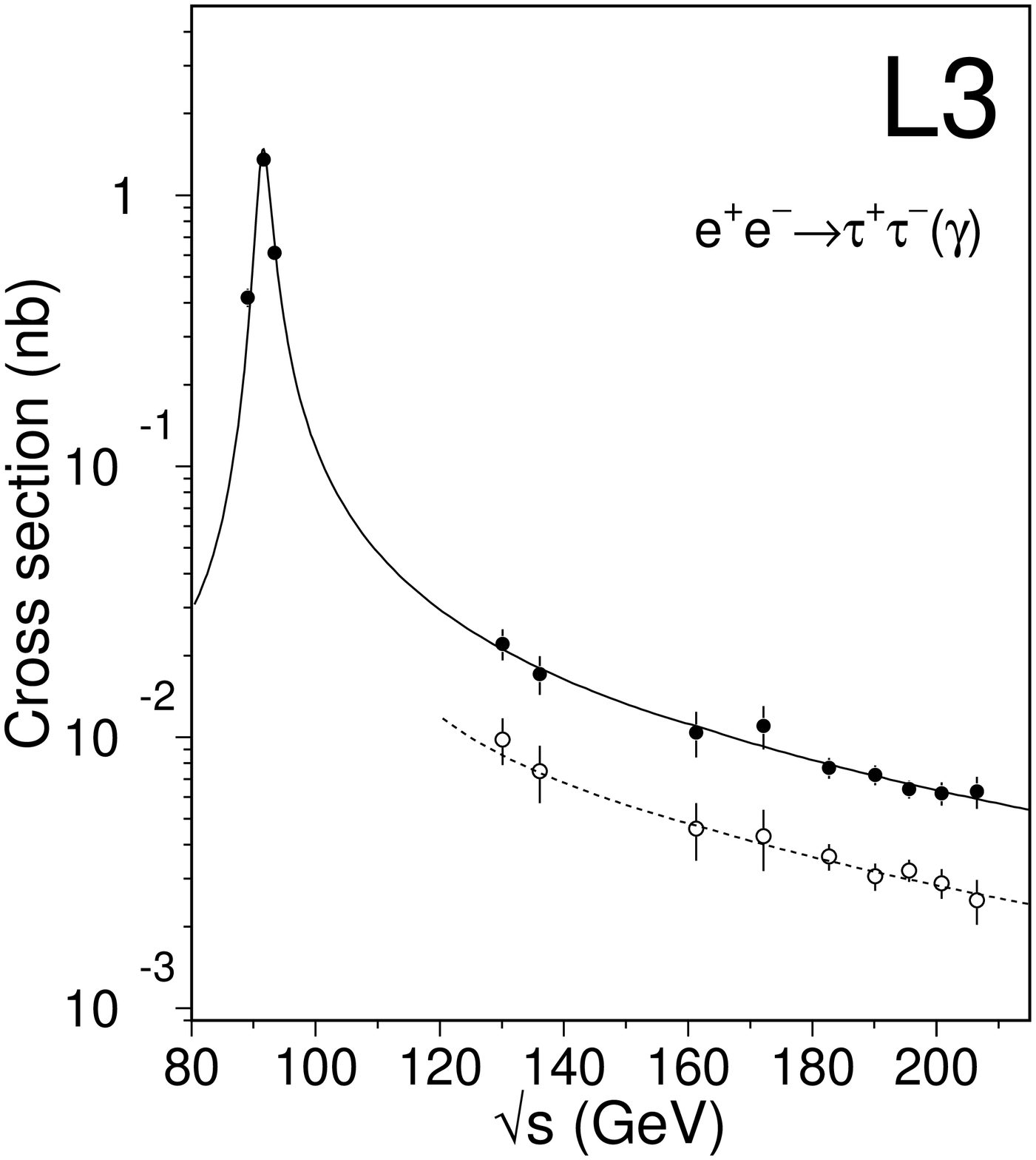}
      \caption[]{ Cross sections of the
         {$\epem\rightarrow\tautau(\gamma)$} process for the inclusive
         sample, solid symbols, and the high-energy sample, open
         symbols. The {\SM} predictions are shown as solid lines for
         the inclusive sample and as dashed lines for the high-energy
         sample.  The entire LEP data-sample is shown. The bars correspond to the sum
        in quadrature of statistical and systematic uncertainties.
      \label{fig:ttm}}
  \end{center}
\end{figure}

\begin{figure}[p]
  \begin{center}
    \includegraphics[width=0.9\textwidth]{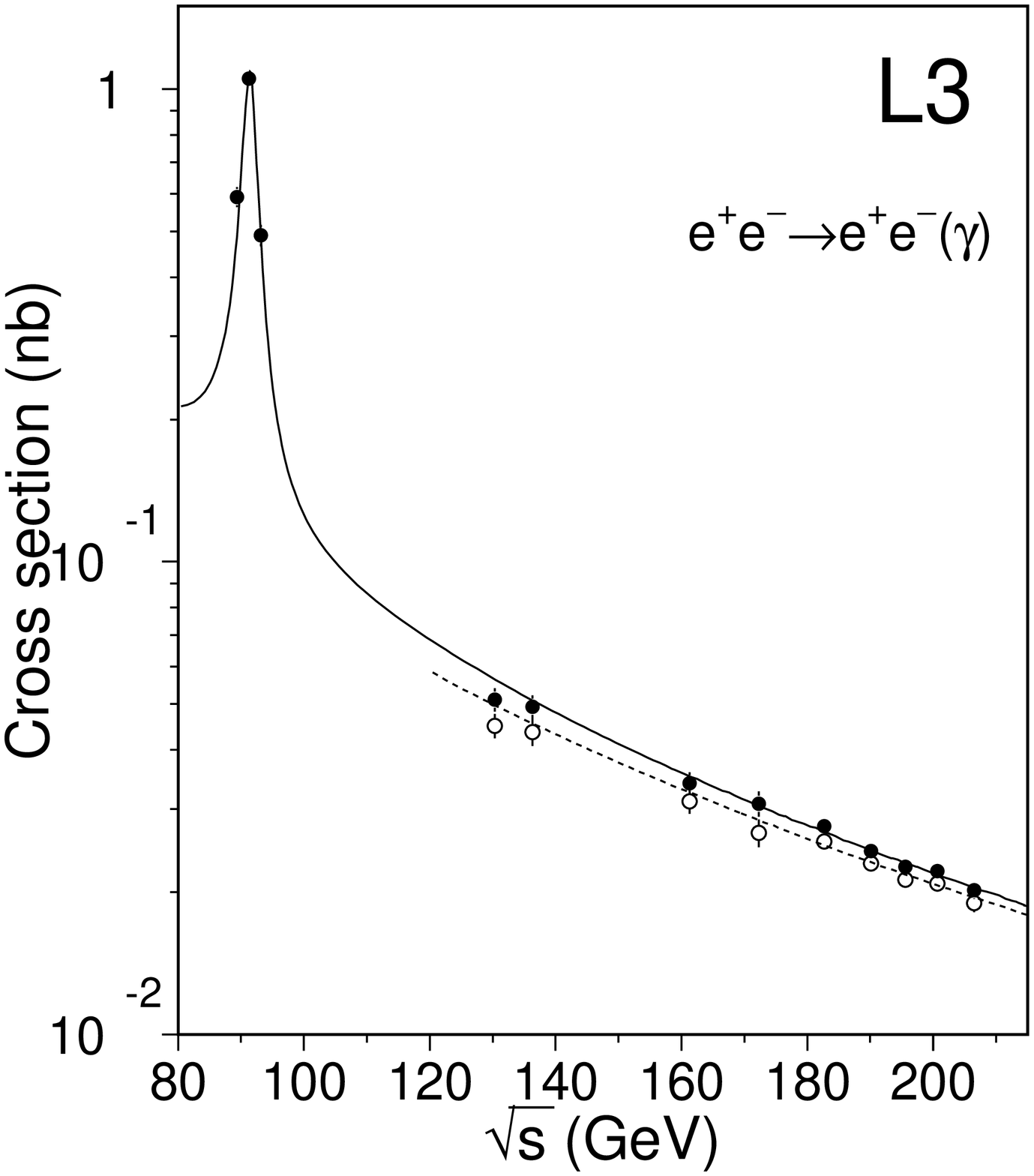}
      \caption[]{ Cross sections  of the
         {$\epem\rightarrow\epem(\gamma)$} process in the angular
         region $|\cos\theta|<0.72$ for the inclusive sample, solid
         symbols, and the high-energy sample, open symbols. The {\SM}
         predictions are shown as solid lines for the inclusive sample
         and as dashed lines for the high-energy sample.  The entire
         LEP data-sample is shown. The bars correspond to the sum
        in quadrature of statistical and systematic uncertainties.
      \label{fig:eem}}
  \end{center}
\end{figure}

\begin{figure}[p]
  \begin{center}
    \includegraphics[width=0.9\textwidth]{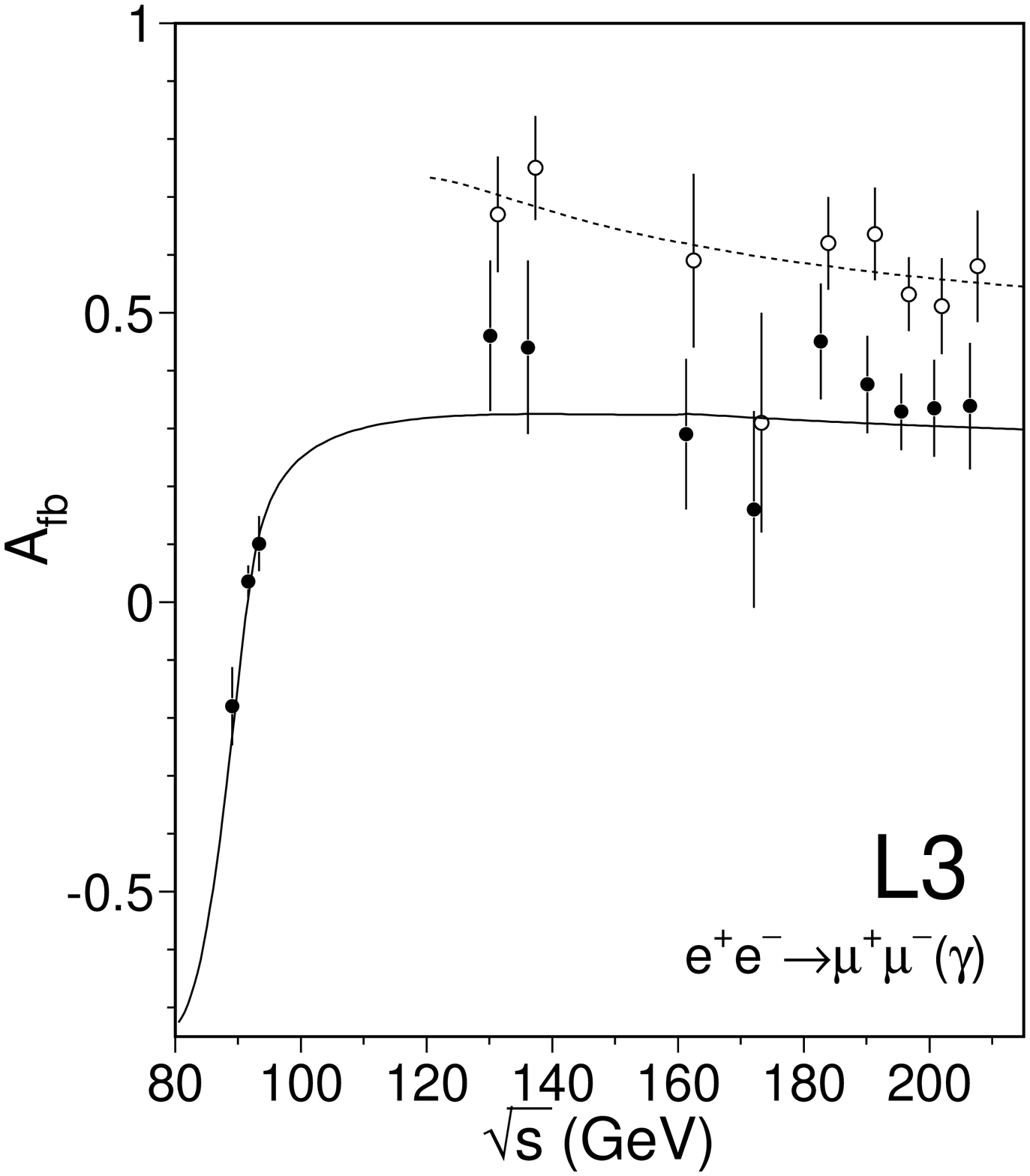}
      \caption[]{ Forward-backward asymmetries of the 
        {$\epem\rightarrow\mumu(\gamma)$} process for the inclusive
        sample, solid symbols, and the high-energy sample, open
        symbols. The {\SM} predictions are shown as solid lines for
        the inclusive sample and as dashed lines for the high-energy
        sample.  The entire LEP data-sample is shown.  For clarity, the solid and open
        symbols are slightly shifted. The bars correspond to the sum
        in quadrature of statistical and systematic uncertainties.
      \label{fig:ammm}}
  \end{center}
\end{figure}

\begin{figure}[p]
  \begin{center}
    \includegraphics[width=0.9\textwidth]{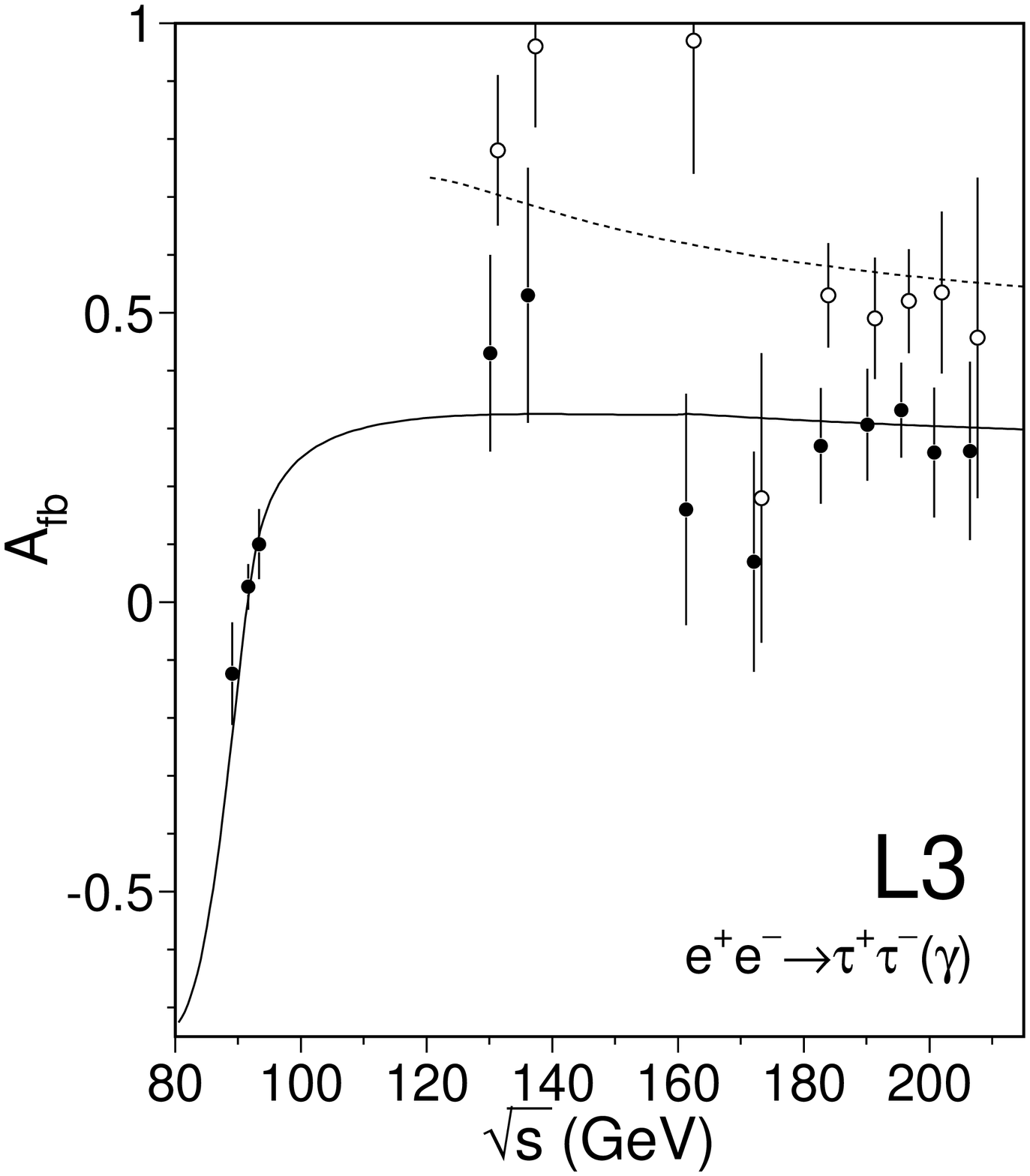}
      \caption[]{ Forward-backward asymmetries of the
         {$\epem\rightarrow\tautau(\gamma)$} process for the inclusive
         sample, solid symbols, and the high-energy sample, open
         symbols. The {\SM} predictions are shown as solid lines for
         the inclusive sample and as dashed lines for the high-energy
         sample.  The entire LEP data-sample is shown.  For clarity, the solid and open
        symbols are slightly shifted. The bars correspond to the sum
        in quadrature of statistical and systematic uncertainties.
      \label{fig:attm}}
  \end{center}
\end{figure}

\begin{figure}[p]
  \begin{center}
    \includegraphics[width=0.9\textwidth]{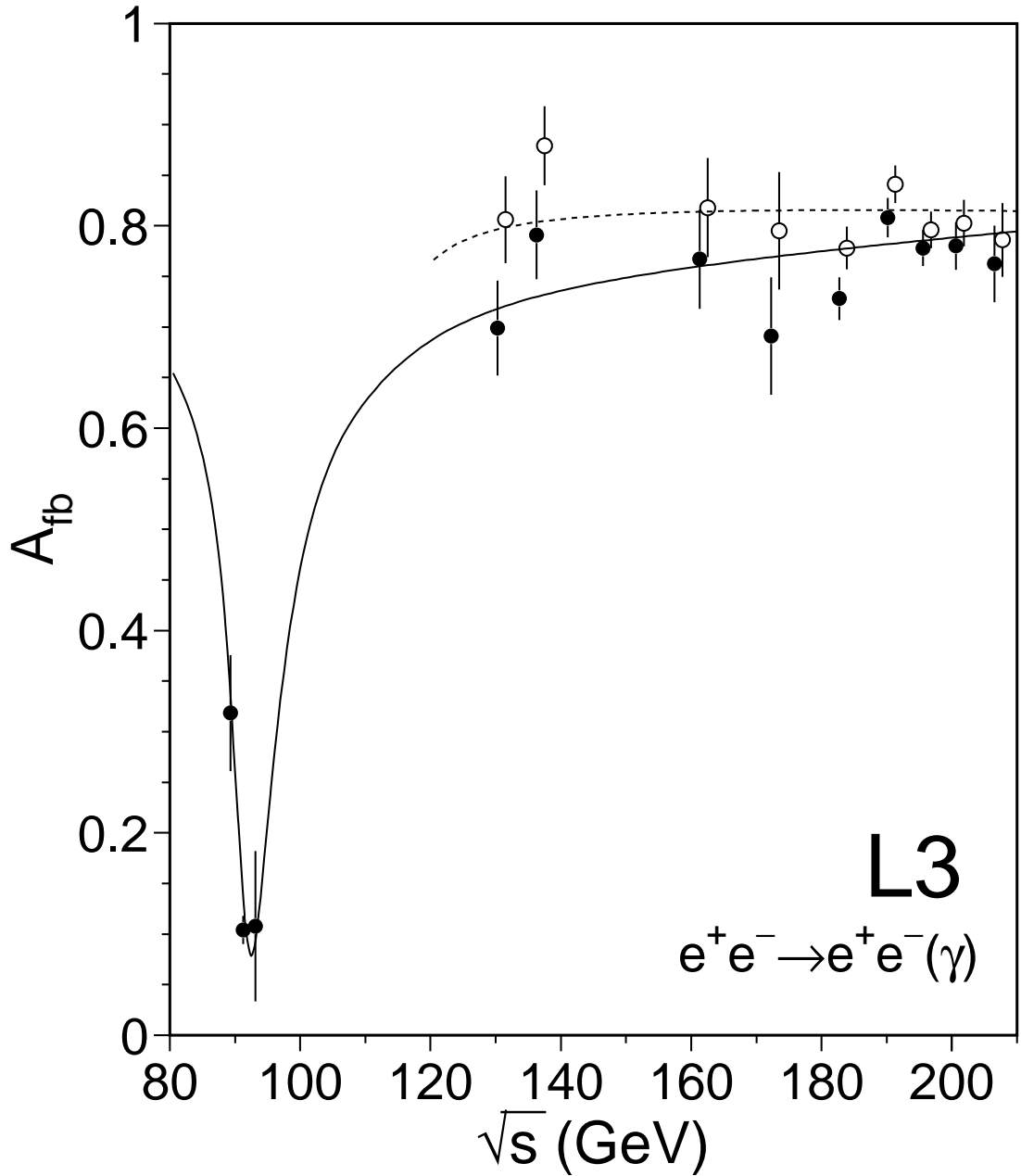}
      \caption[]{ Forward-backward asymmetries of the 
         {$\epem\rightarrow\epem(\gamma)$} process in the angular
         region $|\cos\theta|<0.72$ for the inclusive sample, solid
         symbols, and the high-energy sample, open symbols. The {\SM}
         predictions are shown as solid lines for the inclusive sample
         and as dashed lines for the high-energy sample.  The entire
         LEP data-sample is shown.  For clarity, the solid and open
        symbols are slightly shifted. The bars correspond to the sum
        in quadrature of statistical and systematic uncertainties.
      \label{fig:aeem}}
  \end{center}
\end{figure}

These data complete the picture of fermion-pair production at LEP at
$\sqrt{s} = 90 \GeV - 209 \GeV$:
Figures~\ref{fig:qqm}$-$\ref{fig:aeem} combine the results of this
paper and of previous studies~\cite{l3-202,l3-90+l3-117,l3-196} to
present the cross sections and forward-backward asymmetries measured
with the L3 detector.  Over the whole energy range explored at LEP,
the measurements are well described by the predictions of the Standard
Model.

%%%%%%%%%%%%%%%%%%%%%%%%%%%%%%%%%%%%%%%%%%%%%%%%%%%%%%%%%%%%%%%%%%%%%%%%%%%%%%%%
%                              BIBLIOGRAPHY
%%%%%%%%%%%%%%%%%%%%%%%%%%%%%%%%%%%%%%%%%%%%%%%%%%%%%%%%%%%%%%%%%%%%%%%%%%%%%%%%
\bibliographystyle{l3stylem}
\bibliography{fermionPairs}

%%%%%%%%%%%%%%%%%%%%%%%%%%%%%%%%%%%%%%%%%%%%%%%%%%%%%%%%%%%%%%%%%%%%%%%%%%%%%%%%
%                              AUTHOR LIST
%%%%%%%%%%%%%%%%%%%%%%%%%%%%%%%%%%%%%%%%%%%%%%%%%%%%%%%%%%%%%%%%%%%%%%%%%%%%%%%%
\clearpage
\typeout{   }     
\typeout{Using author list for paper 287 -  }
\typeout{$Modified: Jul 15 2001 by smele $}
\typeout{!!!!  This should only be used with document option a4p!!!!}
\typeout{   }
%
%
%
%  L A T E X  version!!
%
%
% Make sure that the Lep package has been used!
%\input{Lep.sty}%
%
%\ifx\LepCalled\undefined%
%\typeout{     }%
%\typeout{!!!!!!!!!!!!!!!!!!!!!!!!!!!!!!!!!!!!!!!!!!!!!!!!!!!!!!!!!!!}%
%\typeout{Yikes.  You haven't used the Lep package!}%
%\typeout{Please put \protect\usepackage\protect{Lep\protect} in your preamble,
%         followed by}%
%\typeout{\protect\Lep\protect{1\protect} or \protect\Lep\protect{2\protect}}%
%\typeout{     }%
%\typeout{For now you will get a Lep phase 2 authorlist (may not be right!).}%
%\typeout{!!!!!!!!!!!!!!!!!!!!!!!!!!!!!!!!!!!!!!!!!!!!!!!!!!!!!!!!!!!}%
%\typeout{     }%
%\Lep{2}\fi%

\newcount\tutecount  \tutecount=0
\def\tutenum#1{\global\advance\tutecount by 1 \xdef#1{\the\tutecount}}
\def\tute#1{$^{#1}$}
\tutenum\aachen            % 1 
\tutenum\nikhef            % 2 
\tutenum\mich              % 3 
\tutenum\lapp              % 4 
\tutenum\basel             % 5 
\tutenum\lsu               % 6 
\tutenum\beijing           % 7 
\tutenum\bologna           % 8 
\tutenum\tata              % 9 
\tutenum\ne                % 10
\tutenum\bucharest         % 11
\tutenum\budapest          % 12
\tutenum\mit               % 13
\tutenum\panjab            % 14 
\tutenum\debrecen          % 15
\tutenum\dublin            % 16
\tutenum\florence          % 17
\tutenum\cern              % 18
\tutenum\wl                % 19
\tutenum\geneva            % 20
\tutenum\hamburg           % 21
\tutenum\hefei             % 22
\tutenum\lausanne          % 23
\tutenum\lyon              % 24
\tutenum\madrid            % 25
\tutenum\florida           % 26
\tutenum\milan             % 27
\tutenum\moscow            % 29
\tutenum\naples            % 30
\tutenum\cyprus            % 31
\tutenum\nymegen           % 32
\tutenum\caltech           % 33
\tutenum\perugia           % 34
\tutenum\peters            % 35
\tutenum\cmu               % 36
\tutenum\potenza           % 37
\tutenum\prince            % 38
\tutenum\riverside         % 39
\tutenum\rome              % 40
\tutenum\salerno           % 41
\tutenum\ucsd              % 42
\tutenum\sofia             % 43
\tutenum\korea             % 44
\tutenum\taiwan            % 45
\tutenum\tsinghua          % 46
\tutenum\purdue            % 47
\tutenum\psinst            % 48
\tutenum\zeuthen           % 49
\tutenum\eth               % 50

{
\parskip=0pt
\noindent
{\bf The L3 Collaboration:}
\ifx\selectfont\undefined%  old style font selection
 \baselineskip=10.8pt
 \baselineskip\baselinestretch\baselineskip
 \normalbaselineskip\baselineskip
 \ixpt
\else%                      new style font selection
 \fontsize{9}{10.8pt}\selectfont
\fi
\medskip
\tolerance=10000
\hbadness=5000
\raggedright
\hsize=162truemm\hoffset=0mm
\def\r{\rlap,}
\noindent

P.Achard\r\tute\geneva\ 
O.Adriani\r\tute{\florence}\ 
M.Aguilar-Benitez\r\tute\madrid\ 
J.Alcaraz\r\tute{\madrid}\ 
G.Alemanni\r\tute\lausanne\
J.Allaby\r\tute\cern\
A.Aloisio\r\tute\naples\ 
M.G.Alviggi\r\tute\naples\
H.Anderhub\r\tute\eth\ 
V.P.Andreev\r\tute{\lsu,\peters}\
% ...
\newpage
%\rule{\textwidth}{0.4pt}
\begin{list}{A}{\itemsep=0pt plus 0pt minus 0pt\parsep=0pt plus 0pt minus 0pt
                \topsep=0pt plus 0pt minus 0pt}
\item[\aachen]
 III. Physikalisches Institut, RWTH, D-52056 Aachen, Germany$^{\S}$
\item[\nikhef] National Institute for High Energy Physics, NIKHEF, 
     and University of Amsterdam, NL-1009 DB Amsterdam, The Netherlands
\item[\mich] University of Michigan, Ann Arbor, MI 48109, USA
\item[\lapp] Laboratoire d'Annecy-le-Vieux de Physique des Particules, 
     LAPP,IN2P3-CNRS, BP 110, F-74941 Annecy-le-Vieux CEDEX, France
\item[\basel] Institute of Physics, University of Basel, CH-4056 Basel,
     Switzerland
\item[\lsu] Louisiana State University, Baton Rouge, LA 70803, USA
\item[\beijing] Institute of High Energy Physics, IHEP, 
  100039 Beijing, China$^{\triangle}$ 
\item[\bologna] University of Bologna and INFN-Sezione di Bologna, 
     I-40126 Bologna, Italy
\item[\tata] Tata Institute of Fundamental Research, Mumbai (Bombay) 400 005, India
\item[\ne] Northeastern University, Boston, MA 02115, USA
\item[\bucharest] Institute of Atomic Physics and University of Bucharest,
     R-76900 Bucharest, Romania
\item[\budapest] Central Research Institute for Physics of the 
     Hungarian Academy of Sciences, H-1525 Budapest 114, Hungary$^{\ddag}$
\item[\mit] Massachusetts Institute of Technology, Cambridge, MA 02139, USA
\item[\panjab] Panjab University, Chandigarh 160 014, India
\item[\debrecen] KLTE-ATOMKI, H-4010 Debrecen, Hungary$^\P$
\item[\dublin] UCD School of Physics, University College Dublin, 
 Belfield, Dublin 4, Ireland
\item[\florence] INFN Sezione di Firenze and University of Florence, 
     I-50125 Florence, Italy
\item[\cern] European Laboratory for Particle Physics, CERN, 
     CH-1211 Geneva 23, Switzerland
\item[\wl] World Laboratory, FBLJA  Project, CH-1211 Geneva 23, Switzerland
\item[\geneva] University of Geneva, CH-1211 Geneva 4, Switzerland
\item[\hamburg] University of Hamburg, D-22761 Hamburg, Germany
\item[\hefei] Chinese University of Science and Technology, USTC,
      Hefei, Anhui 230 029, China$^{\triangle}$
\item[\lausanne] University of Lausanne, CH-1015 Lausanne, Switzerland
\item[\lyon] Institut de Physique Nucl\'eaire de Lyon, 
     IN2P3-CNRS,Universit\'e Claude Bernard, 
     F-69622 Villeurbanne, France
\item[\madrid] Centro de Investigaciones Energ{\'e}ticas, 
     Medioambientales y Tecnol\'ogicas, CIEMAT, E-28040 Madrid,
     Spain${\flat}$ 
\item[\florida] Florida Institute of Technology, Melbourne, FL 32901, USA
\item[\milan] INFN-Sezione di Milano, I-20133 Milan, Italy
\item[\moscow] Institute of Theoretical and Experimental Physics, ITEP, 
     Moscow, Russia
\item[\naples] INFN-Sezione di Napoli and University of Naples, 
     I-80125 Naples, Italy
\item[\cyprus] Department of Physics, University of Cyprus,
     Nicosia, Cyprus
\item[\nymegen] Radboud University and NIKHEF, 
     NL-6525 ED Nijmegen, The Netherlands
\item[\caltech] California Institute of Technology, Pasadena, CA 91125, USA
\item[\perugia] INFN-Sezione di Perugia and Universit\`a Degli 
     Studi di Perugia, I-06100 Perugia, Italy   
\item[\peters] Nuclear Physics Institute, St. Petersburg, Russia
\item[\cmu] Carnegie Mellon University, Pittsburgh, PA 15213, USA
\item[\potenza] INFN-Sezione di Napoli and University of Potenza, 
     I-85100 Potenza, Italy
\item[\prince] Princeton University, Princeton, NJ 08544, USA
\item[\riverside] University of Californa, Riverside, CA 92521, USA
\item[\rome] INFN-Sezione di Roma and University of Rome, ``La Sapienza",
     I-00185 Rome, Italy
\item[\salerno] University and INFN, Salerno, I-84100 Salerno, Italy
\item[\ucsd] University of California, San Diego, CA 92093, USA
\item[\sofia] Bulgarian Academy of Sciences, Central Lab.~of 
     Mechatronics and Instrumentation, BU-1113 Sofia, Bulgaria
\item[\korea]  The Center for High Energy Physics, 
     Kyungpook National University, 702-701 Taegu, Republic of Korea
\item[\taiwan] National Central University, Chung-Li, Taiwan, China
\item[\tsinghua] Department of Physics, National Tsing Hua University,
      Taiwan, China
\item[\purdue] Purdue University, West Lafayette, IN 47907, USA
\item[\psinst] Paul Scherrer Institut, PSI, CH-5232 Villigen, Switzerland
\item[\zeuthen] DESY, D-15738 Zeuthen, Germany
\item[\eth] Eidgen\"ossische Technische Hochschule, ETH Z\"urich,
     CH-8093 Z\"urich, Switzerland
\item[\S]  Supported by the German Bundesministerium 
        f\"ur Bildung, Wissenschaft, Forschung und Technologie.
\item[\ddag] Supported by the Hungarian OTKA fund under contract
numbers T019181, F023259 and T037350.
\item[\P] Also supported by the Hungarian OTKA fund under contract
  number T026178.
\item[$\flat$] Supported also by the Comisi\'on Interministerial de Ciencia y 
        Tecnolog{\'\i}a.
\item[$\sharp$] Also supported by CONICET and Universidad Nacional de La Plata,
        CC 67, 1900 La Plata, Argentina.
\item[$\triangle$] Supported by the National Natural Science
  Foundation of China.
\item[$\dagger$] Deceased.
\end{list}
}
\vfill

%%% Local Variables: 
%%% mode: latex
%%% TeX-master: t
%%% End:

%%%%%%%%%%%%%%%%%%%%%%%%%%%%%%%%%%%%%%%%%%%%%%%%%%%%%%%%%%%%%%%%%%%%%%%%%%%%%%%%
%                                 
%%%%%%%%%%%%%%%%%%%%%%%%%%%%%%%%%%%%%%%%%%%%%%%%%%%%%%%%%%%%%%%%%%%%%%%%%%%%%%%%
\clearpage

\begin{table}[p]
  \begin{center}
    \begin{tabular}{|c|c|c|}
      \hline
      $\langle\sqrt{s}\rangle$~(\GeV) & $\mathcal{L}$~(\pb) & Named as\\
      \hline
      \hline
      \multicolumn{3}{|c|}{$\epem\rightarrow\mbox{hadrons}\,(\gamma)$} \\
      \hline
      191.6        &  \phantom{0}29.8  & $192\GeV$\\
      195.5        &  \phantom{0}83.7  & $196\GeV$\\
      199.6        &  \phantom{0}83.2  & $200\GeV$\\
      201.8        &  \phantom{0}36.8  & $202\GeV$\\
      204.9        &  \phantom{0}75.9  & $205\GeV$\\
      206.5        & 130.5             & $207\GeV$\\
      208.0        &  \phantom{00}8.3  & $208\GeV$\\
      \hline
      \hline
      \multicolumn{3}{|c|}{$\epem\rightarrow\mumu(\gamma)$} \\
      \hline
      191.6        &  \phantom{0}28.0 & $192\GeV$\\
      195.5        &  \phantom{0}82.1 & $196\GeV$\\
      199.6        &  \phantom{0}80.4 & $200\GeV$\\
      201.9        &  \phantom{0}38.1 & $202\GeV$\\
      205.0        &  \phantom{0}73.5 & $205\GeV$\\
      206.5        &            126.8 & $207\GeV$\\
      208.0        &  \phantom{00}8.1 & $208\GeV$\\
      \hline
      \hline
      \multicolumn{3}{|c|}{$\epem\rightarrow\tautau(\gamma)$} \\
      \hline
      191.6        &  \phantom{0}28.9  & $192\GeV$\\
      195.5        &  \phantom{0}81.7  & $196\GeV$\\
      199.5        &  \phantom{0}72.3  & $200\GeV$\\
      201.7        &  \phantom{0}38.1  & $202\GeV$\\
      205.2        &  \phantom{0}73.5  & $205\GeV$\\
      206.7        &            125.9  & $207\GeV$\\
      208.1        &  \phantom{00}8.1  & $208\GeV$\\
      \hline
      \hline
      \multicolumn{3}{|c|}{$\epem\rightarrow\epem(\gamma)$}\\
      \hline
      191.6        &   \phantom{0}27.5 & $192\GeV$\\
      195.5        &   \phantom{0}82.7 & $196\GeV$\\
      199.5        &   \phantom{0}82.6 & $200\GeV$\\
      201.8        &   \phantom{0}37.0 & $202\GeV$\\
      205.2        &   \phantom{0}66.9 & $205\GeV$\\
      206.7        &             122.7 & $207\GeV$\\
      208.2        &   \phantom{00}7.9 & $208\GeV$\\
      \hline
    \end{tabular}
    \caption[]{Average centre-of-mass energies and corresponding
      integrated luminosities of the data samples, together with the
      names used in the following Tables. Luminosity
      differences across the channels are due to different
      data-quality requirements, which also result in slightly different
      average values.
      \label{tab:roots}}
  \end{center}
\end{table}

\clearpage

\begin{table}[p] 
 \begin{sideways}
   \begin{minipage}[b]{\textheight}% width of page in landscape mode
     \small
     \begin{center}
       \renewcommand{\arraystretch}{1.0}
       \begin{tabular}{|l|r|r|r|r|r|r|r|r|r|r|r|r|r|r|}
	 \hline
	 \multicolumn{1}{|c|}{$\sqrt{s}$ (\GeV)}
	 & \multicolumn{1}{c|}{192}
	 & \multicolumn{1}{c|}{196}
	 & \multicolumn{1}{c|}{200}
	 & \multicolumn{1}{c|}{202}
	 & \multicolumn{1}{c|}{205}
	 & \multicolumn{1}{c|}{207}
	 & \multicolumn{1}{c|}{208} \\
	 \hline
	 \hline
	 \multicolumn{8}{|c|}{$\epem\rightarrow\mbox{\rm{hadrons}}\,(\gamma)$} \\
	 \hline
	 Selection Efficiency   & $89.1 \pm 0.1$ & $88.9 \pm 0.1$ & $87.5  \pm 0.1$
         & $88.5 \pm 0.2$ & $87.6 \pm 0.1$ & $87.5 \pm 0.0$ & $87.4 \pm 0.1$\\
	 Total Background       & \multicolumn{1}{c|}{9.8}&\multicolumn{1}{c|}{10.7}&\multicolumn{1}{c|}{11.1}&\multicolumn{1}{c|}{11.3}&
	 \multicolumn{1}{c|}{11.8}&\multicolumn{1}{c|}{12.2}&\multicolumn{1}{c|}{13.4}\\\hline
	 $\epem  \rightarrow \epem {\rm \; hadrons}$ & 
 	 \multicolumn{1}{c|}{2.9}&\multicolumn{1}{c|}{\phantom{1}3.1}&\multicolumn{1}{c|}{\phantom{1}3.1}&\multicolumn{1}{c|}{\phantom{1}3.0}&
	 \multicolumn{1}{c|}{\phantom{1}3.3}&\multicolumn{1}{c|}{\phantom{1}3.4}&\multicolumn{1}{c|}{\phantom{1}3.7}\\
	 $\epem  \rightarrow \Wp \Wm$ & 
	 \multicolumn{1}{c|}{5.2}&\multicolumn{1}{c|}{\phantom{1}5.7}&\multicolumn{1}{c|}{\phantom{1}6.0}&\multicolumn{1}{c|}{\phantom{1}6.3}&
	 \multicolumn{1}{c|}{\phantom{1}6.4}&\multicolumn{1}{c|}{\phantom{1}6.6}&\multicolumn{1}{c|}{\phantom{1}7.3}\\
	 Other &		\multicolumn{1}{c|}{1.5}&\multicolumn{1}{c|}{\phantom{1}1.7}&\multicolumn{1}{c|}{\phantom{1}1.8}&\multicolumn{1}{c|}{\phantom{1}1.8}&
	 \multicolumn{1}{c|}{\phantom{1}1.9}&\multicolumn{1}{c|}{\phantom{1}2.0}&\multicolumn{1}{c|}{\phantom{1}2.2}\\
	 ISR Contamination &	\multicolumn{1}{c|}{0.2}&\multicolumn{1}{c|}{\phantom{1}0.2}&\multicolumn{1}{c|}{\phantom{1}0.2}&\multicolumn{1}{c|}{\phantom{1}0.2}&
	 \multicolumn{1}{c|}{\phantom{1}0.2}&\multicolumn{1}{c|}{\phantom{1}0.2}&\multicolumn{1}{c|}{\phantom{1}0.2}\\
	 \hline
	 \hline
	 \multicolumn{8}{|c|}{$\epem\rightarrow\mumu(\gamma)$} \\
	 \hline
	 Selection Efficiency    & $60.8 \pm 0.8 $ & $59.1 \pm 0.8 $ & $59.0 \pm 0.8 $
 	 & $59.5 \pm 0.8 $ & $61.1 \pm 0.6 $ & $62.1 \pm 0.6 $ & $61.5 \pm 0.8 $ \\
	 Total Background        & $16.6 \pm 1.4$ &  $18.6 \pm 1.5$ &  $17.1 \pm 1.7$ &  $21.0 \pm 1.8$ 
         & $19.5 \pm 1.1$ &  $24.6 \pm 1.2$ &  $18.9 \pm 1.1$ \\\hline
	 $\epem \rightarrow\epem\ell^+\ell^-$ 
 	 &  $9.8 \pm 1.3$ & $11.1 \pm 1.4$ & $10.0 \pm 1.6$ & $13.3 \pm 1.6$ & $11.7 \pm 0.9$ & $15.3 \pm 1.1$ & $12.2 \pm 0.9$ \\
	 $\epem \rightarrow \Wp\Wm$   
 	 &  $3.4 \pm 0.2$ &  $3.4 \pm 0.2$ &  $3.5 \pm 0.2$ &  $3.3 \pm 0.2$ &  $3.4 \pm 0.4$ &  $4.0 \pm 0.4$ &  $3.5 \pm 0.3$ \\
	 Cosmic Rays      		 &  $0.5 \pm 0.3$ &  $0.3 \pm 0.1$ &  $0.2 \pm 0.1$ &  $0.2 \pm 0.2$ &  $0.8 \pm 0.2$ &  $1.0 \pm 0.2$ &  $0.5 \pm 0.5$ \\
	 Other        	         &  $2.5 \pm 0.3$ &  $3.5 \pm 0.3$ &  $3.0 \pm 0.2$ &  $3.9 \pm 0.7$ &  $3.2 \pm 0.2$ &  $4.2 \pm 0.2$ &  $2.5 \pm 0.2$ \\
	 ISR Contamination       &  $0.4 \pm 0.1$ &  $0.3 \pm 0.1$ &  $0.4 \pm 0.1$&  $0.3 \pm 0.1$ &  $0.4 \pm 0.1$ &  $0.1 \pm 0.1$ &  $0.2 \pm 0.1$ \\
	 \hline
	 \hline
	 \multicolumn{8}{|c|}{$\epem\rightarrow\tautau(\gamma)$} \\
	 \hline
	 Selection Efficiency   &$ 44.0 \pm 0.5$   &$ 43.9 \pm 0.5$   &$ 43.3 \pm 0.6$    &$ 42.9 \pm 0.5$  
         &$ 40.9 \pm 0.5$    &$ 41.8 \pm 0.5$    &$ 41.0 \pm 0.7$  \\
	 
	 Total Background       & $21.0 \pm 1.6$ &  $23.1 \pm 0.9$ &  $22.5 \pm 1.4$ &  $21.9 \pm 1.8$ 
         & $24.8 \pm 1.3$ &  $26.1 \pm 1.8$ &  $23.7 \pm 2.7$ \\\hline
	 $\epem \rightarrow \epem \ell^+\ell^-$  
 	 &$ 10.4 \pm 1.1$   &$ 11.4 \pm 0.6$   &$ 11.3 \pm 1.2$    &$ 10.2 \pm 1.5$   
         &$ 13.9 \pm 0.9$    &$ 14.2 \pm 1.3$    &$ 12.6 \pm 1.2$  \\
	 Other      		&$ 10.1 \pm 1.1$   &$ 11.3 \pm 0.7$   &$ 10.7 \pm 0.7$    &$ 11.5 \pm 1.0$   
         &$ 10.5 \pm 0.9$    &$ 11.4 \pm 1.2$    &$ 10.7 \pm 2.4$  \\
	 ISR Contamination      &$ 0.5 \pm 0.1$   &$  0.4 \pm 0.1$   &$ 0.5 \pm 0.1$     &$ 0.2 \pm 0.1$  
         &$ 0.4 \pm 0.1$    &$ 0.5 \pm 0.1$      &$ 0.4  \pm 0.1$  \\
	 \hline
	 \hline
	 \multicolumn{8}{|c|}{$\epem\rightarrow\epem(\gamma)$} \\
	 \hline
	 Selection Efficiency
	 &$96.7 \pm  0.2$ &$97.2 \pm  0.2$ &$95.7 \pm  0.3$ &$97.0 \pm  0.2$
	 &$96.8 \pm  0.2$ &$97.5 \pm  0.2$ &$96.9 \pm  0.4$ \\
	 Total Background
	 &$ 4.4 \pm  0.2$ &$ 4.1 \pm  0.2$ &$ 4.4 \pm  0.2$ &$ 4.4 \pm  0.2$
	 &$ 4.3 \pm  0.2$ &$ 4.3 \pm  0.2$ &$ 4.4 \pm  0.3$ \\\hline
	 $\epem \rightarrow \tau^+\tau^-$
	 &$ 2.9 \pm  0.2$ &$ 2.7 \pm  0.2$ &$ 2.9 \pm  0.2$ &$ 3.0 \pm  0.2$
	 &$ 2.9 \pm  0.2$ &$ 2.9 \pm  0.2$ &$ 3.1 \pm  0.3$ \\
	 $\epem \rightarrow \Wp\Wm$ 
	 &$ 0.7 \pm  0.1$ &$ 0.5 \pm  0.1$ &$ 0.5 \pm  0.1$ &$ 0.5 \pm  0.1$
	 &$ 0.4 \pm  0.1$ &$ 0.4 \pm  0.1$ &$ 0.3 \pm  0.1$ \\
	 Other
	 &$ 0.8 \pm  0.1$ &$ 0.9 \pm  0.1$ &$ 1.0 \pm  0.1$ &$ 0.9 \pm  0.2$
	 &$ 1.0 \pm  0.1$ &$ 1.0 \pm  0.1$ &$ 1.0 \pm  0.1$ \\
	 \hline
       \end{tabular} 
       \caption[]{
        Selection efficiencies, background fractions and their
        breakdown, all in \%, for the inclusive samples. The
        uncertainties reflect the limited Monte Carlo statistics and
        are negligible for the background to the
        $\epem\rightarrow\mbox{\rm{hadrons}}\,(\gamma)$ channel. Values
        for the {$\epem\rightarrow\epem(\gamma)$} channel refer to the
        angular region $|\cos\theta|<0.72$.
      \label{tab:seleinc}}
  \end{center}
 \normalsize
 \end{minipage}
 \end{sideways}
\end{table}

\clearpage

\begin{table}[p] %--------- Efficiencies and Backgrounds -----------------------
  \begin{sideways}
    \begin{minipage}[b]{\textheight}% width of page in landscape mode
      \small
      \begin{center}
	\renewcommand{\arraystretch}{1.0}
	\begin{tabular}{|l|r|r|r|r|r|r|r|}
	  \hline
	  \multicolumn{1}{|c|}{$\sqrt{s}$ (\GeV)}
	  & \multicolumn{1}{c|}{192}
	  & \multicolumn{1}{c|}{196}
	  & \multicolumn{1}{c|}{200}
	  & \multicolumn{1}{c|}{202}
	  & \multicolumn{1}{c|}{205}
	  & \multicolumn{1}{c|}{207}
	  & \multicolumn{1}{c|}{208} \\
	  \hline
	  \hline
	  \multicolumn{8}{|c|}{$\epem\rightarrow\mbox{\rm{hadrons}}\,(\gamma)$} \\
	  \hline
	  Selection Efficiency  & $85.7 \pm 0.3$ & $85.5 \pm 0.2$ & $84.7 \pm 0.3$
          & $85.8 \pm 0.3$ & $85.2 \pm 0.1$ & $85.2 \pm 0.1$ & $85.0 \pm 0.1$ \\
	  Total Background      & $15.7 \pm 0.3$ &  $17.2 \pm 0.2$ & $17.5 \pm 0.3$ &  $17.0 \pm 0.3$ 
          & $17.9 \pm 0.1$ &  $18.4 \pm 0.1 $ &  $20.8 \pm 0.1$ \\ \hline
	  $\epem  \rightarrow \epem {\rm \; hadrons}$ & 
 	  \multicolumn{1}{c|}{ 0.2}&\multicolumn{1}{c|}{ 0.1}&\multicolumn{1}{c|}{ 0.2}&\multicolumn{1}{c|}{ 0.1}&
	  \multicolumn{1}{c|}{ 0.1}&\multicolumn{1}{c|}{ 0.2}&\multicolumn{1}{c|}{\phantom{1}0.2}\\
	  $\epem  \rightarrow \Wp \Wm$ & 
	  \multicolumn{1}{c|}{ 7.5}&\multicolumn{1}{c|}{ 8.4}&\multicolumn{1}{c|}{ 9.0}&\multicolumn{1}{c|}{ 9.1}&
	  \multicolumn{1}{c|}{ 9.6}&\multicolumn{1}{c|}{ 9.8}&\multicolumn{1}{c|}{11.3}\\
	  Other &		\multicolumn{1}{c|}{ 0.8}&\multicolumn{1}{c|}{ 0.9}&\multicolumn{1}{c|}{ 1.1}&\multicolumn{1}{c|}{ 0.9}&
	  \multicolumn{1}{c|}{ 1.0}&\multicolumn{1}{c|}{ 1.0}&\multicolumn{1}{c|}{\phantom{1}1.2}\\
	  ISR Contamination      & $7.2 \pm 0.3 $ &  $7.8 \pm 0.2 $ & $7.3 \pm 0.2 $ 
          & $6.9 \pm 0.2 $ &  $7.2 \pm 0.1 $ &  $7.4 \pm 0.1$ &  $8.1 \pm 0.1$ \\
	  \hline
	  \hline
	  \multicolumn{8}{|c|}{$\epem\rightarrow\mumu(\gamma)$} \\
	  \hline
	  Selection Efficiency  &  $71.1 \pm 1.1 $ & $71.1 \pm 1.1 $ & $70.5 \pm 1.1 $
	  & $74.0 \pm 1.0 $ & $76.1 \pm 0.7 $ & $76.4 \pm 0.7 $ & $77.7 \pm 0.9 $ \\
	  
	  Total Background       &  $16.1 \pm 5.9$& $13.2 \pm 3.9$&$ 11.7 \pm 3.9$&$ 15.4 \pm 4.3$&$ 14.5 \pm 4.0$&$ 15.8 \pm 4.0$&$ 17.1 \pm 4.5$\\\hline
	  $\epem  \rightarrow \epem\ell^+\ell^-$ 
 	  &   $5.7 \pm 1.8$&  $5.0 \pm 1.5$&$  3.4 \pm 1.5$&$  6.0 \pm 1.7$&$  5.3 \pm 1.1$&$  5.6 \pm 1.1$&$  5.7 \pm 1.1$\\
	  $\epem  \rightarrow \Wp\Wm$ 
 	  &   $2.6 \pm 0.2$&  $2.8 \pm 0.2$&$  3.0 \pm 0.3$&$  2.7 \pm 0.2$&$  3.0 \pm 0.5$&$  3.9 \pm 0.6$&$  4.4 \pm 0.5$\\
	  Cosmic      		&   $0.7 \pm 0.5$&  $0.2 \pm 0.1$&$  0.3 \pm 0.2$&$  0.2 \pm 0.2$&$  0.4 \pm 0.2$&$  0.6 \pm 0.2$&$  1.2 \pm 1.2$\\
	  Other 		        &   $1.5 \pm 0.2$&  $1.6 \pm 0.2$&$  1.5 \pm 0.2$&$  2.6 \pm 1.2$&$  2.0 \pm 0.2$&$  1.9 \pm 0.2$&$  1.5 \pm 0.2$\\
	  ISR Contamination      &   $5.6 \pm 0.7$&  $3.6 \pm 0.5$&$  3.5 \pm 0.5$&$  3.9 \pm 0.5$&$  3.8 \pm 0.2$&$  3.8 \pm 0.3$&$  4.3 \pm 0.5$\\
	  \hline
	  \hline
	  \multicolumn{8}{|c|}{$\epem\rightarrow\tautau(\gamma)$} \\
	  \hline
	  Selection Efficiency  	&$ 51.5 \pm 0.8$   &$51.7 \pm 0.8$   & $52.1 \pm 0.8$   &$50.6 \pm 0.8$  
          &$48.3 \pm 0.8$   &$ 51.0 \pm 0.8$   &$ 46.7 \pm 1.1$ \\
	  Total Background      	&$15.8 \pm 2.1$& $15.1 \pm 1.4$&$ 14.1 \pm 1.5$&$ 14.8 \pm 1.9$&$ 16.4 \pm 1.7$&$ 20.5 \pm 2.5$&$ 22.9 \pm 5.7$\\\hline
	  $\epem  \rightarrow \epem\ell^+\ell^-$ 
 	  &$  2.2 \pm 0.7$   &$ 1.8 \pm 0.3$   & $ 2.2 \pm 0.6$   &$ 1.1 \pm 0.6$ 
          &$ 2.3 \pm 0.5$   &$  3.0 \pm 0.9$   &$  2.8 \pm 0.9$ \\
	  Other       		&$ 10.4 \pm 1.8$   &$ 9.9 \pm 1.1$   & $ 9.3 \pm 1.1$   &$10.6 \pm 1.6$   
          &$10.8 \pm 1.4$   &$ 13.8 \pm 2.2$   &$ 15.9 \pm 5.4$ \\
	  ISR Contamination      &$  3.2 \pm 0.9$   &$ 3.4 \pm 0.8$   & $ 2.6 \pm 0.8$   &$ 3.1 \pm 0.8$   
          &$ 3.3 \pm 0.8$   &$  3.7 \pm 0.9$   &$  4.2 \pm 1.4$ \\
	  \hline
	  \hline
	  \multicolumn{8}{|c|}{$\epem\rightarrow\epem(\gamma)$} \\
	  \hline
	  Selection Efficiency
	  &$97.3 \pm  0.2$ &$97.7 \pm  0.2$ &$96.4 \pm  0.3$ &$97.6 \pm  0.2$
	  &$97.5 \pm  0.2$ &$98.0 \pm  0.2$ &$97.7 \pm  0.3$ \\
	  Total Background
	  &$ 3.9 \pm  0.3$ &$ 3.4 \pm  0.2$ &$ 3.8 \pm  0.3$ &$ 3.8 \pm  0.3$
	  &$ 3.7 \pm  0.2$ &$ 3.8 \pm  0.3$ &$ 3.7 \pm  0.4$ \\\hline
	  $\epem  \rightarrow \tau^+\tau^-$ 
	  &$ 2.8 \pm  0.2$ &$ 2.6 \pm  0.2$ &$ 2.9 \pm  0.2$ &$ 3.0 \pm  0.2$
	  &$ 2.9 \pm  0.2$ &$ 2.9 \pm  0.2$ &$ 3.0 \pm  0.3$ \\
	  $\epem  \rightarrow \Wp\Wm$
	  &$ 0.8 \pm  0.1$ &$ 0.5 \pm  0.1$ &$ 0.6 \pm  0.1$ &$ 0.5 \pm  0.1$
	  &$ 0.4 \pm  0.1$ &$ 0.5 \pm  0.1$ &$ 0.3 \pm  0.1$ \\
	  Other 
	  &$ 0.3 \pm  0.1$ &$ 0.3 \pm  0.1$ &$ 0.3 \pm  0.1$ &$ 0.3 \pm  0.1$
	  &$ 0.4 \pm  0.1$ &$ 0.4 \pm  0.1$ &$ 0.4 \pm  0.1$ \\
	  \hline
	\end{tabular} 
	  \caption[]{
            Selection efficiencies, background fractions and their
            breakdown, all in \%, for the high-energy samples. The
            uncertainties reflect the limited Monte Carlo statistics and
            are negligible for the background to the
            $\epem\rightarrow\mbox{\rm{hadrons}}\,(\gamma)$ channel. Values
            for the {$\epem\rightarrow\epem(\gamma)$} channel refer to the
            angular region $|\cos\theta|<0.72$.
	    \label{tab:sele}}
      \end{center}
      \normalsize
    \end{minipage}
  \end{sideways}
\end{table}

\clearpage

\begin{table}[p] %--------- Selected Events and Cross Sections -----------------
  \begin{center}
    \begin{tabular}{|c|c|c|c|c|c|c|}
      \cline{2-7}
      \multicolumn{1}{ c|}{}
      & \multicolumn{3}{|c|}{Inclusive sample}
      & \multicolumn{3}{c|}{High-energy sample} \\  
      \hline
      $\sqrt{s}$~(\GeV) & \Nsel & $\sigma$~(pb) &
      \sigsm~(pb) &\Nsel & $\sigma$~(pb) &
      \sigsm~(pb) \\
      \hline
      \hline
      \multicolumn{7}{|c|}{$\epem\rightarrow\mbox{hadrons}(\gamma)$} \\
      \hline
      192   &           2767 &   $93.76\pm  1.98\pm 0.99$ & 92.91  & \phantom{0}679 &  $22.38 \pm 1.02 \pm 0.19$ & 21.32 \\
      196   &           7166 &   $86.05\pm  1.14\pm 0.93$ & 88.17  &           1740 &  $20.14 \pm 0.58 \pm 0.16$ & 20.21 \\
      200   &           6753 &   $82.45\pm  1.13\pm 0.88$ & 83.61  &           1629 &  $19.09 \pm 0.57 \pm 0.16$ & 19.13 \\
      202   &           2956 &   $80.51\pm  1.67\pm 0.87$ & 81.32  & \phantom{0}736 &  $19.33 \pm 0.89 \pm 0.16$ & 18.59 \\
      205   &           5949 &   $78.95\pm  1.16\pm 0.82$ & 78.27  &           1452 &  $18.46 \pm 0.59 \pm 0.14$ & 17.87 \\
      207   &           9888 &   $76.07\pm  0.87\pm 0.82$ & 76.77  &           2430 &  $17.87 \pm 0.44 \pm 0.13$ & 17.52 \\
      208   & \phantom{0}578 &   $68.78\pm  3.30\pm 0.83$ & 75.40  & \phantom{0}135 &  $15.09 \pm 1.64 \pm 0.14$ & 17.20 \\
      \hline
      \hline
      \multicolumn{7}{|c|}{$\epem\rightarrow\mumu(\gamma)$} \\
      \hline
      192  &  \phantom{0}131 &  $\phantom{0}6.41 \pm 0.67\pm  0.21$ &  \phantom{0}7.02  & \phantom{00}61 & $\phantom{0}2.54 \pm0.39  \pm0.09$ &  \phantom{0}3.11 \\
      196  &  \phantom{0}397 &  $\phantom{0}6.52 \pm 0.41\pm  0.25$ &  \phantom{0}6.67  & \phantom{0}207 & $\phantom{0}3.05 \pm0.25  \pm0.10$ &  \phantom{0}2.97 \\
      200  &  \phantom{0}349 &  $\phantom{0}6.09 \pm 0.39\pm  0.23$ &  \phantom{0}6.37  & \phantom{0}185 & $\phantom{0}2.85 \pm0.24  \pm0.09$ &  \phantom{0}2.84 \\
      202  &  \phantom{0}175 &  $\phantom{0}6.08 \pm 0.58\pm  0.24$ &  \phantom{0}6.20  & \phantom{00}99 & $\phantom{0}2.97 \pm0.35  \pm0.10$ &  \phantom{0}2.76 \\
      205  &  \phantom{0}358 &  $\phantom{0}6.53 \pm 0.43\pm  0.32$ &  \phantom{0}5.95  & \phantom{0}157 & $\phantom{0}2.37 \pm0.22  \pm0.07$ &  \phantom{0}2.67 \\
      207  &  \phantom{0}521 &  $\phantom{0}5.05 \pm 0.29\pm  0.17$ &  \phantom{0}5.88  & \phantom{0}260 & $\phantom{0}2.24 \pm0.17  \pm0.06$ &  \phantom{0}2.63 \\
      208  &  \phantom{00}44 &  $\phantom{0}7.70 \pm 1.44\pm  0.28$ &  \phantom{0}5.79  & \phantom{00}17 & $\phantom{0}2.49 \pm0.74  \pm0.10$ &  \phantom{0}2.59 \\
      \hline
      \hline
      \multicolumn{7}{|c|}{$\epem\rightarrow\tautau(\gamma)$} \\
      \hline
      192  &  \phantom{0}116 &  $\phantom{0}7.21 \pm 0.85 \pm 0.20$ &  \phantom{0}7.01 & \phantom{00}52 & $\phantom{0}2.93 \pm 0.48 \pm0.06$ &  \phantom{0}3.11 \\
      196  &  \phantom{0}300 &  $\phantom{0}6.42 \pm 0.48 \pm 0.24$ &  \phantom{0}6.68 & \phantom{0}161 & $\phantom{0}3.22 \pm 0.30 \pm0.07$ &  \phantom{0}2.97 \\
      200  &  \phantom{0}263 &  $\phantom{0}6.52 \pm 0.52 \pm 0.23$ &  \phantom{0}6.37 & \phantom{0}131 & $\phantom{0}2.97 \pm 0.30 \pm0.07$ &  \phantom{0}2.84 \\
      202  &  \phantom{0}123 &  $\phantom{0}5.86 \pm 0.68 \pm 0.22$ &  \phantom{0}6.20 & \phantom{00}64 & $\phantom{0}2.81 \pm 0.42 \pm0.07$ &  \phantom{0}2.77 \\
      205  &  \phantom{0}261 &  $\phantom{0}6.51 \pm 0.54 \pm 0.28$ &  \phantom{0}5.98 & \phantom{0}125 & $\phantom{0}2.93 \pm 0.32 \pm0.07$ &  \phantom{0}2.68 \\
      207  &  \phantom{0}406 &  $\phantom{0}5.70 \pm 0.38 \pm 0.23$ &  \phantom{0}5.88 & \phantom{0}189 & $\phantom{0}2.34 \pm 0.21 \pm0.08$ &  \phantom{0}2.63 \\
      208  &  \phantom{00}29 &  $\phantom{0}6.65 \pm 1.62 \pm 0.27$ &  \phantom{0}5.78 & \phantom{00}11 & $\phantom{0}2.23 \pm 0.88 \pm0.07$ &  \phantom{0}2.59 \\
      \hline
      \hline
      \multicolumn{7}{|c|}{$\epem\rightarrow\epem(\gamma)$}\\
      \hline
      192  &  \phantom{0}659 & $23.71 \pm 0.92\pm 0.32$ & 24.00 & \phantom{0}624 &  $22.46 \pm0.90 \pm0.11$ & 22.68 \\
      196  &            1899 & $22.65 \pm 0.52\pm 0.31$ & 23.04 &           1781 &  $21.27 \pm0.50 \pm0.11$ & 21.76 \\
      200  &            1776 & $21.49 \pm 0.51\pm 0.29$ & 21.98 &           1668 &  $20.14 \pm0.49 \pm0.10$ & 20.86 \\
      202  &  \phantom{0}857 & $22.82 \pm 0.78\pm 0.31$ & 21.45 & \phantom{0}811 &  $21.62 \pm0.76 \pm0.11$ & 20.36 \\
      205  &            1483 & $21.94 \pm 0.57\pm 0.30$ & 20.91 &           1380 &  $20.39 \pm0.55 \pm0.10$ & 19.75 \\
      207  &            2572 & $20.60 \pm 0.41\pm 0.28$ & 20.41 &           2418 &  $19.36 \pm0.39 \pm0.10$ & 19.47 \\
      208  &  \phantom{0}144 & $18.00 \pm 1.50\pm 0.25$ & 20.06 & \phantom{0}137 &  $17.10 \pm1.46 \pm0.09$ & 19.16 \\
      \hline
    \end{tabular}
    \caption[]{
      Numbers of selected events, {\Nsel}, measured cross sections,
      $\sigma$, with their statistical and systematic uncertainties and corresponding
      {\SM} predictions for the inclusive and high-energy samples.
      Results for the $\epem\ra\epem(\gamma)$ process refer to the range
      $|\cos\theta|<0.72$.  The theoretical uncertainties on the \SM\
      predictions are estimated to be below 1\% except
      for large-angle Bhabha scattering where they
      reach 1.5\%~\cite{BHABHA-THEORY}.
      \label{tab:xsecincl}}
  \end{center}
\end{table}

\clearpage

\begin{table}[p] %--------- Forward-Backward Asymmetries ---------------
  \begin{center}
    \begin{tabular}{|c|c|c|c|c|c|c|c|c|}
      \cline{2-9} 
      \multicolumn{1}{c}{}
      & \multicolumn{4}{|c|}{Inclusive sample}
      & \multicolumn{4}{|c|}{High-energy sample} \\
      \hline
      $\sqrt{s}$~(\GeV) & \Nf & \Nb & \Afb &
      \AfbSM
      & \Nf & \Nb & \Afb & \AfbSM \\
      \hline
      \hline
      \multicolumn{9}{|c|}{$\epem\rightarrow\mumu(\gamma)$} \\
      \hline
           192   &\phantom{00}82 &\phantom{0}49 & $0.43 \pm 0.13 \pm 0.09$ & 0.308
                 &\phantom{00}48 &\phantom{0}13 & $0.69 \pm 0.12 \pm 0.07$ & 0.569\\
           196   &\phantom{0}259&129& $0.33 \pm 0.07 \pm 0.04$ & 0.306
                 &\phantom{0}151&\phantom{0}46 & $0.53 \pm 0.07 \pm 0.04$ & 0.564\\
           200   &\phantom{0}226&123& $0.31 \pm 0.07 \pm 0.04$ & 0.304
                 &\phantom{0}126&\phantom{0}59 & $0.44 \pm 0.08 \pm 0.04$ & 0.560\\
           202   &\phantom{0}121&\phantom{0}54 & $0.36 \pm 0.10 \pm 0.05$ & 0.303
                 &\phantom{00}75 &\phantom{0}24 & $0.59 \pm 0.09 \pm 0.02$ & 0.557\\
           205   &\phantom{0}236&122& $0.34 \pm 0.07 \pm 0.05$ & 0.302
                 &\phantom{0}110&\phantom{0}47 & $0.48 \pm 0.09 \pm 0.03$ & 0.554\\
           207   &\phantom{0}346&175& $0.32 \pm 0.06 \pm 0.03$ & 0.302
                 &\phantom{0}189&\phantom{0}71 & $0.54 \pm 0.06 \pm 0.02$ & 0.553\\
           208   &\phantom{00}33 &\phantom{0}11 & $0.36 \pm 0.20 \pm 0.10$ & 0.301
                 &\phantom{00}14 &\phantom{00}3  & $0.72 \pm 0.16 \pm 0.04$ & 0.551\\
	   \hline
	   \hline
	   \multicolumn{9}{|c|}{$\epem\rightarrow\tautau(\gamma)$} \\
	   \hline                 
           192   &\phantom{00}54 &\phantom{0}27 & $0.38 \pm 0.13 \pm 0.04$ & 0.311
                 &\phantom{00}35 &\phantom{00}8  & $0.52 \pm 0.12 \pm 0.05$ & 0.569\\
           196   &\phantom{0}142&\phantom{0}80 & $0.33 \pm 0.08 \pm 0.03$ & 0.309
                 &\phantom{00}95 &\phantom{0}34 & $0.44 \pm 0.09 \pm 0.05$ & 0.565\\
           200   &\phantom{0}110&\phantom{0}74 & $0.18 \pm 0.10 \pm 0.03$ & 0.307
                 &\phantom{00}66 &\phantom{0}28 & $0.46 \pm 0.10 \pm 0.05$ & 0.560\\ 
           202   &\phantom{00}60 &\phantom{0}27 & $0.34 \pm 0.13 \pm 0.05$ & 0.305
                 &\phantom{00}37 &\phantom{00}8  & $0.47 \pm 0.13 \pm 0.08$ & 0.557\\
           205   &\phantom{0}123&\phantom{0}57 & $0.44 \pm 0.10 \pm 0.04$ & 0.303
                 &\phantom{00}77 &\phantom{0}20 & $0.56 \pm 0.09 \pm 0.05$ & 0.554\\
           207   &\phantom{0}204&\phantom{0}88 & $0.35 \pm 0.07 \pm 0.04$ & 0.302
                 &\phantom{0}110&\phantom{0}20 & $0.61 \pm 0.07 \pm 0.09$ & 0.553\\		 
           208   &\phantom{00}10 &\phantom{0}10 & $0.00 \pm 0.30 \pm 0.05$ & 0.301
                 &\phantom{000}6  &\phantom{00}3  & $0.08 \pm 0.35 \pm 0.09$ & 0.551\\
	   \hline
	   \hline
	   \multicolumn{9}{|c|}{$\epem\rightarrow\epem(\gamma)$} \\
	   \hline
           192   &\phantom{0}405 &\phantom{0}59  & $0.839 \pm  0.027 \pm 0.010$ & 0.782
                 &\phantom{0}395 &\phantom{0}53  & $0.863 \pm  0.025 \pm 0.007$ & 0.815\\
           196   &1121&203& $0.778 \pm  0.018 \pm 0.010$ & 0.785
	         &1096&191& $0.796 \pm  0.018 \pm 0.007$ & 0.815\\		
           200   &1048&174& $0.801 \pm  0.018 \pm 0.010$ & 0.787  
                 &1047&168& $0.814 \pm  0.018 \pm 0.007$ & 0.816\\
           202   &\phantom{0}480 &\phantom{0}94  & $0.759 \pm  0.029 \pm 0.010$ & 0.789  
                 &\phantom{0}468 &\phantom{0}84  & $0.791 \pm  0.028 \pm 0.007$ & 0.816\\
           205   &\phantom{0}820 &145 & $0.777 \pm  0.022 \pm 0.010$ & 0.791  
                 &\phantom{0}799 &132 & $0.803 \pm  0.021 \pm 0.007$ & 0.817\\
           207   &1430&242& $0.791 \pm  0.016 \pm 0.010$ & 0.792  
                 &1393&228& $0.805 \pm  0.016 \pm 0.007$ & 0.817\\
           208   &\phantom{0}81  &\phantom{0}16  & $0.719 \pm  0.076 \pm 0.010$ & 0.793  
	         &\phantom{0}79  &\phantom{0}14  & $0.750 \pm  0.073 \pm 0.007$ & 0.818\\
	   \hline
    \end{tabular}
    \caption[]{
      Measurements of the forward-backward asymmetries of lepton-pair
      production, \Afb, with statistical and systematic
      uncertainties. The corresponding \SM\ predictions, $A_{\rm
	fb}^{\rm SM}$, are also given, together with the numbers of events
      selected in the forward, $N_{\rm f}$, and backward, $N_{\rm b}$,
      hemispheres.  Results for the $\epem\ra\epem(\gamma)$ process
      refer to $|\cos\theta|<0.72$.
      \label{tab:afb}}
  \end{center}
\end{table}

\clearpage

\begin{table}[p] %--------- Selected Events and Cross Sections -----------------
\begin{center}
\begin{tabular}{|c|c|c|c|c|c|}
\hline
$|\cos\theta^\star|$ &  $N_{\rm data}$ &  $f_{\rm back}$ ($\%$) &    
$\varepsilon$ ($\%$) &  $\frac{\rm d\sigma}{\rm d|\cos\theta^\star|}$ (pb) & ${\frac{\rm d\sigma}{\rm d|\cos\theta^\star|}}_
{\rm SM}$ (pb)\\
\hline
\hline
\multicolumn{6}{|c|}{ $\sqrt{s} = 192 \GeV$} \\
\hline
$ 0.00- 0.09$ & $   \phantom{00}24$ &  $   6.7\pm  4.9$ & $  97.6\pm  1.4$ & $   \phantom{00}9.3 \pm    \phantom{0}1.9 \pm   \phantom{0}0.2$ & $  \phantom{0}10.1$  \\
$ 0.09- 0.18$ & $   \phantom{00}26$ &  $   6.8\pm  4.8$ & $  96.4\pm  1.6$ & $  \phantom{0}10.2 \pm    \phantom{0}2.0 \pm    \phantom{0}0.2$ & $  \phantom{0}11.0$  \\
$ 0.18- 0.27$ & $   \phantom{00}30$ &  $   6.2\pm  4.0$ & $  99.4\pm  0.6$ & $  \phantom{0}11.5 \pm    \phantom{0}2.1 \pm    \phantom{0}0.1$ & $  \phantom{0}13.0$  \\
$ 0.27- 0.36$ & $   \phantom{00}37$ &  $   6.5\pm  3.9$ & $  97.9\pm  1.1$ & $  \phantom{0}14.3 \pm    \phantom{0}2.4 \pm    \phantom{0}0.2$ & $  \phantom{0}16.6$  \\
$ 0.36- 0.45$ & $   \phantom{00}60$ &  $   4.8\pm  2.7$ & $  98.0\pm  0.8$ & $  \phantom{0}23.6 \pm    \phantom{0}3.0 \pm    \phantom{0}0.3$ & $  \phantom{0}22.9$  \\
$ 0.45- 0.54$ & $   \phantom{00}78$ &  $   3.7\pm  2.0$ & $  98.4\pm  0.6$ & $  \phantom{0}30.9 \pm    \phantom{0}3.5 \pm    \phantom{0}0.2$ & $  \phantom{0}34.2$  \\
$ 0.54- 0.63$ & $   \phantom{0}151$ &  $   2.4\pm  1.3$ & $  97.6\pm  0.6$ & $  \phantom{0}61.1 \pm    \phantom{0}5.0 \pm    \phantom{0}0.4$ & $  \phantom{0}55.9$  \\
$ 0.63- 0.72$ & $   \phantom{0}220$ &  $   1.3\pm  0.8$ & $  80.3\pm  1.1$ & $            109.4 \pm    \phantom{0}7.4 \pm    \phantom{0}1.7$ & $            102.6$  \\
$ 0.72- 0.81$ & $   \phantom{00}25$  & $   3.1\pm  3.3$ & $\phantom{0}5.0\pm  0.4$ & $            196.4 \pm              39.3 \pm              16.2$ & $            225.1$  \\
$ 0.81- 0.90$ & $             1325$  & $   2.0\pm  0.4$ & $  73.0\pm  0.5$ & $            720.4 \pm              19.8 \pm    \phantom{0}6.0$ & $            713.5$  \\
\hline
\hline
\multicolumn{6}{|c|}{ $\sqrt{s} = 196 \GeV$} \\
\hline
$ 0.00- 0.09$  & $    \phantom{00}69$ & $   8.0\pm  3.0$ & $  99.1\pm  0.6$ & $   \phantom{00}8.6 \pm   \phantom{0}1.0 \pm   \phantom{0}0.1$ & $   \phantom{00}9.6$  \\
$ 0.09- 0.18$  & $    \phantom{00}82$ & $   7.4\pm  2.7$ & $  97.1\pm  1.1$ &$  \phantom{0}10.5 \pm    \phantom{0}1.2 \pm    \phantom{0}0.2$ & $  \phantom{0}10.5$  \\
$ 0.18- 0.27$  & $    \phantom{00}94$ & $   6.2\pm  2.5$ & $  98.9\pm  0.6$ &$  \phantom{0}12.0 \pm    \phantom{0}1.2 \pm    \phantom{0}0.1$ & $  \phantom{0}12.4$  \\
$ 0.27- 0.36$  & $    \phantom{0}138$ & $   5.0\pm  2.0$ & $  97.7\pm  0.8$ &$  \phantom{0}18.0 \pm    \phantom{0}1.5 \pm    \phantom{0}0.2$ & $  \phantom{0}15.9$  \\
$ 0.36- 0.45$  & $    \phantom{0}159$ & $   4.6\pm  1.6$ & $  98.7\pm  0.5$ &$  \phantom{0}20.6 \pm    \phantom{0}1.6 \pm    \phantom{0}0.2$ & $  \phantom{0}21.9$  \\
$ 0.45- 0.54$  & $    \phantom{0}247$ & $   3.4\pm  1.1$ & $  99.0\pm  0.4$ &$  \phantom{0}32.4 \pm    \phantom{0}2.1 \pm    \phantom{0}0.1$ & $  \phantom{0}32.8$  \\
$ 0.54- 0.63$  & $    \phantom{0}380$ & $   2.3\pm  0.8$ & $  97.4\pm  0.5$ &$  \phantom{0}51.2 \pm    \phantom{0}2.6 \pm    \phantom{0}0.3$ & $  \phantom{0}53.6$  \\
$ 0.63- 0.72$  & $    \phantom{0}616$ & $   1.0\pm  0.4$ & $  82.5\pm  0.8$ &$  \phantom{0}99.3 \pm    \phantom{0}4.0 \pm    \phantom{0}1.2$ & $  \phantom{0}98.5$  \\
$ 0.72- 0.81$  & $    \phantom{00}80$ & $   3.0\pm  1.9$ & $   \phantom{0}4.9\pm  0.3$ & $            211.2 \pm              23.6 \pm              13.9$ & $            216.2$  \\
$ 0.81- 0.90$  & $              3812$ & $   2.0\pm  0.2$ & $  72.7\pm  0.4$ &$            690.4 \pm              11.2 \pm    \phantom{0}5.0$ &    $         685.1$  \\
\hline
\hline
\multicolumn{6}{|c|}{ $\sqrt{s} = 200 \GeV$} \\
\hline
$ 0.00- 0.09$  & $    \phantom{00}83$  & $           11.1\pm  3.7$ & $  96.3\pm  1.3$ & $  \phantom{0}10.3 \pm   \phantom{0}1.1 \pm    \phantom{0}0.2$ & $  \phantom{00}9.2$  \\
$ 0.09- 0.18$  & $    \phantom{00}78$  & $ \phantom{0}9.7\pm  3.4$ & $  96.8\pm  1.2$ & $   \phantom{00}9.8 \pm   \phantom{0} 1.1 \pm    \phantom{0}0.2$ & $  \phantom{0}10.0$  \\
$ 0.18- 0.27$  & $    \phantom{00}93$  & $ \phantom{0}5.2\pm  2.3$ & $  97.0\pm  1.0$ & $  \phantom{0}12.2 \pm    \phantom{0}1.3 \pm    \phantom{0}0.2$ & $  \phantom{0}11.9$  \\
$ 0.27- 0.36$  & $   \phantom{0}111$  & $  \phantom{0}5.7\pm  2.2$ & $  95.7\pm  1.0$ & $  \phantom{0}14.7 \pm    \phantom{0}1.4 \pm    \phantom{0}0.2$ & $  \phantom{0}15.2$  \\
$ 0.36- 0.45$  & $   \phantom{0}154$  & $  \phantom{0}4.9\pm  1.7$ & $  98.2\pm  0.6$ & $  \phantom{0}20.0 \pm    \phantom{0}1.6 \pm    \phantom{0}0.2$ & $  \phantom{0}21.0$  \\
$ 0.45- 0.54$  & $   \phantom{0}211$  & $  \phantom{0}3.8\pm  1.2$ & $  97.3\pm  0.6$ & $  \phantom{0}28.0 \pm    \phantom{0}1.9 \pm    \phantom{0}0.2$ & $  \phantom{0}31.5$  \\
$ 0.54- 0.63$  & $   \phantom{0}355$  & $  \phantom{0}2.1\pm  0.7$ & $  94.6\pm  0.6$ & $  \phantom{0}49.4 \pm    \phantom{0}2.6 \pm    \phantom{0}0.3$ & $  \phantom{0}51.5$  \\
$ 0.63- 0.72$  & $   \phantom{0}588$  & $  \phantom{0}1.0\pm  0.4$ & $  79.2\pm  0.8$ & $  \phantom{0}98.9 \pm    \phantom{0}4.1 \pm    \phantom{0}1.3$ & $  \phantom{0}94.6$  \\
$ 0.72- 0.81$  & $    \phantom{00}74$ & $  \phantom{0}5.8\pm  2.9$ & $   \phantom{0}4.1\pm  0.3$ & $ 231.3 \pm   26.9 \pm   16.1$ & $ 207.7$  \\
$ 0.81- 0.90$  & $              3635$ & $  \phantom{0}2.1\pm  0.2$ & $  71.4\pm  0.4$ & $ 670.3 \pm   11.1 \pm    \phantom{0}4.9$ & $ 658.4$  \\
\hline
\end{tabular}
      \caption[]{Differential cross section for the $\epem\ra\epem(\gamma)$ process,
	$\rm d\sigma/d|cos\theta^\star|$, as a function of the absolute value of the
	scattering angle, $|\cos\theta^\star|$. The first uncertainty is statistical and
	the second systematic.  The numbers of observed events, $N_{\rm
	  data}$, and the background fractions, $f_{\rm back}$, are also given,
	together with the selection efficiency, $\varepsilon$. Both $f_{\rm
	  back}$ and $\varepsilon$ are in \%. The Standard Model predictions,
	as computed with the BHWIDE Monte Carlo program, are also given. Only
	high-energy events with $\zeta<25^\circ$ are considered.
	\label{tab:bhabha1}}
\end{center}
\end{table}

\clearpage

\begin{table}[p] %--------- Selected Events and Cross Sections -----------------
\begin{center}
\begin{tabular}{|c|c|c|c|c|c|}
\hline
$|\cos\theta^\star|$ &  $N_{\rm data}$ &  $f_{\rm back}$ ($\%$) &    
$\varepsilon$ ($\%$) &  $\frac{\rm d\sigma}{\rm d|\cos\theta^\star|}$ (pb) & ${\frac{\rm d\sigma}{\rm d|\cos\theta^\star|}}_
{\rm SM}$ (pb)\\
\hline
\hline
\multicolumn{6}{|c|}{ $\sqrt{s} = 202 \GeV$} \\
\hline
$ 0.00- 0.09$ & $  \phantom{00}40$  & $   8.6\pm  4.7$ & $           100.0\pm  0.0$ & $  \phantom{0}11.0 \pm    \phantom{0}1.7 \pm    \phantom{0}0.1$ & $  \phantom{00}9.0$  \\
$ 0.09- 0.18$ & $  \phantom{00}40$  & $   8.7\pm  4.9$ & $ \phantom{0}99.4\pm  0.6$ & $  \phantom{0}11.0 \pm    \phantom{0}1.7 \pm    \phantom{0}0.2$ & $  \phantom{00}9.8$  \\
$ 0.18- 0.27$ & $  \phantom{00}42$  & $   6.3\pm  3.8$ & $ \phantom{0}99.0\pm  0.7$ & $  \phantom{0}11.9 \pm    \phantom{0}1.8 \pm    \phantom{0}0.1$ & $  \phantom{0}11.6$  \\
$ 0.27- 0.36$ & $  \phantom{00}51$  & $   6.1\pm  3.3$ & $ \phantom{0}97.1\pm  1.0$ & $  \phantom{0}14.8 \pm    \phantom{0}2.1 \pm    \phantom{0}0.2$ & $  \phantom{0}14.9$  \\
$ 0.36- 0.45$ & $  \phantom{00}73$  & $   4.6\pm  2.5$ & $ \phantom{0}98.6\pm  0.6$ & $  \phantom{0}21.2 \pm    \phantom{0}2.5 \pm    \phantom{0}0.2$ & $  \phantom{0}20.6$  \\
$ 0.45- 0.54$ & $  \phantom{0}126$  & $   3.7\pm  1.8$ & $ \phantom{0}98.0\pm  0.6$ & $  \phantom{0}37.2 \pm    \phantom{0}3.3 \pm    \phantom{0}0.2$ & $  \phantom{0}30.9$  \\
$ 0.54- 0.63$ & $  \phantom{0}185$  & $   2.2\pm  1.1$ & $ \phantom{0}98.0\pm  0.5$ & $  \phantom{0}55.4 \pm    \phantom{0}4.1 \pm    \phantom{0}0.3$ & $  \phantom{0}50.5$    \\
$ 0.63- 0.72$ & $  \phantom{0}255$  & $   1.1\pm  0.6$ & $ \phantom{0}82.9\pm  0.9$ & $  \phantom{0}91.4 \pm    \phantom{0}5.7 \pm    \phantom{0}1.2$ & $  \phantom{0}92.7$   \\
$ 0.72- 0.81$ & $  \phantom{00}39$  & $   3.4\pm  3.2$ & $ \phantom{00}4.6\pm  0.4$ & $            243.7 \pm              39.0 \pm              18.5$ & $            203.7$     \\
$ 0.81- 0.90$ & $  1528$ & $   2.2\pm  0.4$ & $  72.7\pm  0.4$ & $            618.3 \pm              15.8 \pm    \phantom{0}4.8$ & $            645.7$   \\
\hline
\hline
\multicolumn{6}{|c|}{ $\sqrt{s} = 205 \GeV$} \\
\hline
$ 0.00- 0.09$ & $  \phantom{00}54$ &  $   6.5\pm  3.2$ & $  \phantom{0}96.3\pm  1.1$ &$  \phantom{00}8.7 \pm    \phantom{0}1.2 \pm    \phantom{0}0.1$ & $  \phantom{00}8.8$  \\
$ 0.09- 0.18$ & $  \phantom{00}84$ &  $   9.3\pm  3.8$ & $  \phantom{0}98.0\pm  0.9$ &$  \phantom{0}12.9 \pm    \phantom{0}1.4 \pm    \phantom{0}0.3$ & $  \phantom{00}9.6$  \\
$ 0.18- 0.27$ & $  \phantom{00}78$ &  $   5.8\pm  2.7$ & $  \phantom{0}99.1\pm  0.5$ &$  \phantom{0}12.3 \pm    \phantom{0}1.4 \pm    \phantom{0}0.1$ & $  \phantom{0}11.4$  \\
$ 0.27- 0.36$ & $  \phantom{0}103$ &  $   6.9\pm  2.6$ & $  \phantom{0}98.7\pm  0.6$ &$  \phantom{0}16.1 \pm    \phantom{0}1.6 \pm    \phantom{0}0.1$ & $  \phantom{0}14.6$  \\
$ 0.36- 0.45$ & $  \phantom{0}124$ &  $   4.1\pm  1.8$ & $  \phantom{0}99.0\pm  0.4$ &$  \phantom{0}20.0 \pm    \phantom{0}1.8 \pm    \phantom{0}0.2$ & $  \phantom{0}20.2$  \\
$ 0.45- 0.54$ & $  \phantom{0}193$ &  $   3.5\pm  1.4$ & $  \phantom{0}97.5\pm  0.6$ &$  \phantom{0}31.7 \pm    \phantom{0}2.3 \pm    \phantom{0}0.2$ & $  \phantom{0}30.2$  \\
$ 0.54- 0.63$ & $  \phantom{0}286$ &  $   2.5\pm  0.9$ & $  \phantom{0}96.6\pm  0.5$ &$  \phantom{0}48.0 \pm    \phantom{0}2.8 \pm    \phantom{0}0.3$ & $  \phantom{0}49.5$  \\
$ 0.63- 0.72$ & $  \phantom{0}462$ &  $   1.3\pm  0.5$ & $  \phantom{0}81.2\pm  0.8$ &$  \phantom{0}93.3 \pm    \phantom{0}4.3 \pm    \phantom{0}1.1$ & $  \phantom{0}90.9$  \\
$ 0.72- 0.81$ & $  \phantom{00}74$ &  $   2.6\pm  2.1$ & $  \phantom{00}4.7\pm  0.3$ &$            252.2 \pm              29.3 \pm            15.3$ & $            199.7$  \\
$ 0.81- 0.90$ & $            2801$ &  $   2.0\pm  0.3$ & $  \phantom{0}72.5\pm  0.3$ &$            628.7 \pm              11.9 \pm    \phantom{0}4.4$ & $            633.3$  \\
\hline
\hline
\multicolumn{6}{|c|}{ $\sqrt{s} = 207 \GeV$} \\
\hline
$ 0.00- 0.09$ & $  \phantom{0}107$ &  $   8.0\pm  2.9$ & $  \phantom{0}99.1\pm  0.9$ &$  \phantom{00}9.0 \pm    \phantom{0}0.9 \pm    \phantom{0}0.1$ & $  \phantom{00}8.6$  \\
$ 0.09- 0.18$ & $  \phantom{0}105$ &  $   9.6\pm  3.0$ & $  \phantom{0}98.3\pm  1.2$ &$  \phantom{00}8.7 \pm    \phantom{0}0.9 \pm    \phantom{0}0.2$ & $  \phantom{00}9.4$  \\
$ 0.18- 0.27$ & $  \phantom{0}120$ &  $   6.4\pm  2.1$ & $  \phantom{0}97.5\pm  1.2$ &$  \phantom{0}10.4 \pm    \phantom{0}1.0 \pm    \phantom{0}0.2$ & $  \phantom{0}11.2$  \\
$ 0.27- 0.36$ & $  \phantom{0}194$ &  $   6.1\pm  1.7$ & $  \phantom{0}98.3\pm  0.9$ &$  \phantom{0}16.8 \pm    \phantom{0}1.2 \pm    \phantom{0}0.2$ & $  \phantom{0}14.3$  \\
$ 0.36- 0.45$ & $  \phantom{0}261$ &  $   4.4\pm  1.4$ & $  \phantom{0}97.8\pm  0.9$ &$  \phantom{0}23.1 \pm    \phantom{0}1.4 \pm    \phantom{0}0.3$ & $  \phantom{0}19.8$  \\
$ 0.45- 0.54$ & $  \phantom{0}330$ &  $   3.6\pm  1.0$ & $  \phantom{0}98.0\pm  0.7$ &$  \phantom{0}29.4 \pm    \phantom{0}1.6 \pm    \phantom{0}0.2$ & $  \phantom{0}29.7$  \\
$ 0.54- 0.63$ & $  \phantom{0}495$ &  $   2.5\pm  0.7$ & $  \phantom{0}98.3\pm  0.5$ &$  \phantom{0}44.5 \pm    \phantom{0}2.0 \pm    \phantom{0}0.2$ & $  \phantom{0}48.5$  \\
$ 0.63- 0.72$ & $  \phantom{0}812$ &  $   1.2\pm  0.4$ & $  \phantom{0}80.8\pm  1.1$ &$  \phantom{0}90.0 \pm    \phantom{0}3.2 \pm    \phantom{0}1.4$ & $  \phantom{0}89.2$  \\
$ 0.72- 0.81$ & $  \phantom{00}94$ &  $   1.9\pm  1.3$ & $  \phantom{00}4.9\pm  0.4$ &$            170.0 \pm              17.5 \pm              14.4$ & $            195.9$  \\
$ 0.81- 0.90$ & $            4871$ &  $   1.8\pm  0.2$ & $  \phantom{0}72.1\pm  0.5$ &$            604.4 \pm    \phantom{0}8.7 \pm    \phantom{0}5.1$ & $            621.2$  \\
\hline
\hline
\multicolumn{6}{|c|}{ $\sqrt{s} = 208\GeV$} \\
\hline
$ 0.00- 0.09$  & $    \phantom{000}3$ &  $   8.0\pm 11.3$ & $  \phantom{0}99.1\pm  0.9$ &$  \phantom{00}3.9 \pm    \phantom{0}2.3 \pm    \phantom{0}0.1$ & $   \phantom{00}8.5$  \\
$ 0.09- 0.18$  & $    \phantom{000}8$ &  $   9.6\pm 11.6$ & $  \phantom{0}98.3\pm  1.2$ &$  \phantom{0}10.4 \pm    \phantom{0}3.7 \pm    \phantom{0}0.2$ & $   \phantom{00}9.2$  \\
$ 0.18- 0.27$  & $    \phantom{000}4$ &  $   6.3\pm  8.4$ & $  \phantom{0}97.5\pm  1.2$ &$  \phantom{00}5.4 \pm    \phantom{0}2.7 \pm    \phantom{0}0.1$ & $  \phantom{0}10.9$  \\
$ 0.27- 0.36$  & $    \phantom{00}10$ &  $   6.1\pm  6.9$ & $  \phantom{0}98.3\pm  0.9$ &$  \phantom{0}13.4 \pm    \phantom{0}4.2 \pm    \phantom{0}0.1$ & $  \phantom{0}14.0$  \\
$ 0.36- 0.45$  & $    \phantom{00}17$ &  $   4.4\pm  5.5$ & $  \phantom{0}97.8\pm  0.9$ &$  \phantom{0}23.4 \pm    \phantom{0}5.7 \pm    \phantom{0}0.3$ & $  \phantom{0}19.4$  \\
$ 0.45- 0.54$  & $    \phantom{00}19$ &  $   3.6\pm  4.1$ & $  \phantom{0}98.0\pm  0.7$ &$  \phantom{0}26.3 \pm    \phantom{0}6.0 \pm    \phantom{0}0.2$ & $  \phantom{0}29.1$  \\
$ 0.54- 0.63$  & $    \phantom{00}27$ &  $   2.5\pm  2.7$ & $  \phantom{0}98.3\pm  0.5$ &$  \phantom{0}37.6 \pm    \phantom{0}7.2 \pm    \phantom{0}0.2$ & $  \phantom{0}47.6$  \\
$ 0.63- 0.72$  & $    \phantom{00}49$ &  $   1.2\pm  1.5$ & $  \phantom{0}80.8\pm  1.1$ &$  \phantom{0}84.3 \pm              12.0 \pm    \phantom{0}1.3$ & $  \phantom{0}87.5$  \\
$ 0.72- 0.81$  & $    \phantom{00}10$ &  $   1.9\pm  5.2$ & $  \phantom{00}4.9\pm  0.4$ &$            280.3 \pm              88.7 \pm              23.8$ & $            192.2$  \\
$ 0.81- 0.90$  & $    \phantom{0}294$ &  $   1.9\pm  0.8$ & $   \phantom{0}72.1\pm  0.5$ &$            564.8 \pm              32.9 \pm    \phantom{0}4.8$ & $            609.5$  \\
\hline
\end{tabular}
      \caption{Differential cross section for the $\epem\ra\epem(\gamma)$ process, continued from Table~\protect{\ref{tab:bhabha1}}
      \label{tab:bhabha2}}
\end{center}
\end{table}

%%%%%%%%%%%%%%%%%%%%%%%%%%%%%%%%%%%%%%%%%%%%%%%%%%%%%%%%%%%%%%%%%%%%%%%%%%%%%%%%
%                                   FIGURES
%%%%%%%%%%%%%%%%%%%%%%%%%%%%%%%%%%%%%%%%%%%%%%%%%%%%%%%%%%%%%%%%%%%%%%%%%%%%%%%%
\clearpage

\end{document}